\documentclass[aps,prc,amsfonts,preprintnumbers,superscriptaddress,showpacs,nofootinbib]{revtex4}
\usepackage{epsfig}
\usepackage{epstopdf}
\usepackage{amssymb,amsmath}
\usepackage{ulem}
\usepackage{mathtools}
\usepackage{color}
\allowdisplaybreaks

\newcommand{\fet}[1]{\mbox{\boldmath $#1$}}

\newcommand{\beq}{\begin{equation}}
\newcommand{\eeq}{\end{equation}}
\newcommand{\beqa}{\begin{eqnarray}}
\newcommand{\eeqa}{\end{eqnarray}}

\newcommand{\ben}{\begin{displaymath}}
\newcommand{\een}{\end{displaymath}}
\newcommand{\be}{\begin{equation}}
\newcommand{\ee}{\end{equation}}
\newcommand{\bea}{\begin{eqnarray}}
\newcommand{\eea}{\end{eqnarray}}
\newcommand{\nn}{\nonumber \\ }

\DeclareMathOperator{\Tr}{Tr}
\begin{document}

\title{Semilocal momentum-space regularized chiral two-nucleon
  potentials \\ up to fifth order}

\author{P.~Reinert}
\email[]{Email: patrick.reinert@rub.de}
\affiliation{Institut f\"ur Theoretische Physik II, Ruhr-Universit\"at Bochum,
 D-44780 Bochum, Germany}
\author{H.~Krebs}
\email[]{Email: hermann.krebs@rub.de}
\affiliation{Institut f\"ur Theoretische Physik II, Ruhr-Universit\"at Bochum,
  D-44780 Bochum, Germany}
\author{E.~Epelbaum}
\email[]{Email: evgeny.epelbaum@rub.de}
\affiliation{Institut f\"ur Theoretische Physik II, Ruhr-Universit\"at Bochum,
  D-44780 Bochum, Germany}

\begin{abstract}
We introduce new semilocal two-nucleon potentials up to fifth order in
the chiral expansion. We employ a simple regularization approach for the
pion-exchange contributions which (i) maintains the
long-range part of the interaction, (ii) is implemented in momentum space
and (iii) can be straightforwardly applied to regularize many-body forces and current
operators. We discuss in detail the two-nucleon contact interactions
at fourth order and demonstrate that three terms out of fifteen used
in previous calculations can be eliminated via suitably chosen 
unitary transformations. The removal of the redundant contact terms
results in a drastic simplification of the fits to scattering data and
leads to interactions which are much softer (i.e.~more perturbative)
than our recent semilocal coordinate-space regularized
potentials.  Using the pion-nucleon low-energy constants from 
matching pion-nucleon Roy-Steiner equations to chiral perturbation
theory,  we perform a comprehensive
analysis of nucleon-nucleon scattering and the deuteron properties up to
fifth chiral order and study the impact of the leading F-wave
two-nucleon contact
interactions which appear at sixth order. The resulting chiral
potentials lead to an outstanding description of the proton-proton
and neutron-proton scattering data from the self-consistent
Granada-2013 database below the  pion production threshold, which is
significantly better than for any other chiral potential. For the first
time, the chiral potentials match in precision and even
outperform the available high-precision phenomenological
potentials, while the number of adjustable parameters is, at the same
time, reduced by about $\sim 40\%$.  Last but not least, we perform a
detailed error analysis and, in particular, quantify for the first
time the statistical uncertainties of the fourth- and the considered sixth-order
contact interactions. 
\end{abstract}


\maketitle

\vspace{-0.2cm}

%%%%%%%%%%%%%%%%%%%%%%%%%%%%%%%%%%%%%%%%%%%%%%%%%%%%%%%%%%%%%%%%%%%%%%%%%%%%%%%%%
\section{Introduction}
\def\theequation{\arabic{section}.\arabic{equation}}
\label{sec:intro}

In recent years, considerable progress has been made towards
developing accurate and precise nuclear forces in the framework of
chiral effective field theory (EFT), see
Refs.~\cite{Epelbaum:2008ga,Machleidt:2011zz} for review articles and
Ref.~\cite{Epelbaum:2012vx} for a pedagogical introduction. In particular, the chiral
expansion of the two-nucleon force was pushed to fifth order (N$^4$LO)
\cite{Entem:2014msa}
using the heavy-baryon formulation with pions and nucleons as the only explicit
degrees of freedom in the effective Lagrangian. Even the  dominant
sixth-order (N$^5$LO) contributions have been worked out in Ref.~\cite{Entem:2015xwa}. In
parallel with these developments, a new generation of chiral
nucleon-nucleon (NN) potentials up to N$^4$LO was introduced by the
Bochum-Bonn  \cite{Epelbaum:2014efa,Epelbaum:2014sza} and
Idaho-Salamanca \cite{Entem:2017gor} groups, see also
Refs.~\cite{Piarulli:2014bda,Carlsson:2015vda,Piarulli:2016vel,Ekstrom:2017koy} 
for related recent studies along similar lines. One important difference between these
potentials concerns the implementation of the regulator(s). It was
argued in Ref.~\cite{Epelbaum:2014efa} that the usage of a local regularization for the
pion-exchange contributions allows one to significantly reduce the amount
of finite-cutoff artefacts. In
\cite{Epelbaum:2014efa,Epelbaum:2014sza}, the one-pion exchange (OPEP)
and two-pion exchange potentials (TPEP) were regularized in coordinate
space via 
\beq
\label{RegCoord}
V_\pi (\vec r \, ) \; \longrightarrow \; V_{\pi,  R} (\vec r \, ) = V_\pi
(\vec r \, ) \left[ 1 - {\rm
    exp} (- r^2/R^2) \right]^n\,,
\eeq
where the cutoff $R$ was chosen in
the range of $R = 0.8 \ldots 1.2~$fm in line with the
estimation of the breakdown distance of the chiral expansion 
of Ref.~\cite{Baru:2012iv}. For the earlier applications of a
similar regularization scheme in the context of nuclear chiral EFT see
Refs.~\cite{Gezerlis:2013ipa,Gezerlis:2014zia}. The exponent $n$ was
set $n=6$, but choosing $n=5$ or $n=7$ was shown in
Ref.~\cite{Epelbaum:2014efa} 
to lead to a comparable description of the phase shifts. For the
contact interactions, a simple Gaussian momentum-space regulator 
${\rm exp} (- (p^2 + {p'} ^2)/\Lambda^2 )$ with the cutoff $\Lambda =
2 R^{-1}$  was used. Here and in what
follows, $\vec p \, '$
and $\vec p$ denote the final and initial momenta of the nucleons in
the center of mass system. On the other
hand, Entem {\it et al.} employ in their recent study \cite{Entem:2017gor} the same
nonlocal momentum-space regulator as used in the first-generation
chiral fourth-order (N$^3$LO) potentials of Refs.~\cite{Epelbaum:2004fk,Entem:2003ft}, namely 
\beq
\label{RegMom}
V ( \vec p \, ', \vec p \,) \; \longrightarrow \; V_\Lambda ( \vec p
\, ',
\vec p \,) = V ( \vec p \, ', \vec p \,) \,  {\rm  exp}\left[ - (p' /
  \Lambda )^{2n} -  (p / \Lambda )^{2n} \right]\,, 
\eeq
for both the long- and short-range contributions. The exponent $n$ is
chosen in such a way that the induced contributions to the potential resulting from the Taylor expansion of
the regulator function appear beyond the considered chiral order
(assuming that the cutoff $\Lambda$ is chosen of the order of the
breakdown scale $\Lambda_{\rm b}$ of the chiral expansion).  Clearly,
being angle-independent, the momentum-space regulator in
Eq.~(\ref{RegMom}) acts in the same multiplicative way in all partial
waves and, therefore,
unavoidably leads to  distortions of phase shifts in channels
with arbitrarily high values of the angular momentum. This indicates
that it affects
the long-range part of the interaction. The induced artefacts are
expected to be
beyond the accuracy of the calculation as long as the cutoff $\Lambda$
is chosen sufficiently large, but they may become an issue if it is lowered
below the breakdown scale $\Lambda_{\rm b}$. In contrast, the 
local regulator in Eq.~(\ref{RegCoord}) does not affect the long-range
interaction as long as $R \lesssim \Lambda_{\rm b}^{-1}$, and the corresponding 
distortions in peripheral NN scattering decrease with increasing
values of the angular momentum (for the case of non-singular
potentials).  
It is also worth mentioning that the nonlocal regulator  in Eq.~(\ref{RegMom}) does not
completely remove unphysical short-distance components of the
chiral TPEP so that an additional spectral function regularization 
had to be employed in Ref.~\cite{Entem:2017gor} along the lines of
Refs.~\cite{Epelbaum:2003gr,Epelbaum:2003xx}. 

The semilocal coordinate-space regularized (SCS) chiral potentials by Epelbaum, Krebs, Mei{\ss}ner (EKM)
developed in Refs.~\cite{Epelbaum:2014efa,Epelbaum:2014sza} are
available for five cutoff values of $R=0.8$, $0.9$, $1.0$, $1.1$ and
$1.2~$fm and were found to provide a very good description of the
phase shifts and mixing angles of the Nijmegen partial wave analysis
(NPWA) \cite{Stoks:1993tb} used to fix the low-energy constants (LECs)
accompanying the NN contact interactions.  In addition, the
significant reduction in the $\chi^2$ per datum for the description of
the Nijmegen neutron-proton (np) and proton-proton (pp) phase shifts
when going from the second chiral order (NLO) to the third one (N$^2$LO) \cite{Epelbaum:2014efa} and from
N$^3$LO to N$^4$LO \cite{Epelbaum:2014sza} provides a clear evidence
of the chiral TPEP, which is completely determined by the spontaneously
broken chiral symmetry of QCD and the experimental/empirical information on the
pion-nucleon ($\pi$N) system. In combination with the algorithm for
quantification of truncation errors formulated in
Ref.~\cite{Epelbaum:2014efa}, that exploits the available
information on the chiral expansion for an observable of interest to
estimate the size of neglected higher-order terms without relying on
cutoff variation, the EKM NN potentials were found to yield promising
results for nucleon-deuteron scattering and the properties of
light nuclei \cite{Binder:2015mbz,Maris:2016wrd}, selected electroweak
processes involving two- and three-nucleon systems
\cite{Skibinski:2016dve,Skibinski:2017vqs} and nuclear matter
properties \cite{Hu:2016nkw}, see also Ref.~\cite{Vorabbi:2017rvk} for a recent
application to derive an optical potential for describing elastic
proton-nucleus scattering. These calculations are based solely on the two-nucleon
potentials and do not provide a complete treatment of the current
operators, so that the next step
is clearly to include the corresponding 
three-nucleon forces
\cite{Weinberg:1991um,vanKolck:1994yi,Epelbaum:2002vt,Bernard:2007sp,Ishikawa:2007zz,Bernard:2011zr}
and exchange current operators
\cite{Kolling:2009iq,Kolling:2011mt,Krebs:2016rqz}.\footnote{Notice
  that the vector and (some parts of the) axial current operators at
  fourth order in the chiral expansion have
  also been worked out by the JLab-Pisa group using time-ordered
  perturbation theory \cite{Pastore:2009is,Piarulli:2012bn,Baroni:2015uza}. Their results disagree with ours for certain
  classes of contributions and correspond to a different
  choice of unitary transformations in the Fock space.} This is, in fact,  
the main goal of the recently formed LENPIC Collaboration, see
Ref.~\cite{Golak:2017jlt} for an overview of the range of activities within
LENPIC. 

While the novel EKM NN potentials lead to very promising results in
all applications considered so far, there is still room for
improvement. First, the contact interactions of the EKM potentials
were determined from fits to the NPWA rather than to the scattering
data which may introduce some model dependence. This also
prevents one from making a direct comparison with other chiral and
phenomenological potentials fitted to the scattering
data in terms of precision. Secondly, truncation errors estimated in 
Refs.~\cite{Epelbaum:2014efa,Epelbaum:2014sza} represent only one source of uncertainties in the
calculations. While they can be assumed to dominate the theoretical
error, see the discussion in Ref.~\cite{Epelbaum:2014efa}, it is important to
quantify and propagate the statistical uncertainties of the LECs
accompanying the NN contact interactions, see
Ref.~\cite{Ekstrom:2014dxa} for a related study at N$^2$LO, and to 
address other sources of uncertainties including the
ones associated with the pion-nucleon LECs and with the
choice of energy range used in the determination of the NN contact
interactions. However, the most pressing issue for the applications of
the EKM potentials beyond the two-nucleon system is the implementation
of a consistent local regularization for many-body forces and
current operators. Here, the main technical obstacle is 
the need to perform the regularization in coordinate space. Given the
rather complicated expressions for certain kinds of contributions to the
three-nucleon force (3NF) at N$^3$LO and beyond and the dependence of
the 3NF on two relative distances, a direct approach by (numerically)
performing the Fourier transforms to coordinate space and, after
regularization, back to
momentum space appears to be challenging. Alternatively,
regularization can also be carried out directly in terms of the
momentum-space matrix elements in the partial wave basis, which can be
obtained for arbitrary 3NF momentum-space expressions using the
methods of Refs.~\cite{Golak:2009ri,Hebeler:2015wxa}, but this
approach requires additional work to 
achieve numerically stable results \cite{Hebeler:InPrep}. Similar
problems occur for the nuclear current operators, where, in addition, one
needs to pay special attention to maintain the corresponding
symmetries upon introducing the regulator.  

The purpose of this paper is to address the issues mentioned above
and, in particular, to remedy the situation with the
(technically) complicated implementation of the regulator in
coordinate space.  To this aim, we introduce a simple local regularization
scheme in momentum space by an appropriate modification of the
propagators of pions exchanged between different nucleons. Our
approach shares some similarities with the higher-derivative
regularization scheme
\cite{Slavnov:1971aw,Djukanovic:2004px,Djukanovic:2007zz,Behrendt:2016nql}
carried out at the Lagrangian level and 
is well suited for applications to many-body forces and electroweak
processes. Moreover, it does not require a recalculation of the TPEP
NN potential as it can be formulated in terms of a particular form of
the spectral function regularization. The important feature of the
regulator is that it does, per construction, not affect the left-hand
singularities in 
the potential due to single or multi-pion exchange  {\it at any
finite order in $1/\Lambda$-expansion}, which is quite different
from the regulator in Eq.~(\ref{RegMom}). As we will demonstrate, this
feature has important phenomenological implications, especially for
soft choices of $\Lambda$.  

Another important insight into the construction of nuclear
forces from our
analysis concerns the contact interactions at and beyond N$^3$LO. 
When performing fits to NN scattering data, we found the convergence
towards a minimum of the $\chi^2$ in the parameter space to be very poor starting from
N$^3$LO. This unpleasant feature is found to be related to the appearance of
three redundant contact operators in the set of fifteen N$^3$LO
contact terms used in
\cite{Epelbaum:2004fk,Entem:2003ft,Epelbaum:2014efa,Epelbaum:2014sza,Entem:2017gor}.\footnote{We
  are grateful to Dick Furnstahl for drawing our 
attention to a possible over-fitting issue in the NN S-waves.} 
This redundancy was mentioned earlier, in particular in the context of
pionless EFT, where these terms were included perturbatively
\cite{Beane:2000fi}. Here, we explicitly construct the
corresponding short-range unitary transformations which can be used to
eliminate the redundant terms from the potential. We demonstrate
upon performing explicit calculations that the redundancy of the contact
interactions remains intact in the non-perturbative regime as
relevant for the case at hand. The elimination of this redundancy
turns out to be crucial for obtaining stable and physically 
meaningful results for the NN LECs when
performing fits to scattering data. 

Building upon these two approaches, we develop in this paper a new
family of semilocal momentum-space regularized (SMS) chiral potentials
up to fifth order in the chiral expansion and perform a comprehensive
partial wave analysis of the neutron-proton and proton-proton
scattering data including uncertainty quantification. 
Our paper is organized as follows. In section \ref{sec:Cont}, we
demonstrate the redundancy of three out of fifteen fourth-order
contact interactions. The impact of their removal on the softness of
the potentials is discussed in section
\ref{Sec:RedundantFits}. Section \ref{sec:reg} is devoted to the
description of our new regularization approach for pion exchange
contributions. Our treatment of the long-range electromagnetic
interactions is detailed in section
\ref{sec2}, followed by the description of the fitting procedure in
section \ref{sec3}. In section \ref{sec4}, we present the results for
phase shifts, deuteron properties and the effective range parameters
based on the new family of semilocal momentum-space regularized chiral
potentials. This section also includes a comprehensive error analysis
for all considered observables. Next, in section  \ref{sec:Redundant},
we discuss an alternative choice of the order-four contact interactions.  
A comparison of our new potentials at
the highest considered chiral order with the modern high-precision
phenomenological and nonlocal chiral EFT potentials is presented in
section \ref{sec:ComparisonPot}, while the main results of our work
are summarized in section \ref{summary}. The appendices provide
further details on the treatment of the short-range part of the
interaction and calculation of scattering observables and show the
neutron-proton and proton-proton phase shifts and mixing angles as
obtained from our analysis. 

\section{Redundant contact interactions at fourth order}
\def\theequation{\arabic{section}.\arabic{equation}}
\label{sec:Cont}

Up to and including fourth order in
the chiral expansion, the short-range potential in the  center--of--mass
system (cms) in the limit of exact isospin symmetry takes the form 
\beq
V_{\rm cont} = V^{(0)}_{\rm cont} + V^{(2)}_{\rm cont} + V^{(4)}_{\rm cont}~, \nn
\eeq
where the individual terms at orders $Q^0$, $Q^2$ and $Q^4$, with $Q
\in \{ p/\Lambda_{\rm b}, \;  M_\pi/\Lambda_{\rm b} \}$ being the
chiral expansion parameter, can be
chosen as 
\beqa 
\label{Vcon}
V^{(0)}_{\rm cont} &=& C_S  + C_T  \vec{\sigma}_1 \cdot  \vec{\sigma}_2\,,\nn
V^{(2)}_{\rm cont} &=& C_1 q^2 + C_2 k^2 +
( C_3  q^2 + C_4  k^2 ) ( \vec{\sigma}_1 \cdot \vec{\sigma}_2)
+  \frac{i}{2} C_5  ( \vec{\sigma}_1 + \vec{\sigma}_2) \cdot ( \vec{k} \times
\vec{q})    + C_6  (\vec{q}\cdot \vec{\sigma}_1 )(\vec{q}\cdot \vec{\sigma}_2 ) 
+ C_7  (\vec{k}\cdot \vec{\sigma}_1 )(\vec{k}\cdot \vec{\sigma}_2 )\,, \nn
V^{(4)}_{\rm cont} &=& D_1  q^4 + D_2  k^4 + D_3 q^2  k^2
+ D_4 \, (\vec q \times \vec k )^2 +  \big( D_5 q^4 + D_6 k^4 + D_7 q^2 k^2 
+ D_8  (\vec q \times \vec k )^2 \big)  (\vec \sigma_1 \cdot \vec \sigma_2 ) \nn
&+&   \frac{i}{2} \big(  D_{9} q^2 + D_{10}  k^2 \big) (\vec \sigma_1
+ \vec \sigma_2 ) \cdot (\vec k \times \vec q ) + \big( D_{11}  q^2 +
D_{12}  k^2 \big) (\vec \sigma_1 \cdot \vec q \, ) 
(\vec \sigma_2 \cdot \vec q \, ) 
+ \big( D_{13}  q^2 + D_{14}  k^2 \big) (\vec \sigma_1 \cdot \vec k ) 
(\vec \sigma_2 \cdot \vec k ) \nn
&+& D_{15} \vec \sigma_1 \cdot ( \vec q \times \vec k ) \, \vec 
\sigma_2 \cdot (\vec q \times \vec k)  \,,
\eeqa
see e.g.~Ref.~\cite{Epelbaum:2004fk} and references
therein. Here,  $\vec q$ denotes the momentum transfer of the nucleon $\vec{q} =
\vec{p}\,'-\vec{p}$ with $\vec{p}$ and $\vec{p}\, '$ 
being the  initial and final nucleon momenta in the cms, Further,   
$\vec k = (\vec p\, ' + \vec p \, )/2$ is the average nucleon
momentum, $\vec \sigma_i$ are the Pauli spin matrices of the nucleon
$i$ while $C_S$, $C_T$, $C_{1, \ldots , 7}$ and $D_{1, \ldots , 15}$ refer to the
corresponding LECs. Here and in what follows, we use the notation $q \equiv
| \vec q \, | $, $k \equiv | \vec k \, | $. Our treatment of
isospin-breaking (IB) effects is the same as in
Refs.~\cite{Epelbaum:2014efa,Epelbaum:2014sza}, and we refer the
reader to these papers for the explicit form of included IB contact
terms. Finally, all
momentum-space expressions for the potentials throughout this 
paper are to be understood as matrix elements with respect to
momenta. 

The choice of the operator basis to represent $V_{\rm
	cont}$ given above is ambiguous. In particular, terms involving
the isospin matrices are not listed as they do not provide additional information by
virtue of the Pauli principle. Alternatively,  quasi-local operator
bases involving isospin matrices are employed in 
Refs.~\cite{Gezerlis:2013ipa,Gezerlis:2014zia,Piarulli:2014bda,Piarulli:2016vel}.  When expressed in the partial wave
basis, the contact interactions have the form
\beqa
\langle i_S, \, p' | V_{\rm cont} | i_S, \, p \rangle  &=& \tilde
C_{i_S} + C_{i_S} (p^2 + p'^2) + D_{i_S}^1 p^2 p'^2 + D_{i_S}^2 (p^4 +
p'^4 )\,, \nn [3pt]
\langle i_P, \, p' | V_{\rm cont} | i_P, \, p \rangle  &=& C_{i_P} p
p' +  D_{i_P} p p' (p^2 + p'^2)\,, \nn [3pt]
\langle i_D, \, p' | V_{\rm cont} | i_D, \, p \rangle  &=&  D_{i_D}
p^2 p'^2 \,, \nn [3pt]
\langle ^3S_1 , \, p' | V_{\rm cont} | ^3D_1 , \, p \rangle  &=&
C_{\epsilon 1} p^2 + D_{\epsilon 1}^1 p^2 p'^2 + D_{\epsilon 1}^2 p^4
\,, \nn [3pt]
\langle ^3P_2 , \, p' | V_{\rm cont} | ^3F_2 , \, p \rangle  &=&
D_{\epsilon 2} p^3 p' \,, 
\eeqa
where $i_S = \big\{1S0, \; 3S1 \big\}$, $i_P = \big\{1P1, \; 3P0 , \;
3P1, \; 3P2 \big\}$ and $i_D = \big\{1D2, \; 3D1 , \;
3D2, \; 3D3 \big\}$. The 24 LECs $\{ \tilde
C_{i_S} \}$, $\{ C_{i_S} , \; C_{i_P} , \; C_{\epsilon 1} \}$ and 
$\{ D_{i_S}^j , \; D_{i_P} , \; D_{i_D} , \; D_{\epsilon 1}^j,  \;
D_{\epsilon 2} \}$ with $j=1,2$ are given by linear combinations of
the LECs $C_S$, $C_T$, $C_{1, \ldots , 7}$ and $D_{1, \ldots , 15}$ in
Eq.~(\ref{Vcon}), whose explicit form can be found e.g.~in Ref.~\cite{Epelbaum:2004fk}.  

Obviously, the $D_{i_S}^1$- and $D_{i_S}^2$-terms as well as
the $D_{\epsilon 1}^1$- and $D_{\epsilon 1}^2$-terms become
indistinguishable on
the energy shell, i.e.~for  $p' = p$, and the corresponding LECs cannot
be disentangled from each other if treated
perturbatively. The appearance of linear combinations formed out
of the LECs in the $^1S_0$, $^3S_1$ and $\epsilon_1$ channels that 
vanish on the energy shell suggests that some of these terms are
redundant. We
now show that this is indeed the case and demonstrate explicitly that
such redundant terms
can be eliminated via suitably chosen unitary transformations. 

Consider unitary transformations (UTs) 
acting on the purely nucleonic subspace of the pion-nucleon Fock
space. When deriving the
nuclear forces and current operators using the method of unitary
transformation \cite{Epelbaum:1998ka,Epelbaum:1999dj} we, in fact, have to employ a broad class of such
transformations constructed out of the vertices from the  effective pion-nucleon
Hamiltonian and the corresponding energy denominators in
order to ensure renormalizability of the resulting nuclear potentials, see
Refs.~\cite{Epelbaum:2007us,Bernard:2007sp,Bernard:2011zr,Krebs:2012yv,Krebs:2013kha,Krebs:2016rqz} 
for more details.  We now consider all possible UTs whose generators
are given by the order-$Q^2$ two-nucleon contact operators. Specifically, we have  
\beq
\label{ShortRangeUT}
U = e^{ \gamma_1 T_1 +  \gamma_2 T_2 +  \gamma_3 T_3}\,, 
\eeq
where $\gamma_i$ are dimensionless transformation angles and the
anti-hermitian, Galilean-invariant generators $T_i$ have the form:
\beqa
T_1 &=& \frac{m_N}{8 \Lambda_{\rm b}^4} \big( p_1'^2 + p_2'^2 - p_1^2 -
p_2^2 \big) = \frac{m_N}{2\Lambda_{\rm b}^4} \vec k \cdot \vec q\,, \nn
T_2 &=& \frac{m_N}{8 \Lambda_{\rm b}^4} \big( p_1'^2 + p_2'^2 - p_1^2 -
p_2^2 \big) \vec \sigma_1 \cdot \vec \sigma_2 =
\frac{m_N}{2\Lambda_{\rm b}^4} \vec k \cdot \vec q \; \vec \sigma_1 \cdot
\vec \sigma_2 \,, \nn
T_3 &=& \frac{m_N}{16 \Lambda_{\rm b}^4} \Big(  \vec \sigma_1 \cdot (
\vec p_1 - \vec p_2 + \vec p_1{}' - \vec p_2{}') \; \vec \sigma_2 \cdot (
\vec p_1{}' - \vec p_2{}' - \vec p_1 + \vec p_2 ) 
+ 
\vec \sigma_1 \cdot (\vec p_1{}' - \vec p_2{}' - \vec p_1 + \vec p_2 )
\; \vec \sigma_2 \cdot ( \vec p_1 - \vec p_2 + \vec p_1{}' - \vec
p_2{}' 
) \Big) \nn
&=& \frac{m_N}{2\Lambda_{\rm b}^4}  \Big(  \vec \sigma_1 \cdot \vec k \;
\vec \sigma_2 \cdot \vec q +  \vec \sigma_1 \cdot \vec q \;
\vec \sigma_2 \cdot \vec k \Big)\,. 
\eeqa
The reason why such unitary transformations have escaped our consideration
when deriving the nuclear forces and currents is that the
generators $T_i$ cannot be written in terms of the vertices entering the
effective pion-nucleon Hamiltonian.  Notice further that the first
non-vanishing anti-hermitian contact interactions appear at order
$Q^2$. Further, we do not consider the generators involving the
isospin Pauli matrices for the reasons already mentioned. 
Finally, the factors of the nucleon mass $m_N$ included in the definition of
the generators ensure that the induced terms in the
nuclear Hamiltonian appear at order $Q^4$ rather than $Q^5$ by virtue
of the employed counting scheme for $m_N$ \cite{Weinberg:1991um}, $m_N \sim \mathcal{O}
(\Lambda_{\rm b}^2/Q)$, while the factors of $\Lambda_{\rm b}^{-4}$ are required
for dimensional reasons. 

When applied to the LO nuclear Hamiltonian, the UTs induce short-range
interactions through
\beq
\label{InducedTerms}
\delta \hat H = \hat U^\dagger \hat H^{(0)} \hat U - \hat H^{(0)} = \sum_i \gamma_i
\, \Big[  \hat H^{(0)} , \;
\hat T_i \Big] + \ldots = \sum_i \gamma_i \, \Big[  \Big( \hat H_{\rm kin}^{(0)}  +
\hat V_{1 \pi}^{(0)} + \hat V_{\rm cont}^{(0)} \Big) , \;
\hat T_i \Big] + \ldots \,,
\eeq
where ellipses refer to terms beyond the accuracy of our
calculations and $\hat X$ means that the quantity $X$ is to be
regarded as an operator rather that matrix element with respect to
momenta of the nucleons. The commutator on the right-hand side gives
rise to two- and three-nucleon operators. Consider first the induced
two-nucleon terms. It is easy to see that the commutator $\big[  \big( 
\hat V_{1 \pi}^{(0)} + \hat V_{\rm cont}^{(0)} \big) , \;
\hat T_i \big]$ generates, after evaluating the corresponding loop
integrals, shifts to the order-$Q^0$ and $Q^2$ contact interactions
and thus needs not to be considered explicitly. On the other hand, the
commutator with the kinetic energy term generates the additional
order-$Q^4$ contact interactions: 
\beq
\big\langle \vec p_1{}' \vec p_2{}' \big|
\Big[ \hat H_{\rm kin}^{(0)} , \; \hat T_i \Big] \big| \vec p_1 \vec p_2 \big\rangle= \gamma_1
\frac{1}{\Lambda_{\rm b}^4} (\vec k \cdot \vec q\, )^2 + \gamma_2
\frac{1}{\Lambda_{\rm b}^4} (\vec k \cdot \vec q\, )^2 \; \vec \sigma_1
\cdot \vec \sigma_2 +  \gamma_3
\frac{1}{\Lambda_{\rm b}^4} \vec k \cdot \vec q\; ( \vec \sigma_1 \cdot
\vec k \; \vec \sigma_2 \cdot \vec q + \vec \sigma_1 \cdot
\vec q \; \vec \sigma_2 \cdot \vec k \, \big)\,.
\eeq 
The $\gamma_1$- and $\gamma_2$-transformations induce the following
shifts in the LECs $D_i$: 
\beq
\delta D_3 = - \delta D_4 = \frac{\gamma_1}{\Lambda_{\rm b}^4}\,, \quad \quad
\delta D_7 = - \delta D_8 = \frac{\gamma_2}{\Lambda_{\rm b}^4}\,.
\eeq
Using the identity 
\beq
\vec k \cdot \vec q \; \big(
\vec \sigma_1 \cdot \vec k \; \vec \sigma_2 \cdot \vec q + \vec
\sigma_1 \cdot \vec q \; \vec \sigma_2 \cdot \vec k \big) = - ( \vec q
\times \vec k \, )^2 \; \vec \sigma_1 \cdot \vec \sigma_2 + q^2 \,
\vec \sigma_1 \cdot \vec k \; \vec \sigma_2 \cdot \vec k + k^2 \,
\vec \sigma_1 \cdot \vec q \; \vec \sigma_2 \cdot \vec q + \vec
\sigma_1 \cdot ( \vec q \times \vec k ) \;\sigma_2 \cdot ( \vec q \times \vec k ) \,,
\eeq
one can read off the shifts in the LECs $D_i$ induced by the $\gamma_3$-transformation:
\beq
-\delta D_8 = \delta D_{12} =  \delta D_{13} = \delta D_{15}  = \frac{\gamma_3}{\Lambda_{\rm b}^4}\,.
\eeq
In the partial wave basis, the induced terms can obviously only affect
off-the-energy-shell linear combinations of the order-$Q^4$ terms,
i.e.~the structures proportional to $p'^2 - p^2$. Introducing the
corresponding linear combinations in the $^1S_0$, $^3S_1$ and $\epsilon_1$
channels e.g.~via
\beqa
\label{ConventionOffShell1}
D_{1S0}^1\,  p^2 p'^2 + D_{1S0}^2 \, (p^4 + p'^4 ) &=:& D_{1S0} \, p^2 p'^2 +
D_{1S0}^{\rm off}\,  (p'^2 - p^2)^2\,,\nn
D_{3S1}^1\,  p^2 p'^2 + D_{3S1}^2 \, (p^4 + p'^4 ) &=:& D_{3S1} \, p^2 p'^2 +
D_{3S1}^{\rm off}\,  (p'^2 - p^2)^2\,, \nn
D_{\epsilon 1}^1\,  p^2 p'^2 + D_{\epsilon 1}^2 \, p^4 &=:&
D_{\epsilon 1} \, p^2 p'^2 +
D_{\epsilon 1}^{\rm off}\,  p^2 (p'^2 - p^2)\,,
\eeqa
we adopt a {\it convention} by requiring the UTs to be chosen in such
a way that 
\beq
\label{ConventionOffShell2}
D_{1S0}^{\rm off} = D_{3S1}^{\rm off} = D_{\epsilon 1}^{\rm off} = 0\,.
\eeq
This corresponds to choosing the phases of the UTs as
\beqa
\frac{\gamma_1}{\Lambda_{\rm b}^4} &=& - 4 D_1 - \frac{1}{4} D_2 - D_3\,,\nn
\frac{\gamma_2}{\Lambda_{\rm b}^4} &=& - 4 D_5 - \frac{1}{4} D_6 - D_7\,, \nn
\frac{\gamma_3}{\Lambda_{\rm b}^4} &=& -\frac{1}{8} \big( 16 D_{11} + 4 D_{12}
+ 4 D_{13} + D_{14} \big)  \,.
\eeqa
For the sake of completeness, we give the form of the 
contact potential $V_{\rm cont}^{(4)}$ after eliminating the redundant
terms, subject to the adopted convention specified in
Eqs.~(\ref{ConventionOffShell1}), (\ref{ConventionOffShell2}),
and after renaming the LECs:
\beqa
\label{ContQ4Min}
V^{(4)}_{\rm cont} &=& D_1' \left( q^4 - 4 (\vec k \cdot \vec q \, )^2
\right) + D_2' \left( k^4 - \frac{1}{4} (\vec k \cdot \vec q \, )^2 
\right) +
D_3' (\vec q \times \vec k)^2 \nn
&+& \left(   D_4' \left( q^4 - 4 (\vec k \cdot \vec q \, )^2
\right) + D_5' \left( k^4 - \frac{1}{4} (\vec k \cdot \vec q \, )^2 
\right) +
D_6' (\vec q \times \vec k)^2  \right) \vec \sigma_1 \cdot \vec
\sigma_2 +  \frac{i}{2} \big(  D_{7}' q^2 + D_{8}'  k^2 \big) (\vec \sigma_1
+ \vec \sigma_2 ) \cdot (\vec k \times \vec q ) \nn
&+& D_9' \left( - \frac{1}{4} q^2 \vec \sigma_1 \cdot \vec q \; \vec
\sigma_2 \cdot \vec q + 4 k^2 \vec \sigma_1 \cdot \vec k \; \vec
\sigma_2 \cdot \vec k \right) + D_{10}' k^2 \left(\vec \sigma_1 \cdot \vec q \; \vec
\sigma_2 \cdot \vec q - 4 \vec \sigma_1 \cdot \vec k \; \vec
\sigma_2 \cdot \vec k \right) 
+D_{11}'  (q^2 - 4 k^2) \vec \sigma_1 \cdot \vec k \; \vec
\sigma_2 \cdot \vec k 
\nn
&+& D_{12}' \vec \sigma_1 \cdot ( \vec q \times \vec k ) \, \vec 
\sigma_2 \cdot (\vec q \times \vec k) \,.
\eeqa
The relations between 21 LECs $C_{S,T}$, $C_{1, \ldots , 7}$, $D_{1
	, \ldots 12}'$ and the ones in the spectroscopic basis  $\{ \tilde
C_{i_S} \}$, $\{ C_{i_S} , \, C_{i_P} , \, C_{\epsilon 1} \}$ and 
$\{ D_{i_S} , \, D_{i_P} , \, D_{i_D} , \, D_{\epsilon 1},  \,
D_{\epsilon 2} \}$ are provided in appendix \ref{appCont}. 

In addition to the two-nucleon potential, the commutator of $\hat V_{1
	\pi}^{(0)}$ and $\hat V_{\rm cont}^{(0)}$ with the generators $\hat
T_i$ in Eq.~(\ref{InducedTerms}) gives rise to the induced
three-nucleon forces. 
It is easy to see that the generated terms have the same form
as the N$^4$LO contributions to the one-pion-exchange-contact and
purely contact-type 3NFs, albeit with the numerical coefficients
enhanced by one power of $Q^{-1}$ due to the appearance of the factor
of $m_N/\Lambda_{\rm b}$. For example, 
\beq
\big \langle \vec p_1{}' \vec p_2{}' \vec p_3{}' \big| \Big[ ( C_S + C_T \vec \sigma_1 \cdot \vec \sigma_2 ) , \;
\hat T_1 \Big] \big| \vec p_1 \vec p_2 \vec p_3
\big\rangle=  \frac{m_N}{4 \Lambda_{\rm b}^4} q_3^2 (C_S + C_T \vec \sigma_1
\cdot \vec \sigma_2)\,,
\eeq  
where $q_3 \equiv | \vec q_3 | = | \vec p_3{} ' - \vec p_3 | $. Given
that the coupling constants of the corresponding short-range 3NFs are
unknown and have to be determined from experimental data anyway, there
is no need to explicitly keep track of the contributions induced by
the considered UTs. 

To summarize, the considered UTs affect the off-shell behavior of the
two-nucleon force starting from N$^3$LO at short distances and the strength of the
short-range three-nucleon forces (and current operators), but have no
effect on observable quantities. Their appearance is a manifestation 
of the redundancy of the basis of $15$ contact
interactions at N$^3$LO used in all previous calculations at this and
higher chiral orders
\cite{Epelbaum:2004fk,Entem:2003ft,Epelbaum:2014efa,Epelbaum:2014sza,Entem:2017gor}.\footnote{
	In Ref.~\cite{Piarulli:2014bda}, minimally nonlocal potentials
	involving TPEP with and without explicit $\Delta$(1232) degrees of
	freedom up to N$^2$LO, accompanied by short-range operators up to N$^3$LO, were
	constructed. The short-range part of the potentials developed in that
	paper also involves 15 isospin-invariant operators at N$^3$LO. More recently, fully
	local versions of these interactions have been constructed, which
	involve 11 isospin-invariant short-range operators at N$^3$LO \cite{Piarulli:2016vel}.  } 
The impact of keeping/removing
the redundant terms when performing the fits will be discussed in section
\ref{Sec:RedundantFits}. Finally, one should keep in mind that the
convention of Eq.~(\ref{ConventionOffShell2}) employed in the present work corresponds to
one particular choice of the unitary transformations or, equivalently,
off-shell behavior of the NN potential. There seems to be no obvious
criterion for finding preferred choices other than that of the naturalness of
the resulting LECs. In principle, one may think of exploiting the freedom in  
the choice of the UTs to e.g.~enforce perturbativeness of the nuclear
potentials in order to make them more suitable for many-body
applications. In section \ref{sec:Redundant} we will discuss an alternative choice
of the basis of independent contact operators at order $Q^4$ which,
however, will be shown to lead to less perturbative potentials.  

We conclude this section by specifying the form of the leading contact
interactions in F-waves whose impact will be discussed in the
following sections. Such
interactions contribute at sixth order in the chiral expansion,
i.e.~at N$^5$LO, and the corresponding partial wave matrix elements
take the form 
\beq
\label{LECsF}
\langle i_F, \, p' | V_{\rm cont} | i_F, \, p \rangle  =  E_{i_F}
p^3 p'^3 \,,
\eeq
where $i_F = \big\{1F3, \; 3F2 , \;
3F3, \; 3F4 \big\}$ and $E_{i_F}$ are the corresponding LECs.

\section{Impact of removing redundant fourth-order contact interactions}
\label{Sec:RedundantFits}

We now discuss the impact of removing redundant contact interactions at
order $Q^4$ for the already published EKM SCS potentials of
Ref.~\cite{Epelbaum:2014sza}. To this aim, we have constructed a
version of the corresponding N$^4$LO potentials with the redundant
contact interactions being removed according to the convention of
Eqs.~(\ref{ConventionOffShell1}), (\ref{ConventionOffShell2}). The
remaining LECs in the $^1$S$_0$ and $^3$S$_1-^3$D$_1$ channels are refitted to
the Nijmegen PWA using the procedure of Ref.~\cite{Epelbaum:2014sza}. 
While the $\chi^2$ for the description of the Nijmegen
phase shifts  $^1$S$_0$ and $^3$S$_1$ and the mixing angle
$\epsilon_1$ does get considerably reduced upon including the redundant contact
interactions, we do not observe any significant improvement in
the description of scattering data. This indicates that the
contributions of the redundant
interactions likely go beyond the accuracy of our
work in agreement with the discussion of section \ref{sec:Cont}.
In section \ref{sec:Redundant}, we will provide further numerical
evidence to support this statement for the case of the SMS potentials. 

We would also like to point out that the removal of the redundant contact interactions
simplifies the data fitting procedure considerably. Including those
interactions results in a very slow 
convergence of the fit once other terms have been adjusted properly and the parameters
have entered the "valley" of the $\chi^2$ surface formed by correlated values of the
parameters, which result in almost equally good $\chi^2$
values. The flatness of the $\chi^2$ function poses a challenge to fitting
algorithms, and it becomes extremely difficult to find a minimum irrespective of
its statistical significance. 
%We were able to perform fits involving the
%redundant interactions within a reasonable time only 
%using Interior Point OPTimizer
%(IPOPT) \cite{IPOPT} and employing the second-derivative information.  

\begin{figure}[tb]
	\vskip 1 true cm
	\begin{center}
		\includegraphics[width=0.85\textwidth,keepaspectratio,angle=0,clip]{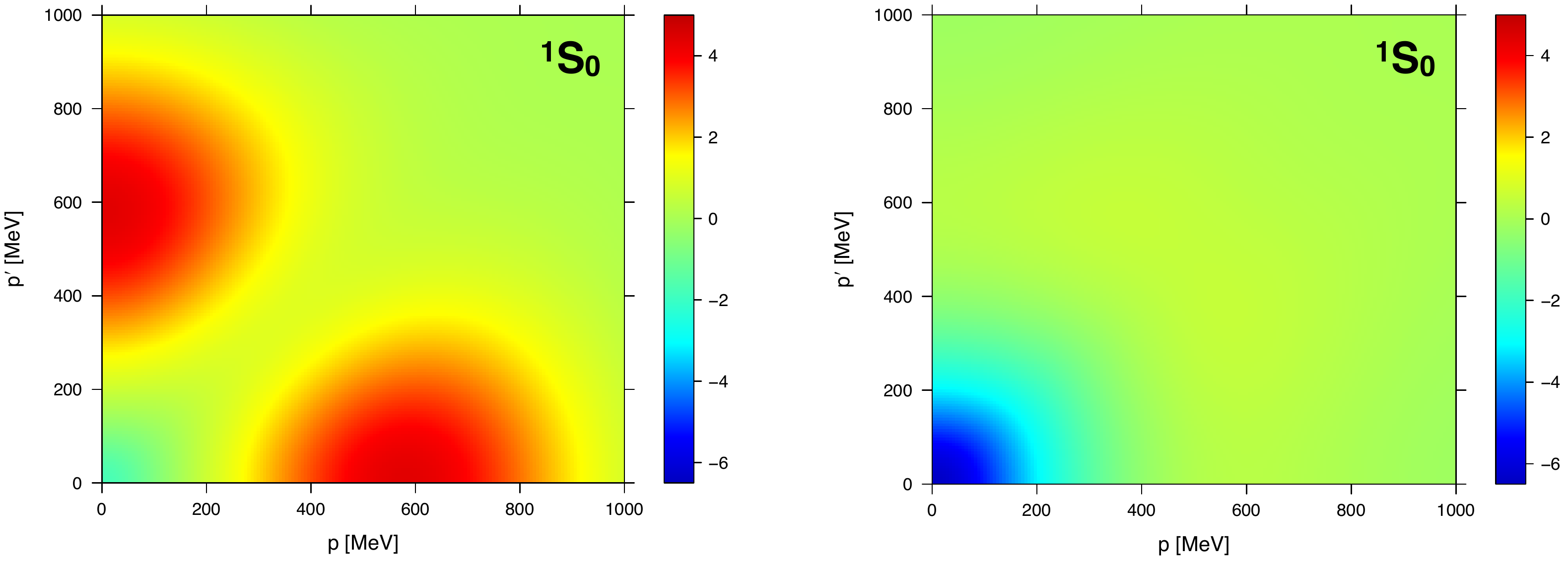}
		\includegraphics[width=0.85\textwidth,keepaspectratio,angle=0,clip]{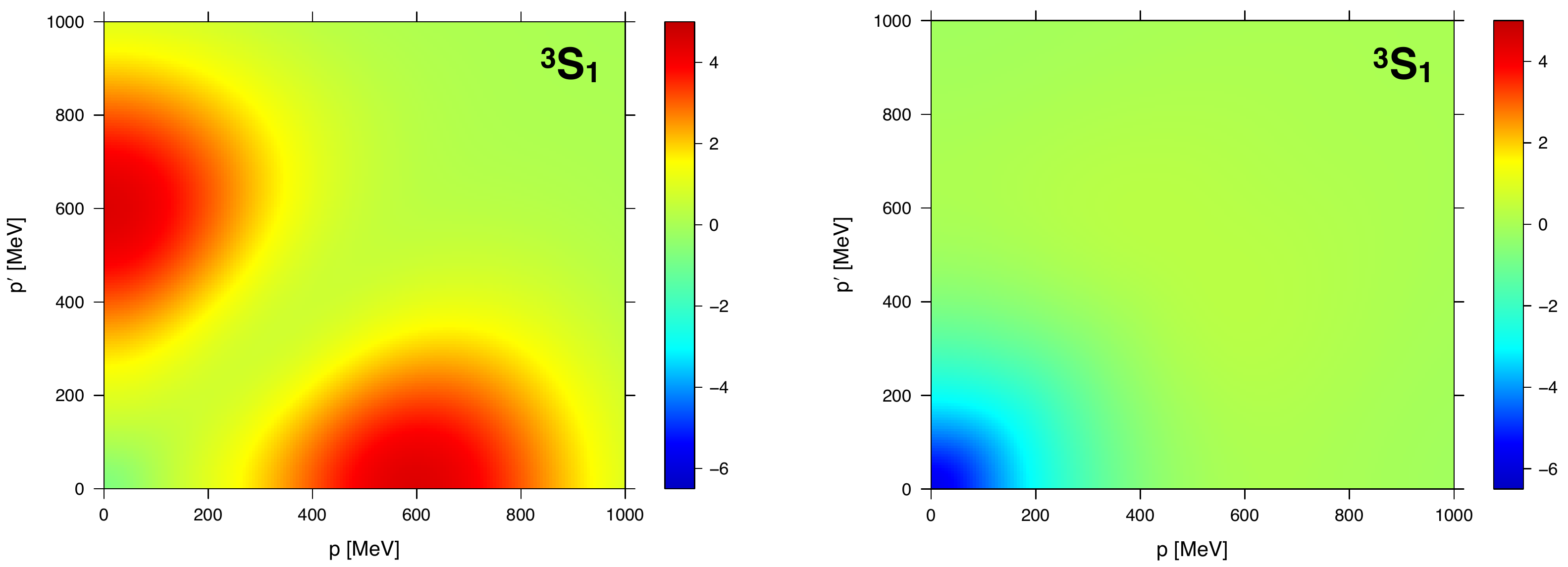}
		\includegraphics[width=0.85\textwidth,keepaspectratio,angle=0,clip]{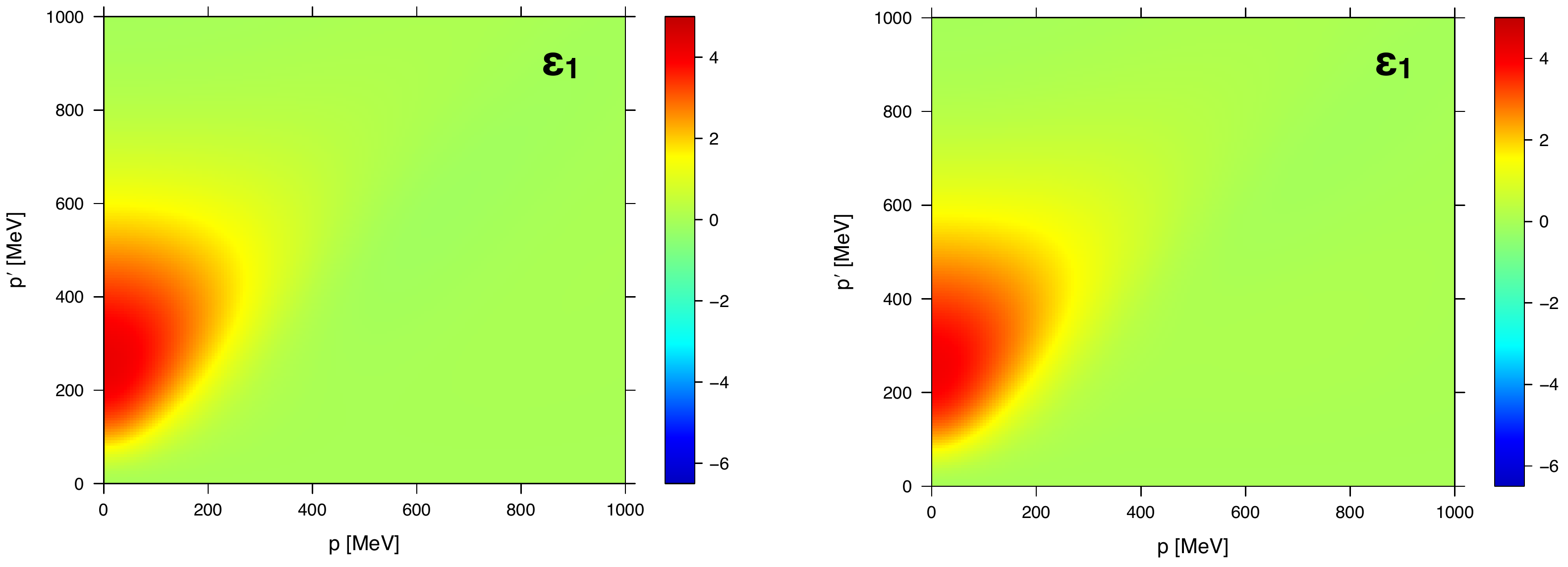}
	\end{center}
	\caption{(Color online) Left panel: Matrix elements of the N$^4$LO potential
		of Ref.~\cite{Epelbaum:2014sza} in the $^1$S$_0$, $^3$S$_1$ and
		$^3$S$_1$-$^3$D$_1$ channels in units of GeV$^{-2}$. Right panel: the corresponding
		matrix elements after eliminating the redundant contact
		interactions by setting $D_{1S0}^{\rm off} = D_{3S1}^{\rm off} =
		D_{\epsilon 1}^{\rm off}  = 0$. In both cases, the cutoff is
		chosen to be $R=0.9$~MeV.  
		\label{fig:IchiralDensity}
	}
\end{figure}

It is instructive to compare the partial wave momentum-space matrix elements of the
N$^4$LO potential with and without redundant contact interactions as
visualized in Fig.~\ref{fig:IchiralDensity}. Clearly, the peaks in the
off-diagonal matrix elements in the
$^1$S$_0$, $^3$S$_1$ channels for momenta of the order of $\sim
600$~MeV are driven by the off-shell contact interactions. Notice that
the corresponding LECs 
$\propto D_{1S0}^{\rm off}$ and $\propto D_{3S1}^{\rm off}$ were found in
Ref.~\cite{Epelbaum:2014efa} to be rather large.  
The behavior of these matrix elements suggests that the removal of the
redundant interactions should lead to softer and more perturbative
interactions. 
The standard method for quantifying perturbativeness of a potential
is based on the Weinberg eigenvalue analysis
\cite{Weinberg:1962hj,Weinberg:1963zza}. In this method, the
convergence of the Born series of the Lippmann-Schwinger equation for
the transfer matrix, written schematically in the operator form via 
\beq
T (E + i \epsilon) =  V + V \, G_0 (E + i \epsilon ) \, T (E + i
\epsilon)  = \sum_{n=0}^\infty V \, (G_0 (E + i \epsilon ) \, V)^n\,,
\eeq
where $E  > 0 $ is the cms energy, $\epsilon \to 0^+$ and $G_0$ denotes the free resolvent operator, 
is analyzed by looking at the eigenvalues of the operator $G_0 (E + i
\epsilon ) \, V$
\beq
G_0 (E + i
\epsilon ) \, V | \Psi ( E + i
\epsilon ) \rangle = \eta_i ( E + i
\epsilon )  | \Psi ( E + i
\epsilon ) \rangle \,,
\eeq
which are complex (for $E > 0$) and form a discret set for a given value of $E$.  
The Born series converges if and only if the condition $\max \big( | \eta_i ( E + i
\epsilon ) | \big) < 1$ is fulfilled
\cite{Weinberg:1962hj,Weinberg:1963zza}. It is, therefore, instructive
to look at the trajectories of the largest in magnitude Weinberg
eigenvalues $\eta $ in the complex plane as a function of energy
$E$. For $E \leq 0$, all Weinberg eigenvalues are real. 
Positive and negative (for $E \leq 0$) eigenvalues $\eta_i $ are
referred to as attractive and repulsive, respectively. For NN
potentials without spurious deeply bound states, the largest in
magnitude attractive Weinberg eigenvalues are typically dominated by
the deuteron in the $^3$S$_1 $-$^3$D$_1$ channel and the virtual
state in  the $^1$S$_0 $ partial wave. The corresponding attractive
eigenvalues in these channels are of the order $|\eta | \sim 1$ at low energy 
and typically decrease in magnitude with increasing energies. 
On the other hand, the magnitude of repulsive eigenvalues can be
significantly larger than $1$ for potentials featuring a  repulsive
core at short distances. It is, therefore, instructive to look at the
trajectories of the largest repulsive eigenvalues to quantify
perturbativeness of a given potential. We refer the reader to
Ref.~\cite{Hoppe:2017lok} and references therein for more details on
the Weinberg 
eigenvalue analysis and applications to NN potentials
derived in chiral EFT. 

\begin{figure}[tb]
	\vskip 1 true cm
	\begin{center}
		\includegraphics[width=0.9\textwidth,keepaspectratio,angle=0,clip]{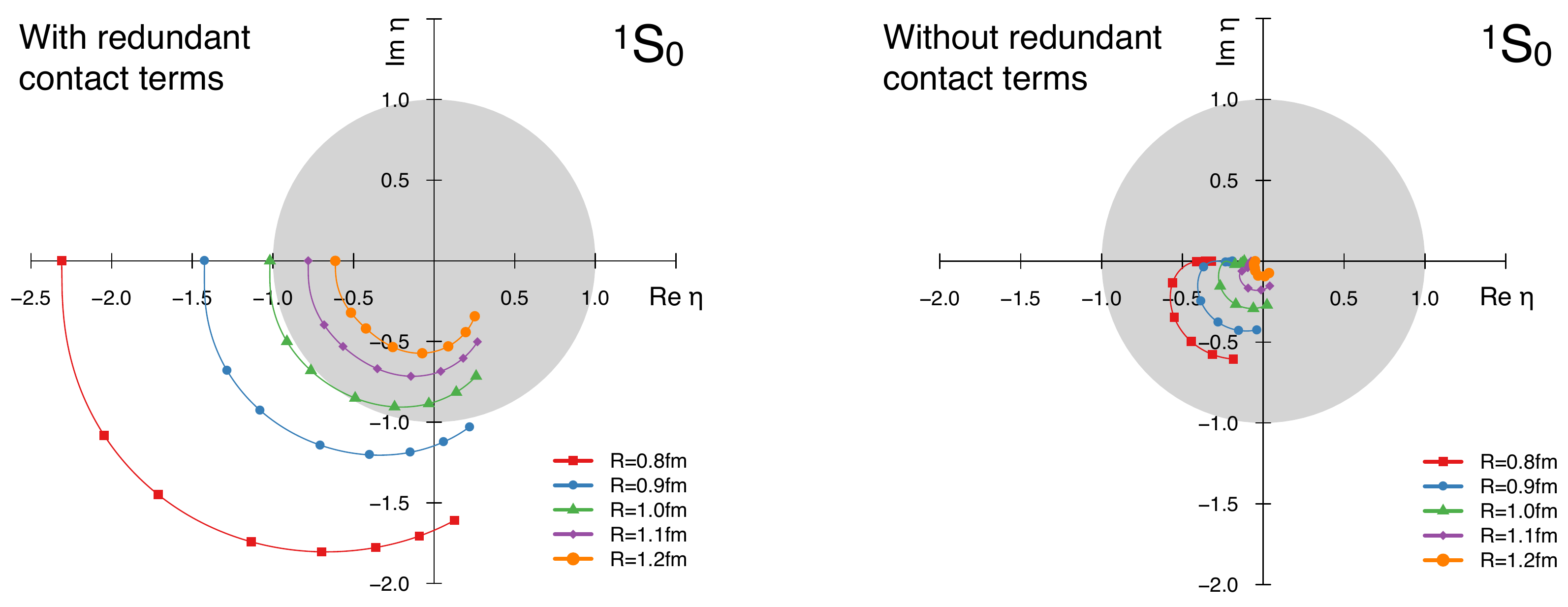}
	\end{center}
	\caption{(Color online) The largest in magnitude repulsive Weinberg eigenvalues $\eta$ from the
		N$^4$LO SCS chiral potentials of
                Ref.~\cite{Epelbaum:2014sza}
                at the cms
		energies from $0$ to $300$~MeV in the $^1$S$_0$ partial
		wave. Left (right) panel shows the results based on the original set of
		the order-$Q^4$ contact interactions as defined in
		Ref.~\cite{Epelbaum:2014efa} (minimal set of independent contact interactions
		with the redundant terms being removed according to
                Eqs.~(\ref{ConventionOffShell1}), (\ref{ConventionOffShell2})). Solid red squares, blue
		dots,
		green triangles, violet diamonds and orange circles refer to the
		cutoff values $R$ of $0.8$, $0.9$, $1.0$, $1.1$ and $1.2$~fm, respectively.  The symbols starting
		from the negative real axis in the counterclockwise direction
		mark the cms energies of $0$, $25$, $50$, $100$, $150$, $200$, $250$
		and $300$~MeV, in order.   
		\label{fig:WeinbergEigenvaluesSCSChiral1S0}
	}
\end{figure}
\begin{figure}[tb]
	\vskip 1 true cm
	\begin{center}
		\includegraphics[width=0.9\textwidth,keepaspectratio,angle=0,clip]{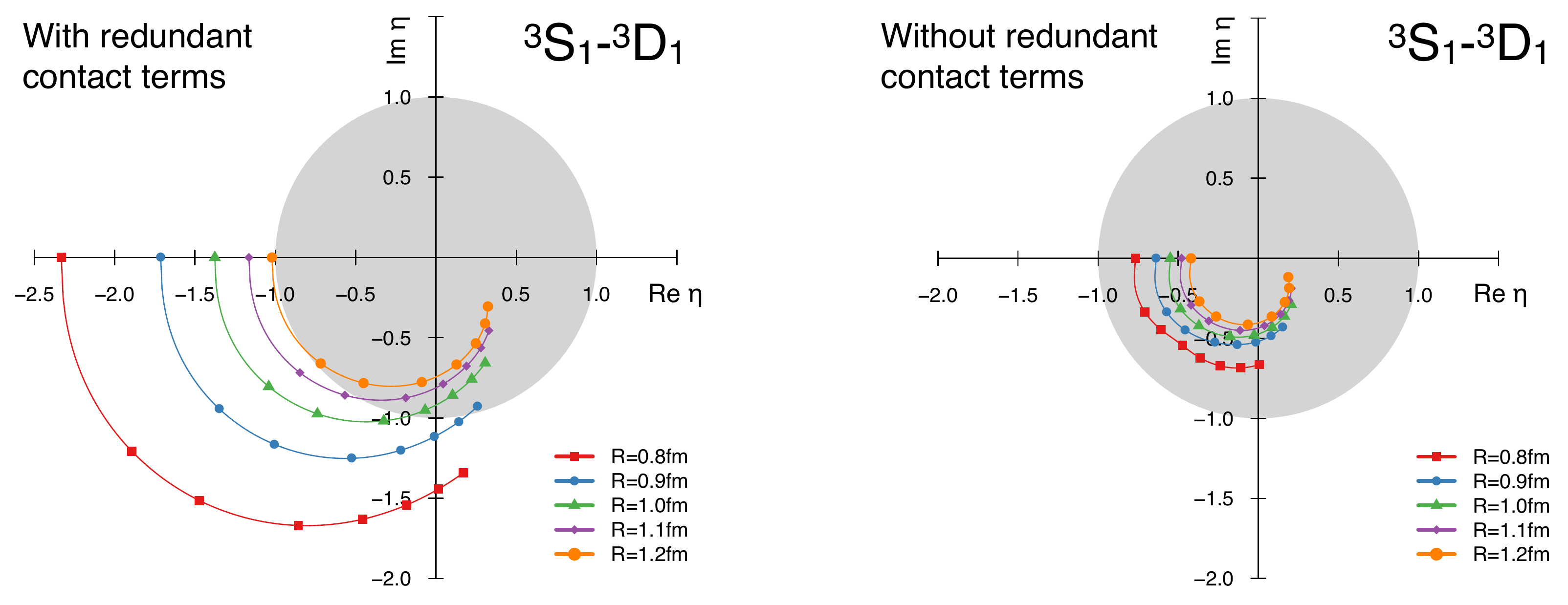}
	\end{center}
	\caption{(Color online) Same as
		Fig.~\ref{fig:WeinbergEigenvaluesSCSChiral1S0} but for the
		$^3$S$_1$-$^3$D$_1$-channel.    
		\label{fig:WeinbergEigenvaluesSCSChiral3S1}
	}
\end{figure}

In the left panel of Figs.~\ref{fig:WeinbergEigenvaluesSCSChiral1S0}
and \ref{fig:WeinbergEigenvaluesSCSChiral3S1}, we show the largest repulsive Weinberg
eigenvalues in the $^1$S$_0$ and $^3$S$_1$-$^3$D$_1$ channels, respectively, for the
N$^4$LO SCS potentials of Ref.~\cite{Epelbaum:2014sza}.  
The
appearance of the repulsive eigenvalues with magnitude larger than $1$
reflects the strongly nonperturbative nature of these interactions
except for the two softest versions corresponding to $R=1.1$ and
$1.2$~fm which, however, suffer from significant  finite-cutoff
artefacts. As demonstrated in Figs.~\ref{fig:WeinbergEigenvaluesSCSChiral1S0}
and \ref{fig:WeinbergEigenvaluesSCSChiral3S1}, the appearance of
large repulsive Weinberg eigenvalues in the potentials of
Ref.~\cite{Epelbaum:2014sza} is a consequence of the large LECs of
the redundant contact interactions at order $Q^4$, as one already may 
expect from looking at the corresponding momentum-space matrix
elements plotted in Fig.~\ref{fig:IchiralDensity}.

\section{Local Regularization of Long-Range Forces in Momentum Space}
\def\theequation{\arabic{section}.\arabic{equation}}
\label{sec:reg}

As already mentioned in the introduction, we employ a local
regularization of the long-range interactions in momentum
space. Specifically, we follow here the approach introduced in
Ref.~\cite{Rijken:1990qs}, in which regularization is achieved by replacing the
Feynman propagators for pions, exchanged between different nucleons, by
the spectral integrals via
\beq
\frac{1}{l^2 - M_\pi^2} \longrightarrow \int_0^\infty d \mu^2
\frac{\rho (\mu ^2)}{l^2 - \mu^2 + i \epsilon}\,,
\eeq 
where $\rho (\mu^2)$ is a spectral function while $l$ is a
four-momentum of the exchanged pion.  The spectral function is
chosen to ensure that the static pion propagators get regularized:
\beq
\label{regul1}
\int_0^\infty d \mu^2
\frac{\rho (\mu ^2)}{\vec l \, ^2 +\mu^2}  \longrightarrow \frac{F(\vec l \, ^2)}{\vec
  l \, ^2 +M_\pi^2} \,.
\eeq
Here and in what follows, we choose the form factor $F (\vec l \, ^2
)$ to be of a Gaussian type
\beq
\label{regul2}
F  (\vec l \, ^2 ) = e^{- \frac{{\vec l}^2 + M_\pi^2}{\Lambda^2}}\,,
\eeq
where $\Lambda$ is an ultraviolet cutoff.  Notice that the $\vec l\,
^2$-dependence of the form factor fixes unambiguously its
$M_\pi$-dependence by the requirement that the residuum of the static
pion propagator  at the pion pole is unchanged. This ensures that the
long-range part of the pion exchange is unaffected by the
regularization (provided $\Lambda$ is chosen of the order of the hard
scale). Obviously, at any finite order in the $1/\Lambda^2$-expansion,
the employed regularization amounts to just adding to the unregularized one-pion
exchange a sequence of contact interactions without inducing any
long-range finite-cutoff artefacts. 

Here and in what follows, we use the
standard decomposition of the long-range potentials in momentum space
\beqa
\label{2PEdecMom}
V (\vec q, \, \vec k\, ) &=& V_C (q) + \fet \tau_1 \cdot \fet \tau_2
\, W_C (q) + \left[   
V_S (q)+ \fet \tau_1 \cdot \fet \tau_2 \, W_S (q)\right] \, \vec \sigma_1 \cdot \vec \sigma_2 
+ \left[ V_T (q)+ \fet \tau_1 \cdot \fet \tau_2 \, W_T (q) \right] 
\, \vec \sigma_1 \cdot \vec q \, \vec \sigma_2 \cdot \vec q \nn
&+& \left[ V_{LS} (q)+ \fet \tau_1 \cdot \fet \tau_2 \, W_{LS} (q)\right] 
\, i (\vec \sigma_1 + \vec \sigma_2 ) \cdot ( \vec q \times \vec k ) \,,
\eeqa
where $\fet \tau_i$ denote the Pauli isospin matrices of the nucleon
$i$. 
Notice that at the order we are working, the scalar
functions $V_i$ and $W_i$ generated by the two-pion
exchange potential depend only on the momentum transfer $\vec q$. More
generally, there may also be dependence on the average momentum $\vec k$.  
We also employ a similar decomposition in coordinate space with 
\beqa
\label{2PEdecCoord}
V (\vec r \, ) &=& V_C (r) + \fet \tau_1 \cdot \fet \tau_2
\, W_C (r) + \left[   
V_S (r)+ \fet \tau_1 \cdot \fet \tau_2 \, W_S (r)\right] \, \vec \sigma_1 \cdot \vec \sigma_2 
+ \left[ V_T (r)+ \fet \tau_1 \cdot \fet \tau_2 \, W_T (r) \right] 
\, S_{12} \nn
&+& \left[ V_{LS} (r)+ \fet \tau_1 \cdot \fet \tau_2 \, W_{LS} (r)\right] 
\, \vec L \cdot \vec S \,,
\eeqa
where $S_{12} = \vec \sigma_1 \cdot \hat r \, \vec \sigma_2 \cdot \hat
r - (1/3) \vec \sigma_1 \cdot \vec \sigma_2$, $\vec S = (\vec \sigma_1
+ \vec  \sigma_2)/2$ and $\vec L  = - i \vec r \times \vec \nabla$. 

Consider first the regularized expressions for the static one-pion exchange potential  (OPEP) 
\beqa
\label{ope_iso}
V_{1\pi, \, \Lambda}^{pp} &=& V_{1\pi, \, \Lambda}^{nn} = V_{1\pi, \, \Lambda} (M_{\pi^0}) \,, \nn
V_{1\pi, \, \Lambda}^{np} &=& -V_{1\pi, \, \Lambda} (M_{\pi^0}) + 2
(-1)^{I+1} V_{1\pi, \, \Lambda} (M_{\pi^\pm}) \,,
\eeqa
where $I$ denotes the total isospin of the two-nucleon system. 
The above expressions include the IB correction due to the different pion masses 
which is the dominant long-range IB effect, see 
Refs.~\cite{vanKolck:1997fu,Friar:1999zr,Walzl:2000cx,Friar:2003yv,Epelbaum:2005fd,Kaiser:2006ck} 
for more details on  the isospin dependence of 
the NN force. Notice that charge dependence of the 
pion-nucleon coupling constant is consistent with zero
\cite{deSwart:1997ep} and for this reason will not be taken into
account in the present analysis. 
The
potential $V_{1\pi, \, \Lambda} (M_{\pi})$ in Eq.~(\ref{ope_iso}) has the following form in momentum space:
\beq
\label{temp0}
V_{1\pi, \, \Lambda} (M_{\pi},  \vec q \, ) = - \frac{g_A^2}{4
  F_\pi^2} \bigg( \frac{\vec \sigma_1 \cdot
  \vec q \, \vec \sigma_2 \cdot \vec q}{q^2 + M_\pi^2}    + C(M_\pi) \, \vec
\sigma_1 \cdot \vec \sigma_2 \bigg)
e^{- \frac{q^2 + M_\pi^2}{\Lambda^2}}\,,
\eeq
where  $g_A$, $F_\pi$, $M_{\pi^0}$ and $M_{\pi^\pm}$
are the axial-vector coupling constant of the nucleon, pion decay
constant, neutral and charged pion mass, respectively. 
Notice that as a matter of convention, we include in the definition of
the OPEP a leading-order
contact interaction chosen in such a way that it minimizes the amount
of short-range contributions in the regularized OPEP. More precisely, the constant
$C(M_\pi)$ is determined by the requirement that the spin-spin part of
the corresponding coordinate-space potential vanishes for $r \to 0$,
which leads to 
\beq
C (M_\pi ) = -\frac{\Lambda \left(\Lambda ^2-2 M_\pi^2 \right) + 2 \sqrt{\pi } M_\pi^3 e^{\frac{M_\pi^2}{\Lambda ^2}}
   \text{erfc}\left(\frac{M_\pi}{\Lambda }\right)}{3 \Lambda ^3} \,,
\eeq
where $\text{erfc} (x)$ is the complementary error function 
\beq
\text{erfc} (x) = \frac{2}{\sqrt{\pi}} \int_x^\infty dt \, e^{-t^2}\,.
\eeq
Using the coordinate-space expression for the regularized Yukawa
potential 
\beq
U_\Lambda (M_\pi , r) = \int \frac{d^3 q}{(2 \pi)^3} \, e^{i \vec q \cdot \vec
r} \, \frac{1}{q^2 + M_\pi^2}\, e^{-\frac{q^2 +
      M_\pi^2}{\Lambda^2}} = \frac{e^{-M_\pi r} \text{erfc}\left(\frac{M_\pi}{\Lambda
   }-\frac{\Lambda  r}{2}\right)-e^{M_\pi r}
   \text{erfc}\left(\frac{M_\pi}{\Lambda }+\frac{\Lambda  r}{2}\right)}{8
   \pi  r}\,,
\eeq
it is easy to Fourier transform the potential $V_{1\pi, \, \Lambda}
(M_{\pi},  \vec q \, )$ to configuration space leading to:
\beqa
\label{OPEPcoord}
V_{1\pi, \, \Lambda} (M_{\pi},  \vec r \, )  &=& \int \frac{d^3 q}{(2 \pi)^3} \, e^{i \vec q \cdot \vec
r} \, V_{1\pi, \, \Lambda} (M_{\pi},  \vec q \, ) \\
&=&  \frac{g_A^2}{4 F_\pi^2} \left( S_{12} \,  r\frac{\partial
  }{\partial r}
  \left( \frac{1}{r} \frac{\partial }{\partial r} \right) U_\Lambda
  (M_\pi , r) + \vec \sigma_1 \cdot \vec \sigma_2 \left[
    \frac{M_\pi^2}{3} U_\Lambda
  (M_\pi , r) - \left(C(M_\pi) + \frac{1}{3}\right) \frac{\Lambda^3}{8 \pi^{3/2}} e^{- \frac{\Lambda^2
      r^2}{4} - \frac{M_\pi^2}{\Lambda^2}}\right] 
\right).
\nonumber
\eeqa
Up to regularization, the employed form
of the static OPEP coincides with the one used in
Refs.~\cite{Epelbaum:2014efa,Epelbaum:2014sza}. 

\begin{figure}[tb]
\vskip 1 true cm
\begin{center}
\includegraphics[width=0.9\textwidth,keepaspectratio,angle=0,clip]{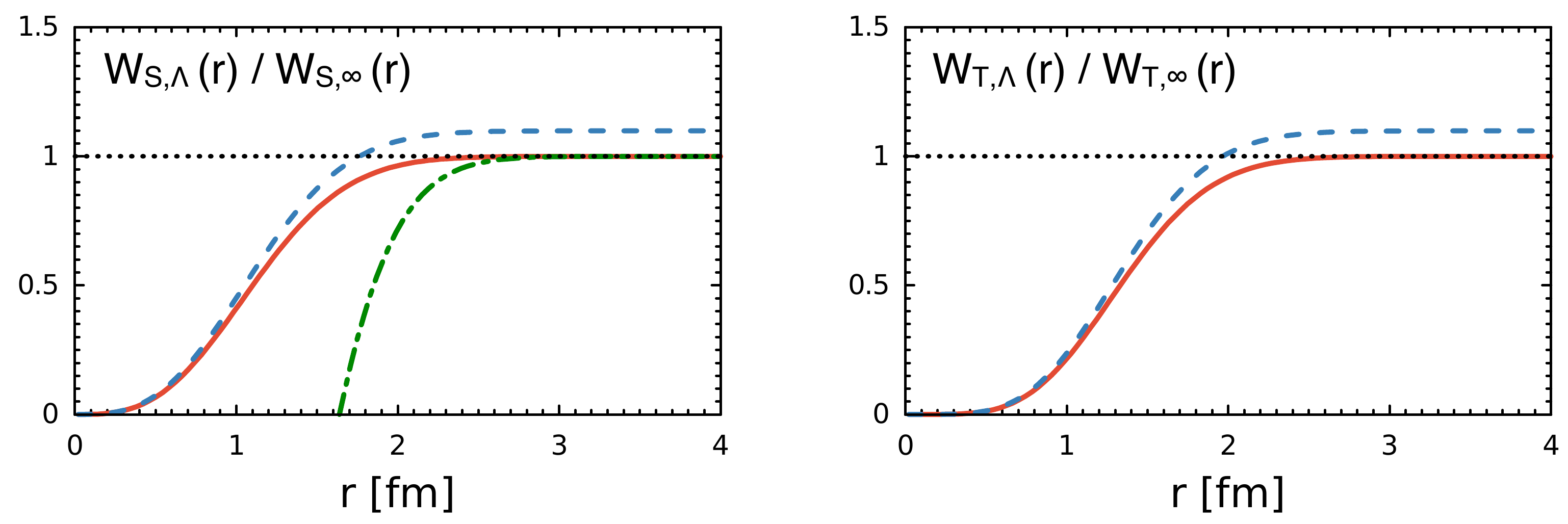}
\end{center}
    \caption{(Color online) Ratio of the regularized to
unregularized spin-spin (left panel) and tensor (right panel) potentials due to one-pion
exchange as a function of the distance between two nucleons for the
cutoff of $\Lambda = 450$~MeV. Red solid lines correspond to the
employed regularization approach as given in
Eq.~(\ref{OPEPcoord}). Dashed-dotted green line shows the result for
the spin-spin potential without performing the additional subtraction,
i.e.~with $C (M_\pi) = 0$. Dashed blue lines show the results based on
the regulator $\exp (- q^2/\Lambda^2)$ instead of $\exp (- (q^2 + M_\pi^2)/\Lambda^2)$.
\label{fig:OPEP}
}
\end{figure}

In Fig.~\ref{fig:OPEP}, we plot the ratio of the regularized to
unregularized spin-spin and tensor potentials of the one-pion
exchange (in the limit of exact isospin symmetry) for  $\Lambda
= 450$~MeV, the intermediate cutoff value employed in our analysis, 
in comparison with the unregularized potential.  
One observes that the employed regulator in momentum space is indeed qualitatively
similar to the coordinate-space regulator used in
Ref.~\cite{Epelbaum:2014efa}. Further, as already pointed out, it is important to keep
the $M_\pi$-dependent term in the exponent of the regulator in
Eq.~(\ref{regul2}). Dropping this term as done e.g.~in Ref.~\cite{Rijken:1990qs} changes the strength of the one-pion
exchange at the pion pole and will require renormalization of the
pion-nucleon coupling constant to restore the long-range part of the
OPEP. This is visualized in Fig.~\ref{fig:OPEP} with dashed blue
lines.

It is straightforward to work out the regularized expressions for the two-pion
exchange potential (TPEP), see also Ref.~\cite{Rijken:1990qs}. 
We start with a generic three-momentum loop
integral emerging in the context of the method of unitary
transformation  \cite{Epelbaum:1998ka,Epelbaum:1999dj,Epelbaum:2007us}, time-ordered perturbation theory
\cite{Weinberg:1990rz,Weinberg:1991um,Ordonez:1995rz} or $S$-matrix-based methods \cite{Kaiser:1997mw} after performing the loop
integrations over the $0$-th momentum components. The corresponding integrands are expressed  
in terms of energies of the two exchanged pions
$\omega_{1,2} = \sqrt{\vec l_{1,2}^2 + M_\pi^2}$.  Using the
identities given in Ref.~\cite{Rijken:1990qs}, it is possible to rewrite the
integrals in the form which resembles a product of two static pion
propagators, generally at the cost
of introducing an additional integral over a mass parameter
\beq
I (\vec q \, ) = \int d \lambda \, \frac{d^3 l_1}{(2 \pi)^3} \, \frac{d^3 l_2}{(2 \pi)^3}
\, (2 \pi)^3 \delta (\vec q - \vec l_1 - \vec l_2 ) \, \frac{1}{(\omega_1^2 +
  \lambda^2 ) (\omega_2 + \lambda^2)} \times \ldots \,,
\eeq
where $\vec q$ is the nucleon momentum transfer and the ellipses refer to the isospin-spin-momentum structure
emerging from vertices. Replacing the pion propagators by the
their regularized form as given in Eqs.~(\ref{regul1}),
(\ref{regul2}),  the regularized integral takes the form 
\beq
\label{temp1}
I_\Lambda (\vec q \, ) = e^{-\frac{q^2}{2 \Lambda^2}} \, 2\int d \lambda \, \frac{d^3 l}{(2 \pi)^3} 
\, \frac{e^{-
    \frac{l^2 + 4 M_\pi^2 + 4 \lambda^2}{2 \Lambda^2}}}{(\omega_+^2 +
 4 \lambda^2 ) (\omega_-^2 + 4 \lambda^2)} \times \ldots \,,
\eeq
where $\vec l = \vec l_1 - \vec l_2$ and $\omega_{\pm} = \sqrt{(\vec q
  \pm \vec l)^2 + 4 M_\pi^2} $. Here and in what follows, we
use the notation $q \equiv | \vec q \, | $, $l \equiv | \vec l \,
|$.   In practical calculations, the two-pion exchange contributions
to the nucleon-nucleon (NN) potential are usually expressed in terms of
the corresponding spectral integrals of the form 
\beq
\label{V}
V(q) = \frac{2}{\pi} \int_{2 M_\pi^2}^\infty  \mu \, d \mu \frac{\rho
  (\mu)}{ q^2 + \mu^2} \,,
\eeq
with the spectral function given by $\rho (\mu ) = \Im \big( V (q)
\big) \big|_{q = 0^+ - i \mu}$.\footnote{One needs to introduce
  subtractions to render the spectral integrals finite. The  required  number of
  subtractions depends on the chiral order. 
}  The explicit expressions for the 
spectral functions of the
central, spin-spin, tensor and spin-orbit potentials up to and
including N$^4$LO can be found in
Refs.~\cite{Kaiser:1997mw,Epelbaum:1998ka,Kaiser:2001pc,Entem:2014msa},
see also \cite{Epelbaum:2004fk,Epelbaum:2014efa}. It is easy to see 
for a general case of the TPEP 
that the regularization of the pion propagators introduced above
leads to a particular form of the spectral function regularization,
which amounts to (modulo a finite number of contact interactions) 
replacing $V(q)$ in Eq.~(\ref{V}) by $V_\Lambda (q)$ given by 
\beq
\label{SpectralReg}
V_\Lambda (q) = e^{-\frac{q^2}{2 \Lambda^2}} \; \frac{2}{\pi} \int_{2
  M_\pi^2}^\infty  \mu \, d \mu \,\frac{\rho
  (\mu)}{ q^2 + \mu^2} \, e^{- \frac{\mu^2}{2
    \Lambda^2}} \,. 
\eeq
The functional form of the regulator in $q^2$ follows from the employed 
regulator for the exchanged pion propagators in Eqs.~(\ref{regul1}),
(\ref{regul2}) as shown in Eq.~(\ref{temp1}). Since the
modification of the pion propagator does not affect the pion pole
contributions at any
order in the $1/\Lambda$ expansion, the
regularized expression for the TPEP must  feature the same
discontinuity across the left-hand cut as the original one in
Eq.~(\ref{V}). Together with the form of the regulator in $q^2$, namely 
$\exp (- q^2/(2 \Lambda^2))$, this then fixes unambiguously  the  form
of the regulator in $\mu$ entering the spectral integral. 
Clearly, this can also be verified upon performing
explicit calculations. Consider, for example, the leading two-pion exchange
potential at order $Q^2$ \cite{Kaiser:1997mw,Epelbaum:1998ka} 
\beqa
V^{(2)} (\vec q \, ) &=& \frac{g_A^2}{(2 F_\pi)^4} \fet \tau_1 \cdot \fet
\tau_2 \int \frac{d^3l}{(2 \pi )^3} \, \frac{l^2 - q^2}{\omega_+
  \omega_- (\omega_+ + \omega_-)} \nn
&-& 
\frac{1}{8 (2 F_\pi)^4} \fet \tau_1 \cdot \fet \tau_2 \int
\frac{d^3l}{(2 \pi )^3} \,  \frac{(\omega_+ - \omega_-)^2}{\omega_+
  \omega_- (\omega_+ + \omega_-)} \nn
&-&  \frac{g_A^4}{2 (2 F_\pi)^4}  \int \frac{d^3l}{(2 \pi )^3} \,
\frac{\omega_+^2  + \omega_+ \omega_- + \omega_-^2}{\omega_+^3
  \omega_-^3 (\omega_+ + \omega_-)}  \Big( \fet \tau_1 \cdot \fet
\tau_2 (l^2 - q^2)^2 + 6 \, \vec \sigma_1 \cdot [\vec q \times \vec l ]
\, \vec \sigma_2 \cdot [\vec q \times \vec l ] \Big)\,.
\eeqa   
The above integrals can be calculated
in a closed form using dimensional regularization \cite{Kaiser:1997mw,Epelbaum:1999dj}. The
long-range behavior of the potential is determined by the
non-polynomial  (in momentum space) terms, which are independent of
the scale introduced through dimensional regularization and have the form 
\beqa
\label{2pi_nlo_np}
V^{(2)}_{\rm non-pol.}  (\vec q \, ) 
&=& - \frac{ \fet{\tau}_1 \cdot \fet{\tau}_2 }{384 \pi^2 F_\pi^4}\,
L(q) \, \biggl\{4M_\pi^2 (5g_A^4 - 4g_A^2 -1)
+ q^2(23g_A^4 - 10g_A^2 -1)
+ \frac{48 g_A^4 M_\pi^4}{4 M_\pi^2 + q^2} \biggr\}\nn
&-&  \frac{3 g_A^4}{64 \pi^2 F_\pi^4} \,L(q)  \, 
\Bigl[ ( \vec \sigma_1 \cdot \vec q \, ) \,
( \vec \sigma_2 \cdot \vec q \, ) 
 - ( \vec \sigma_1 \cdot\vec
\sigma_2 ) \, q^2 \Bigr] \,,
\eeqa
with the loop function $L(q)$ given by 
\beq\label{Lq}
L(q) = \frac{\sqrt{q^2 + 4 M_\pi^2}}{q}\, 
\ln\frac{\sqrt{q^2 + 4 M_\pi^2} + q}{2M_\pi}\,.
\eeq
These expressions can also be represented (modulo contact terms) by
means of the subtracted spectral
integrals
\beq
\label{disp_int}
W_C^{(2)} (q) = \frac{2q^4}{\pi} \int_{2 M_\pi}^\infty \, \frac{d \mu}{\mu^3}  \frac{\eta_{C}^{(2)} (\mu)}
{\mu^2 + q^2} \,,  \quad 
V_S^{(2)} (q) = \frac{2q^4}{\pi} \int_{2 M_\pi}^\infty \, \frac{d \mu}{\mu^3} \, \frac{\rho_{S}^{(2)} (\mu)}
{\mu^2 + q^2} \,, \quad 
V_T^{(2)} (q) = -\frac{2q^2}{\pi} \int_{2 M_\pi}^\infty \, \frac{d \mu}{\mu} \, \frac{\rho_{T}^{(2)} (\mu)}
{\mu^2 + q^2} \,,
\eeq
where  the corresponding spectral functions are given by 
\beqa
\label{spectr_nlo}
\eta_{C}^{(2)}  (\mu) &=& \Im \Big (W_C^{(2)} \big|_{q=0^+ - i \mu } \Big) =
 \frac{ 1}{768 \pi F_\pi^4}\,
 \, \biggl(4M_\pi^2 (5g_A^4 - 4g_A^2 -1) 
- \mu^2(23g_A^4 - 10g_A^2 -1)
+ \frac{48 g_A^4 M_\pi^4}{4 M_\pi^2 - \mu^2} \biggr) \, \frac{\sqrt{\mu^2 - 4 M_\pi^2}}{\mu} \,,\nn
\rho_{T}^{(2)} (\mu)  &=& \Im \Big (V_T^{(2)} \big|_{q=0^+ - i \mu } \Big)
= \frac{1}{\mu^2}\, \rho_{S}^{(2)}  (\mu) 
=  \frac{3 g_A^4}{128 \pi F_\pi^4} \,
\frac{\sqrt{\mu^2 - 4 M_\pi^2}}{\mu} \,.
\eeqa

Consider now the regularized expressions for the TPEP. 
Using the identities 
\beqa
 \frac{(\omega_+ - \omega_-)^2}{\omega_+
  \omega_- (\omega_+ + \omega_-)} &=& \frac{1}{\omega_+} +
\frac{1}{\omega_-} - \frac{4}{\omega_+ + \omega_-} \,, \nn
\frac{\omega_+^2  + \omega_+ \omega_- + \omega_-^2}{\omega_+^3
  \omega_-^3 (\omega_+ + \omega_-)} &=& -\frac{4}{M_\pi}
\frac{\partial }{\partial M_\pi}  \frac{1}{\omega_+ \omega_- (\omega_+
  + \omega_-)} \,,\nn
\frac{1}{\omega_+ + \omega_-} &=& \frac{2}{\pi} \int_0^\infty d \lambda
\frac{\lambda^2}{(\omega_+^2 + \lambda^2) (\omega_-^2 + \lambda^2)}
\,, \nn
 \frac{1}{\omega_+ \omega_- (\omega_+
  + \omega_-)}  &=& \frac{2}{\pi} \int_0^\infty d \lambda
\frac{1}{(\omega_+^2 + \lambda^2) (\omega_-^2 + \lambda^2)}
\,, \nn
\Im \bigg[ \int \frac{d^3 l}{(2 \pi)^3} \frac{f (l^2,
  q^2)}{(\omega_+^2 + \lambda^2) (\omega_-^2 + \lambda^2)} \bigg|_{q =
0^+ - i \mu} \bigg]&=& \frac{1}{16 \mu} f\big( \mu^2 - 4 M_\pi^2 -
\lambda^2, - \mu^2 \big) \, \theta \Big( \mu - \sqrt{4 M_\pi^2 +
  \lambda^2} \Big)\,,
\eeqa
where $f$ is a real function of $l^2$ and $q^2$, it is easy to verify
by explicitly calculating $\Im \big( \exp ( q^2/(2 \Lambda^2) )  V^{(2)}_\Lambda (q)
\big) \big|_{q = 0^+ - i \mu}$
that the regularized two-pion exchange
potential at NLO (modulo polynomial terms)  
indeed takes the form 
\beqa
\label{disp_int_reg}
W_{C, \, \Lambda}^{(2)} (q) &=& e^{-\frac{q^2}{2 \Lambda^2}} \; \frac{2q^4}{\pi} \int_{2 M_\pi}^\infty \, \frac{d \mu}{\mu^3}  \frac{\eta_{C}^{(2)} (\mu)}
{\mu^2 + q^2} \, e^{- \frac{\mu^2}{2
    \Lambda^2}} \,, \nn
V_{S, \, \Lambda}^{(2)} (q) &=& e^{-\frac{q^2}{2 \Lambda^2}} \; \frac{2q^4}{\pi} \int_{2 M_\pi}^\infty \, \frac{d \mu}{\mu^3} \, \frac{\rho_{S}^{(2)} (\mu)}
{\mu^2 + q^2} \, e^{- \frac{\mu^2}{2
    \Lambda^2}} \,, \nn
V_{T, \, \Lambda}^{(2)} (q) &=& - e^{-\frac{q^2}{2 \Lambda^2}} \; \frac{2q^2}{\pi} \int_{2 M_\pi}^\infty \, \frac{d \mu}{\mu} \, \frac{\rho_{T}^{(2)} (\mu)}
{\mu^2 + q^2} \, e^{- \frac{\mu^2}{2
    \Lambda^2}} \,.
\eeqa
Thus, the regularized expressions for the two-pion exchange potential
can be easily obtained by using the known analytical results for the
corresponding spectral functions and employing the
exponential regulator when performing the spectral integrals. As
already pointed out above, this fixes the expressions for the two-pion
exchange modulo contact interactions allowed at the given order in the
chiral expansion. Here and in what follows, we eliminate this
ambiguity by imposing a convention that 
the corresponding long-range potentials in coordinate
space and derivatives thereof vanish at the origin. Such an approach
qualitatively resembles the behavior of the
coordinate-space regulator used in Ref.~\cite{Epelbaum:2014efa}. 

To be specific we provide below the explicit expressions for the
regularized long-range part of the two-pion exchange potential at
different orders in the chiral expansion in terms of the corresponding
spectral integrals.
\begin{itemize}
\item NLO\\
\beqa
\label{disp_int_reg_NLO}
W_{C, \, \Lambda}^{(2)} (q) &=& e^{-\frac{q^2}{2 \Lambda^2}} \;
\frac{2}{\pi} \int_{2 M_\pi}^\infty \, \frac{d \mu}{\mu^3}
\eta_{C}^{(2)} (\mu) \bigg(\frac{q^4} 
{\mu^2 + q^2} + C_{C, 1}^2 (\mu ) +
C_{C, 2}^2 (\mu ) \, q^2\bigg) 
\, e^{- \frac{\mu^2}{2
    \Lambda^2}} \,,\nn
V_{S, \, \Lambda}^{(2)} (q) &=& e^{-\frac{q^2}{2 \Lambda^2}} \;
\frac{2}{\pi} \int_{2 M_\pi}^\infty \, \frac{d \mu}{\mu^3} \,
\rho_{S}^{(2)} (\mu) \bigg( \frac{q^4}{q^2 + \mu^2} + C_{S, 1}^2 (\mu ) +
C_{S, 2}^2 (\mu ) \, q^2\bigg) \, e^{- \frac{\mu^2}{2
    \Lambda^2}} \,, \nn
V_{T, \, \Lambda}^{(2)} (q) &=& - e^{-\frac{q^2}{2 \Lambda^2}} \;
\frac{2}{\pi} \int_{2 M_\pi}^\infty \, \frac{d \mu}{\mu^3} \,
\rho_{S}^{(2)} (\mu) \bigg(
\frac{q^2} 
{\mu^2 + q^2} + C_{T}^1 (\mu) \bigg)\, e^{- \frac{\mu^2}{2
    \Lambda^2}} \,,
\eeqa
where the superscript $a$ of the functions $C_{C, i}^a (\mu)$, $C_{S,
  i}^a (\mu)$ and $C_{T}^a (\mu)$ refers
to the number of subtractions. The 
corresponding spectral functions are specified in
Eq.~(\ref{spectr_nlo}), while the 
functions $C_{C, i}^2 (\mu)$, $C_{S, i}^2 (\mu)$,
and $C_{T}^1 (\mu)$ are determined by the
required short-distance behavior of the coordinate-space
potentials: 
\beq
W_{C, \, \Lambda}^{(2)} (0) = \frac{d^2 }{d r^2} W_{C, \, \Lambda}^{(2)} (r )
\bigg|_{r = 0} = V_{S, \, \Lambda}^{(2)} (0) = \frac{d^2 }{d r^2} V_{S, \, \Lambda}^{(2)} (r )
\bigg|_{r = 0} = \frac{d^2 }{d r^2} V_{T, \, \Lambda}^{(2)} (r )
\bigg|_{r = 0}  = 0\,.
\eeq
Their explicit form can be found in appendix \ref{app1}. 
\item N$^2$LO\\
\beqa
\label{disp_int_reg_N2LO}
V_{C, \, \Lambda}^{(3)} (q) &=& e^{-\frac{q^2}{2 \Lambda^2}} \;
\frac{2}{\pi} \int_{2 M_\pi}^\infty \, \frac{d \mu}{\mu^3}
\rho_{C}^{(3)} (\mu) \bigg(\frac{q^4} 
{\mu^2 + q^2} + C_{C, 1}^2 (\mu ) +
C_{C, 2}^2 (\mu ) \, q^2\bigg) 
\, e^{- \frac{\mu^2}{2
    \Lambda^2}} \,,\nn
W_{S, \, \Lambda}^{(3)} (q) &=& e^{-\frac{q^2}{2 \Lambda^2}} \;
\frac{2}{\pi} \int_{2 M_\pi}^\infty \, \frac{d \mu}{\mu^3} \,
\eta_{S}^{(3)} (\mu) \bigg( \frac{q^4}{q^2 + \mu^2} + C_{S, 1}^2 (\mu ) +
C_{S, 2}^2 (\mu ) \, q^2\bigg) \, e^{- \frac{\mu^2}{2
    \Lambda^2}} \,, \nn
W_{T, \, \Lambda}^{(3)} (q) &=& - e^{-\frac{q^2}{2 \Lambda^2}} \;
\frac{2}{\pi} \int_{2 M_\pi}^\infty \, \frac{d \mu}{\mu^3} \,
\eta_{S}^{(3)} (\mu) \bigg(
\frac{q^2} 
{\mu^2 + q^2} + C_{T}^1 (\mu) \bigg)\, e^{- \frac{\mu^2}{2
    \Lambda^2}} \,,
\eeqa
where the corresponding spectral functions are given by
\beqa
\rho_C^{(3)} (\mu ) &=& - \frac{3 g_A^2}{64 F_\pi^4} \big( 2 M_\pi^2 (2 c_1 -
c_3 ) + c_3 \mu^2 \big) \frac{\big(2 M_\pi^2 - \mu^2\big)}{\mu}\,, \nn
\eta_S^{(3)} (\mu ) &=& - \frac{g_A^2}{128 F_\pi^4} c_4 \mu \big(4 M_\pi^2 - \mu^2\big)\,.
\eeqa
\item N$^3$LO\\
The static contributions to the TPEP are given by 
\beqa
\label{disp_int_reg_N3LO}
X_{C, \, \Lambda}^{(4)} (q) &=& - e^{-\frac{q^2}{2 \Lambda^2}} \;
\frac{2}{\pi} \int_{2 M_\pi}^\infty \, \frac{d \mu}{\mu^5}
x_{C}^{(4)} (\mu) \bigg(\frac{q^6} 
{\mu^2 + q^2} + C_{C, 1}^3 (\mu ) +
C_{C, 2}^3 (\mu ) \, q^2 +
C_{C, 3}^3 (\mu ) \, q^4\bigg) 
\, e^{- \frac{\mu^2}{2
    \Lambda^2}} \,,\nn
X_{S, \, \Lambda}^{(4)} (q) &=& - e^{-\frac{q^2}{2 \Lambda^2}} \;
\frac{2}{\pi} \int_{2 M_\pi}^\infty \, \frac{d \mu}{\mu^5} \,
x_{S}^{(4)} (\mu) \bigg( \frac{q^6}{q^2 + \mu^2} + C_{S, 1}^3 (\mu ) +
C_{S, 2}^3 (\mu ) \, q^2 +
C_{S, 3}^3 (\mu ) \, q^4\bigg) \, e^{- \frac{\mu^2}{2
    \Lambda^2}} \,, \nn
X_{T, \, \Lambda}^{(4)} (q) &=&  e^{-\frac{q^2}{2 \Lambda^2}} \;
\frac{2}{\pi} \int_{2 M_\pi}^\infty \, \frac{d \mu}{\mu^5} \,
x_{S}^{(4)} (\mu) \bigg(
\frac{q^4} 
{\mu^2 + q^2} + C_{T, 1}^2 (\mu) + C_{T, 2}^2 (\mu) \, q^2\bigg)\, e^{- \frac{\mu^2}{2
    \Lambda^2}} \,,
\eeqa
where $X$ and $x$ refer to either $V$ and $\rho$ or to  $W$ and
$\eta$, respectively. The 
functions $C_{C, i}^3 (\mu)$, $C_{S, i}^3 (\mu)$,
and $C_{T, i}^2 (\mu)$ are determined by the conditions
\beqa
X_{C, \, \Lambda}^{(4)} (0) = \frac{d^2 }{d r^2} X_{C, \, \Lambda}^{(4)} (r )
\bigg|_{r = 0} &=& \frac{d^4 }{d r^4} X_{C, \, \Lambda}^{(4)} (r )
\bigg|_{r = 0} = 0\,, \nn
X_{S, \, \Lambda}^{(4)} (0) = \frac{d^2 }{d r^2} X_{S, \, \Lambda}^{(4)} (r )
\bigg|_{r = 0} &=& \frac{d^4 }{d r^4} X_{S, \, \Lambda}^{(4)} (r ) 
\bigg|_{r = 0} = 0 \,, \nn  
\frac{d^2 }{d r^2} X_{T, \, \Lambda}^{(4)} (r )
\bigg|_{r = 0} &=& \frac{d^4}{d r^4} X_{T, \, \Lambda}^{(4)} (r )
\bigg|_{r = 0} = 0\,,
\eeqa
and are listed in appendix \ref{app1}. The spectral functions
$\rho_C^{(4)} (\mu)$, $\rho_S^{(4)} (\mu)$, $\eta_C^{(4)} (\mu)$ and
$\eta_S^{(4)} (\mu)$ have been derived in
Ref.~\cite{Kaiser:2001pc}. The two-loop contributions are also listed
in Eq.~(12) of Ref.~\cite{Epelbaum:2014efa}. The spectral functions corresponding to the one-loop
contributions proportional to the LECs $c_i^2$ can be extracted from
the TPEP contributions given in Eq.~(10) of this reference via
replacements $L (q) \to - \pi/(2 \mu) \sqrt{\mu^2 - 4 M_\pi^2}$ and
$q^2 \to - \mu^2$. 

In addition to the static contributions, one also needs to account for
the leading $1/m_N$-corrections to the TPEP. Here and in what follows,
$m_N$ refers to the nucleon mass, for which we use the value $m_N = 2
m_p m_n /(m_p + m_n)$ in terms of the proton and neutron masses $m_p$
and $m_n$, respectively.  Throughout this work, we employ the
same treatment of the relativistic corrections as in our previous
papers \cite{Epelbaum:2014efa,Epelbaum:2014sza}. Specifically, we use
the Schr\"odinger equation as given in Eq.~(20) of
Ref.~\cite{Epelbaum:2014efa} at all considered orders and adopt the ``minimal nonlocality''
choice for the two unitary transformations which affect the
$1/m_N^2$-contributions  to the OPEP
and $1/m_N$-contributions to the TPEP at N$^3$LO.  Thus, all
unregularized expressions for the relativistic corrections to the TPEP
given in Ref.~\cite{Epelbaum:2014efa}  remain valid in our case. The
regularized expressions for the leading relativistic corrections to
the TPEP are written in terms of the spectral integrals
\beqa
\label{disp_int_reg_N3LO_rel}
X_{C, \, \Lambda}^{(4)} (q) &=&  e^{-\frac{q^2}{2 \Lambda^2}} \;
\frac{2}{\pi} \int_{2 M_\pi}^\infty \, \frac{d \mu}{\mu^3}
x_{C}^{(4)} (\mu) \bigg(\frac{q^4} 
{\mu^2 + q^2} + C_{C, 1}^2 (\mu ) +
C_{C, 2}^2 (\mu ) \, q^2 \bigg) 
\, e^{- \frac{\mu^2}{2
    \Lambda^2}} \,,\nn
X_{S, \, \Lambda}^{(4)} (q) &=&  e^{-\frac{q^2}{2 \Lambda^2}} \;
\frac{2}{\pi} \int_{2 M_\pi}^\infty \, \frac{d \mu}{\mu^3} \,
x_{S}^{(4)} (\mu) \bigg( \frac{q^4}{q^2 + \mu^2} + C_{S, 1}^2 (\mu ) +
C_{S, 2}^2 (\mu ) \, q^2 \bigg) \, e^{- \frac{\mu^2}{2
    \Lambda^2}} \,, \nn
X_{T, \, \Lambda}^{(4)} (q) &=& - e^{-\frac{q^2}{2 \Lambda^2}} \;
\frac{2}{\pi} \int_{2 M_\pi}^\infty \, \frac{d \mu}{\mu^3} \,
x_{S}^{(4)} (\mu) \bigg(
\frac{q^2} 
{\mu^2 + q^2} + C_{T}^1 (\mu) \bigg)\, e^{- \frac{\mu^2}{2
    \Lambda^2}} \,, \nn
X_{LS, \, \Lambda}^{(4)} (q) &=& - e^{-\frac{q^2}{2 \Lambda^2}} \;
\frac{2}{\pi} \int_{2 M_\pi}^\infty \, \frac{d \mu}{\mu} \,
x_{LS}^{(4)} (\mu) \bigg(
\frac{q^2} 
{\mu^2 + q^2} + C_{LS}^1 (\mu) \bigg)\, e^{- \frac{\mu^2}{2
    \Lambda^2}} \,, 
\eeqa
where, again, $X$ and $x$ stay for either $V$ and $\rho$ or  $W$
and $\eta$, respectively. The function $C_{LS}^1 (\mu)$ is determined
by the requirement 
\beq
\frac{d}{d r} X_{LS, \, \Lambda}^{(4)}  (r )
\bigg|_{r = 0} = 0
\eeq
and can be found in appendix \ref{app1}. The corresponding spectral functions can be
read off from the expressions for the potentials derived in
Refs.~\cite{Kaiser:1997mw,Friar:1999sj} and given in Eqs.~(19) and (22) of Ref.~\cite{Epelbaum:2014efa}
upon performing the replacements  $A (q) \to \pi/(4 \mu)$ and $q^2 \to
- \mu^2$ in the non-pole contributions (i.e.~in the contributions
proportional to the loop function $A (q)$). The regularized
expressions for the pole contributions to $V_{C}^{(4)}$ and
$W_{C}^{(4)}$ in Eq.~(19) of Ref.~\cite{Epelbaum:2014efa} take
the form 
\beq
\label{TPEPRelPoles}
V_{C, \, \Lambda}^{(4)} (q) = \frac{1}{2} W_{C, \, \Lambda}^{(4)} (q)  = \frac{3 g_A^4 M_\pi^5}{256 \pi
  m_N F_\pi^4} \bigg( \frac{1}{q^2 + 4 M_\pi^2} + C^2_{1} + C^2_{2}
q^2 \bigg) e^{- \frac{q^2 + 4 M_\pi^2}{2 \Lambda^2}} \,, 
\eeq
where the constants $C_1^2$ and $C_2^2$ are determined by the
requirement that the corresponding $r$-space potentials and their
second derivatives vanish at the origin and are specified in appendix
\ref{app1}. 
\item N$^4$LO\\
Finally, regularized N$^4$LO contributions to the TPEP have the form 
\beqa
\label{disp_int_reg_N4LO}
X_{C, \, \Lambda}^{(5)} (q) &=& - e^{-\frac{q^2}{2 \Lambda^2}} \;
\frac{2}{\pi} \int_{2 M_\pi}^\infty \, \frac{d \mu}{\mu^5}
x_{C}^{(5)} (\mu) \bigg(\frac{q^6} 
{\mu^2 + q^2} + C_{C, 1}^3 (\mu ) +
C_{C, 2}^3 (\mu ) \, q^2 +
C_{C, 3}^3 (\mu ) \, q^4\bigg) 
\, e^{- \frac{\mu^2}{2
    \Lambda^2}} \,,\nn
X_{S, \, \Lambda}^{(5)} (q) &=& - e^{-\frac{q^2}{2 \Lambda^2}} \;
\frac{2}{\pi} \int_{2 M_\pi}^\infty \, \frac{d \mu}{\mu^5} \,
x_{S}^{(5)} (\mu) \bigg( \frac{q^6}{q^2 + \mu^2} + C_{S, 1}^3 (\mu ) +
C_{S, 2}^3 (\mu ) \, q^2 +
C_{S, 3}^3 (\mu ) \, q^4\bigg) \, e^{- \frac{\mu^2}{2
    \Lambda^2}} \,, \nn
X_{T, \, \Lambda}^{(5)} (q) &=&  e^{-\frac{q^2}{2 \Lambda^2}} \;
\frac{2}{\pi} \int_{2 M_\pi}^\infty \, \frac{d \mu}{\mu^5} \,
x_{S}^{(5)} (\mu) \bigg(
\frac{q^4} 
{\mu^2 + q^2} + C_{T, 1}^2 (\mu) + C_{T, 2}^2 (\mu) \, q^2\bigg)\, e^{- \frac{\mu^2}{2
    \Lambda^2}} \,, \nn
X_{LS, \, \Lambda}^{(5)} (q) &=&  e^{-\frac{q^2}{2 \Lambda^2}} \;
\frac{2}{\pi} \int_{2 M_\pi}^\infty \, \frac{d \mu}{\mu^3} \,
x_{LS}^{(5)} (\mu) \bigg(
\frac{q^4} 
{\mu^2 + q^2} + C_{LS, 1}^2 (\mu) + C_{LS, 2}^2 (\mu) \, q^2\bigg)\, e^{- \frac{\mu^2}{2
    \Lambda^2}} \,,
\eeqa
where the functions $C_{LS, \, 1}^2 (\mu)$ and $C_{LS, \, 2}^2 (\mu)$ are determined
by the requirements 
\beq
\frac{d}{d r} X_{LS, \, \Lambda}^{(5)}  (r )
\bigg|_{r = 0} = \frac{d^3}{d r^3} X_{LS, \, \Lambda}^{(5)}  (r )
\bigg|_{r = 0} = 0
\eeq
and are given in appendix \ref{app1}. The spectral functions
$\rho_C^{(5)}$, $\rho_S^{(5)}$, $\eta_C^{(5)}$ and $\eta_S^{(5)}$
receive contributions from the static terms which are given in
Ref.~\cite{Entem:2014msa}. In addition, one has to account for the relativistic
corrections $\propto c_i/m_N$. The corresponding contributions to the
spectral functions can be read off from Eqs.~(13)-(17) of
Ref.~\cite{Kaiser:2001pc} by performing the replacements $L (q) \to - \pi/(2 \mu) \sqrt{\mu^2 - 4 M_\pi^2}$ and
$q^2 \to - \mu^2$.\footnote{Notice that Refs.~\cite{Entem:2014msa,Kaiser:2001pc} use a different sign convention in the
  definition of the potential.} 
\end{itemize}

It is instructive to compare various regularization approaches for the
TPEP with each other. To this end, we consider the isovector central
TPEP at NLO  $W_C^{(2)}$ as a representative example. Its non-regularized
momentum-space expression $W_{C, \, \infty}^{(2)}$  is given in the first line of Eq.~(\ref{2pi_nlo_np})
up to irrelevant polynomial terms. While the Fourier transform of
$W_{C, \, \infty}^{(2)} (q)$ does not exist, one can define its coordinate-space
representation via the limit 
\beq
W_{C, \, \infty}^{(2)} (r) = \lim_{\Lambda \to \infty} \int \frac{d^3 q}{(2
  \pi)^3}\, 
e^{i \vec q \cdot \vec r} \, W_{C, \, \infty}^{(2)} (q) \, e^{-\frac{q^2}{\Lambda^2}}\,,
\eeq
which exists for $r > 0$. The resulting potential can be obtained in a
more convenient way by making use of the spectral representation which
leads to \cite{Kaiser:1997mw}
\beqa
W_{C, \, \infty}^{(2)} (r) &=& \frac{1}{2 \pi^2 r} \int_{2 M_\pi}^\infty
d \mu \, \mu \, e^{- \mu r} \, \eta_C^{(2)} (\mu)  \nn
&=& \frac{M_\pi}{128 \pi^3 F_\pi^4 r^4} \left[ \left(1 + 2 g_A^2 (5 +
    2 x^2) - g_A^4 (23 + 12 x^2) \right) K_1 (2 x) + x \left( 1 + 10
    g_A^2 - g_A^4 (23 + 4 x^2) \right) K_0 (2 x) \right] \,,
\nonumber
\eeqa
where $x \equiv M_\pi r$ and $K_i (x)$ denote the modified
Bessel-functions. We regard this expression as our reference potential
in coordinate space. We now consider four different ways to implement
regularization as described below. 
\begin{enumerate}
\item
First, we retain only the momentum-independent part of the Gaussian regulator in Eq.~(\ref{SpectralReg}) responsible for
the spectral function regularization. The corresponding
coordinate-space TPEP can be calculated by means of the spectral
integral 
\beq
\label{WC1}
W_{C, \, \Lambda ,  \, 1}^{(2)} (r) =\frac{1}{2 \pi^2 r} \int_{2 M_\pi}^\infty
d \mu \, \mu \, e^{- \mu r} \, \eta_C^{(2)} (\mu) \,  e^{-
  \frac{\mu^2}{2 \Lambda^2}}\,.
\eeq 
\item
Next, we follow the opposite approach and retain only the
momentum-dependent part of the regulator without introducing
spectral function regularization. The regularized potential is defined 
by means of the twice subtracted spectral integral 
\beq
\label{WC2}
W_{C, \, \Lambda , \, 2}^{(2)} (q) =e^{-\frac{q^2}{2 \Lambda^2}} \;
\frac{2}{\pi} \int_{2 M_\pi}^\infty \, \frac{d \mu}{\mu^3}
\eta_{C}^{(2)} (\mu) \, \frac{q^4} 
{\mu^2 + q^2} 
\,, \quad \quad W_{C, \, \Lambda , \, 2}^{(2)} (r) =\frac{1}{2 \pi^2} \int q^2 dq \, j_0 (q
r) \, W_{C, \, \Lambda , \, 2}^{(2)}  (q)\,.
\eeq
Alternatively, one can simply multiply  $W_{C, \, \infty}^{(2)} (q) $ by
the regulator $e^{-\frac{q^2}{2 \Lambda^2}}$, which leads to a
different admixture of contact interactions. We found, however,  that this definition
leads to larger distortions  at short distances as compared to the one in
Eq.~(\ref{WC2}). 
\item
In the third approach, the regularized potential is defined according
to Eq.~(\ref{disp_int_reg_NLO}) but without explicitly subtracting the
short-range terms, 
i.e.:
\beq
\label{WC3}
W_{C, \, \Lambda , \, 3}^{(2)} (q) = e^{-\frac{q^2}{2 \Lambda^2}} \;
\frac{2}{\pi} \int_{2 M_\pi}^\infty \, \frac{d \mu}{\mu^3}
\eta_{C}^{(2)} (\mu) \, \frac{q^4} 
{\mu^2 + q^2} 
\, e^{- \frac{\mu^2}{2
    \Lambda^2}} \,,
\eeq
and the Fourier transform to coordinate space can be performed using
the second relation in Eq.~(\ref{WC2}).  
\item
Finally, the approach to define the regularized potential $W_{C, \,
  \Lambda}^{(2)} (q)$ adopted in the present analysis is
\beq
\label{WC4}
W_{C, \, \Lambda , \, 4}^{(2)} (q) = e^{-\frac{q^2}{2 \Lambda^2}} \;
\frac{2}{\pi} \int_{2 M_\pi}^\infty \, \frac{d \mu}{\mu^3}
\eta_{C}^{(2)} (\mu) \bigg(\frac{q^4} 
{\mu^2 + q^2} + C_{C, 1}^2 (\mu ) +
C_{C, 2}^2 (\mu ) \, q^2\bigg) 
\, e^{- \frac{\mu^2}{2
    \Lambda^2}} \,,
\eeq
where the functions $C_{C, i}^2 (\mu )$ are determined as described
above and given in appendix \ref{app1}. 
\end{enumerate}
In Fig.~\ref{fig:RatiosWC}, 
\begin{figure}[tb]
\vskip 1 true cm
\begin{center}
\includegraphics[width=0.7\textwidth,keepaspectratio,angle=0,clip]{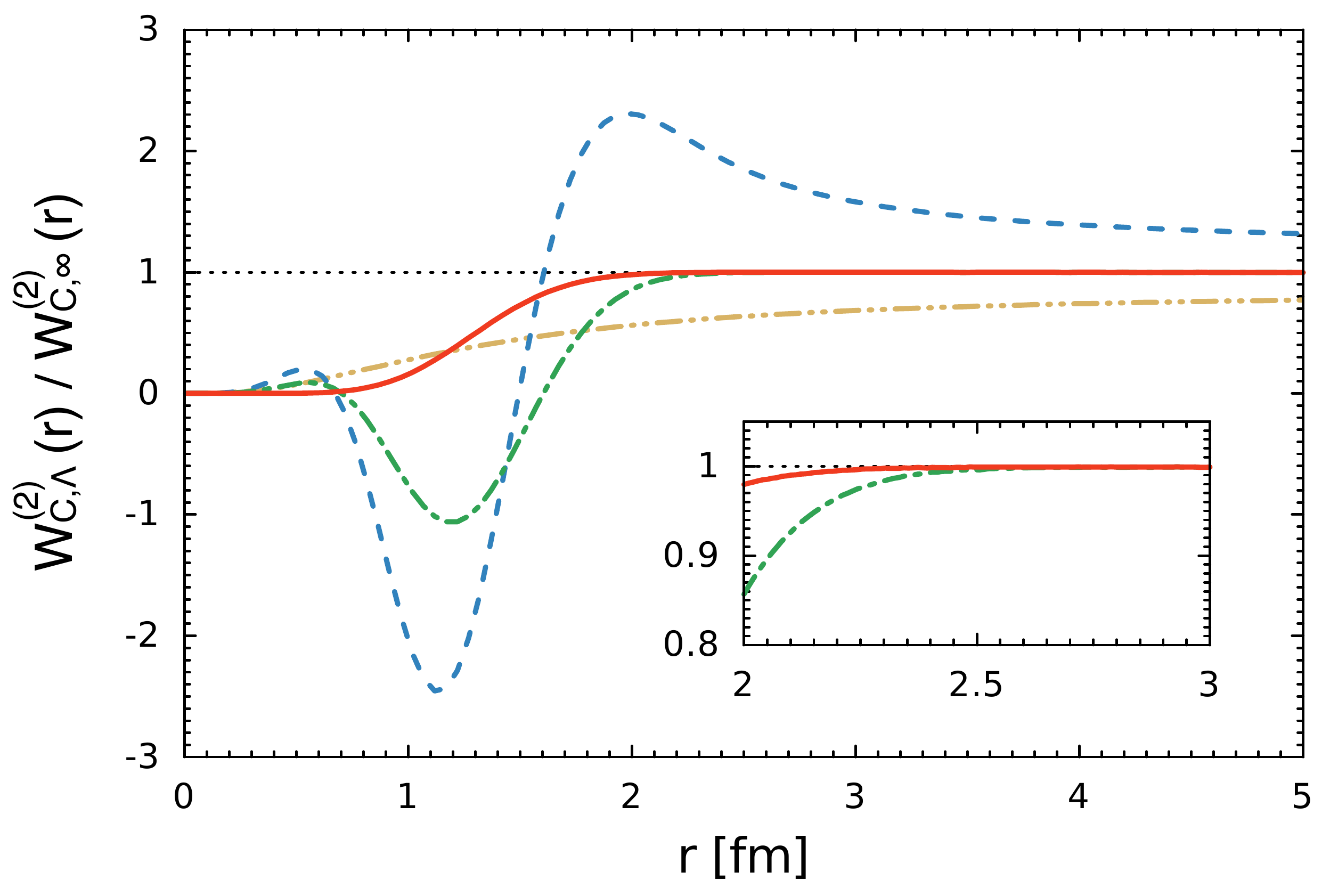}
\end{center}
    \caption{(Color online) Ratios $W_{C, \, \Lambda , \, i}^{(2)} (r)
      / W_{C, \, \infty}^{(2)} (r)$ for different implementations
      of the regularization $i=1, \ldots, 4$ defined in the
      text as a function of the relative distance between the nucleons. Dashed-double-dotted light-brown, dashed blue,
      dashed-dotted green and solid red lines refer to $i=1$, $2$, $3$
      and $4$, respectively. The cutoff $\Lambda$ is set to be
      $\Lambda = 450$~MeV. The dotted horizontal line corresponds to
      the 
      unregularized potential, i.e., the ratio is equal to $1$. 
\label{fig:RatiosWC}
}
\end{figure}
we show the ratios of the potentials
$W_{C, \, \Lambda , \, i}^{(2)} (r)$, with $i = 1, \ldots , 4$,  to the
unregularized expression $W_{C, \, \infty}^{(2)} (r)$
As before, we use the intermediate value of the cutoff of $\Lambda = 450$~MeV. 
Retaining either only the momentum-transfer- or the $\mu$-dependent part
of the regulator in Eq.~(\ref{SpectralReg}), i.e.~refraining from performing the
spectral function or momentum regularization, strongly affects the
long-range part of the TPEP and results in large deviations from
the unregularized TPEP even at rather large distances. While both
the third and fourth approaches maintain the long-range part of the TPEP,
one observes that our final approach, which minimizes the amount of
short-range contributions in the long-range potentials, leads to the
smoothest behavior at intermediate distances with the smallest amount of
distortions, similarly to what has been observed in the case of the
OPEP in Fig.~\ref{fig:OPEP}. Notice further that the resulting
coordinate-space behavior of the regularized potentials is
qualitatively similar to that of potentials in Ref.~\cite{Epelbaum:2014efa}. 

\begin{figure}[tb]
\vskip 1 true cm
\begin{center}
\includegraphics[width=1.0\textwidth,keepaspectratio,angle=0,clip]{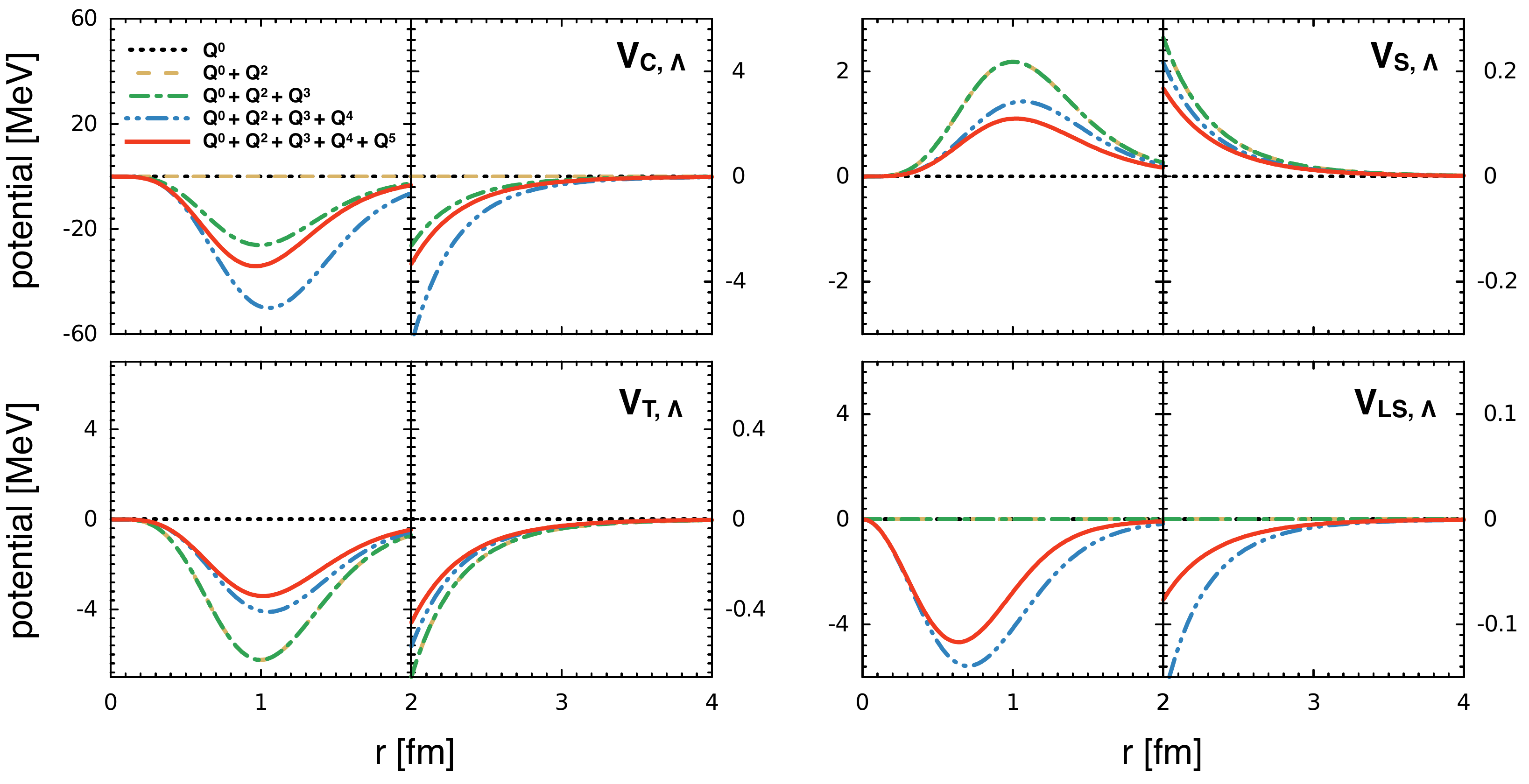}
\end{center}
    \caption{(Color online) Chiral expansion of the regularized
      long-range isoscalar potentials $V_{i, \, \Lambda} (r)$ for the cutoff value of $\Lambda =
      450$~MeV. Black dotted, orange dashed, green dashed-dotted, blue
      dashed-double-dotted and red solid lines refer to LO, NLO,
      N$^2$LO, N$^3$LO and N$^4$LO, respectively. 
\label{fig:V}
}
\end{figure}

\begin{figure}[tb]
\vskip 1 true cm
\begin{center}
\includegraphics[width=1.0\textwidth,keepaspectratio,angle=0,clip]{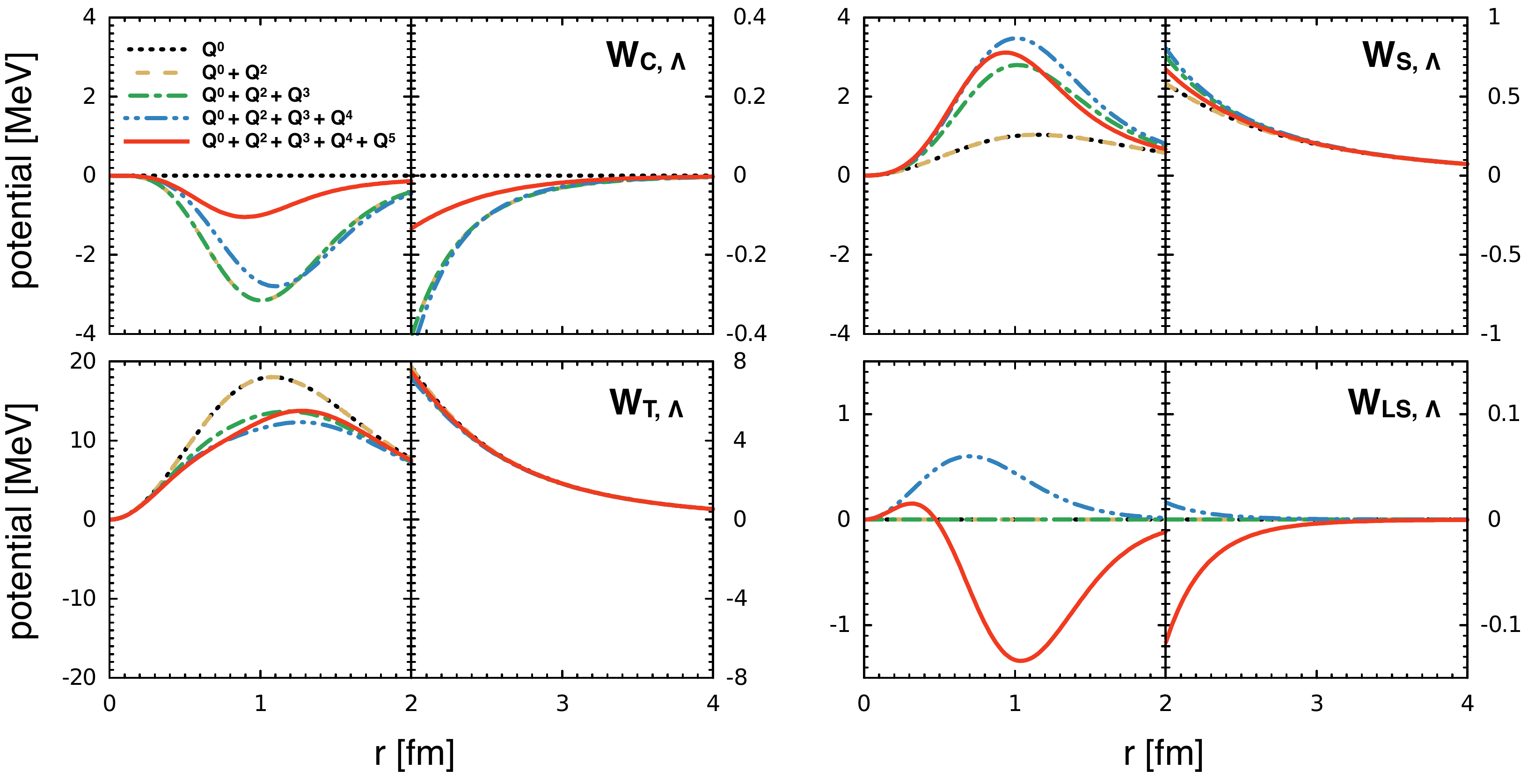}
\end{center}
    \caption{(Color online) Chiral expansion of the regularized
      long-range isovector potentials $W_{i, \, \Lambda} (r)$ for the cutoff value of $\Lambda =
      450$~MeV. Black dotted, orange dashed, green dashed-dotted, blue
      dashed-double-dotted and red solid lines refer to LO, NLO,
      N$^2$LO, N$^3$LO and N$^4$LO, respectively. 
\label{fig:W}
}
\end{figure}

After these preparations, we have all the necessary tools to address the
convergence of the chiral expansion for the long-range interactions in
a meaningful way. 
In Figs.~\ref{fig:V} and \ref{fig:W}, we show the chiral expansion of
the long-range isoscalar and isovector potentials $V_{i, \, \Lambda} (r)$ and $W_{i, \,
  \Lambda} (r)$. 
While we employ in this work the cutoff
values in the range of $\Lambda = 350 \ldots 550$~MeV, here we only
show the results for the intermediate cutoff value of $\Lambda =
450$~MeV as representative examples. The values of the pion-nucleon
($\pi N$)
LECs are taken
from the recent analysis of $\pi N$ scattering in the framework
of the Roy-Steiner equation \cite{Hoferichter:2015tha} as discussed in section \ref{sec:Parameters}. 
Notice that the vanishing of the
corresponding central and spin-spin potentials at short distances is
enforced by the adopted convention in our definition of the
long-range contributions as discussed above.  It should be understood
that the short-distance behavior of  $V_{i, \, \Lambda} (r)$ and $W_{i, \,
  \Lambda} (r)$ is scheme dependent which manifests itself e.g.~in their
strong dependence on $\Lambda$. On the other hand, we found that the profile
functions are largely insensitive to the considered cutoff variation at
distances of $r \gtrsim 2$~fm.   The strongest long-range potentials
are generated by the OPEP in the isovector tensor channel and the TPEP
in the isoscalar singlet channel, the feature which follows from 
the large-$N_c$ analysis of nuclear forces \cite{Kaplan:1996rk} and is also
supported by phenomenological studies. Generally, one observes a
fairly good convergence of the chiral expansion except for the
cases where the corresponding potentials are weak such as especially the
isovector spin-orbit and, to a lesser extent, the isovector central potentials $W_{LS, \,
  \Lambda} (r)$ and $W_{C, \,  \Lambda} (r)$, respectively. Notice that while
in these cases the N$^4$LO contribution appears to be much larger than
the N$^3$LO one, its absolute size is comparable in magnitude with the
size of N$^4$LO contributions in other channels. Clearly, final
conclusions on the convergence of the chiral expansion can only be
drawn from looking at observables rather than profile
functions. Further discussion is thus
relegated to section \ref{sec4}. 


To summarize the main results of this section, we have introduced a
momentum-space regularization framework for pion exchange
contributions by regularizing the static propagators of pions
exchanged between different nucleons. Our approach maintains the
long-range part of the interaction and can be easily implemented at
the level of the spectral representation of the TPEP with no
need to recalculate loop integrals entering the definition of the
potential. The method is highly flexible and by no means restricted to
the cases where the spectral function representation is available. In
particular, it can be easily applied to the ring diagrams appearing in
the three-nucleon force starting from N$^3$LO \cite{Bernard:2007sp,Krebs:2013kha} by switching to
coordinate space. It can also be employed  straightforwardly to regularize the 
exchange electroweak charge and current operators derived using the
method of unitary transformation 
\cite{Kolling:2009iq,Kolling:2011mt,Krebs:2016rqz} or based on
time-ordered perturbation theory \cite{Pastore:2009is,Piarulli:2012bn,Baroni:2015uza}. 
Notice that in contrast with the simple coordinate-space
regularization scheme of Ref.~\cite{Epelbaum:2014efa}, the new approach, being less
phenomenological, determines unambiguously the form of the regulator
in e.g.~the ring contributions to the three-nucleon force.

\section{Scattering amplitude in the presence of electromagnetic interactions}
\def\theequation{\arabic{section}.\arabic{equation}}
\label{sec2}

In addition to the strong nuclear force, the inclusion of
electromagnetic interactions is mandatory for an accurate description
of $NN$ observables. We follow the treatment of the Nijmegen PWA
\cite{Stoks:1993tb} and include for np scattering the magnetic
moment potential  
\begin{equation}
  V_{MM}^{np}(r) = -\frac{\alpha\kappa_n}{2m_nr^3}\left[ \frac{3\mu_p}{2m_p}S_{12} + \frac{1}{m_r}\left( \vec{L} \cdot \vec{S} + \vec{L} \cdot \vec{A} \right) \right],
\end{equation}
where $\vec{A} =
(\vec{\sigma}_1-\vec{\sigma}_2)/2$ . Furthermore,
$m_r$ is the the np reduced mass, $\mu_p$ is the proton
magnetic moment and $\kappa_n$ denotes the neutron anomalous magnetic
moment. For pp scattering, we include the so-called
improved Coulomb potentials $V_{C1}$, $V_{C2}$, the magnetic
moment potential $V_{MM}^{pp}$ and the vacuum polarization potential
$V_{VP}$: 
\begin{align}
  V_{C1}(r) =& \frac{\alpha'}{r},
  \\
  V_{C2}(r) =& -\frac{1}{m_p^2} \left[ (\Delta+q^2) \frac{\alpha}{r} + \frac{\alpha}{r}(\Delta+q^2) \right] \approx -\frac{\alpha \alpha'}{m_p^2} \frac{1}{r^2},
  \\
  V_{VP}(r) =& \frac{2\alpha}{3\pi} \frac{\alpha'}{r} \int_1^\infty e^{-2m_erx} \left(1+\frac{1}{2x^2} \right) \frac{\sqrt{x^2-1}}{x^2} dx,
  \\
  V_{MM}^{pp}(r) =& -\frac{\alpha}{4m_p^2r^3}\left[ 3\mu_p^2 S_{12} + (6+8\kappa_p)\vec{L}\cdot\vec{S} \right].
\end{align}
The energy-dependent factor $\alpha'$ is defined as
\begin{equation}
  \alpha' = \alpha \left(1+ \frac{2q^2}{m_p^2} \right) \left(1+ \frac{q^2}{m_p^2}\right)^{-\frac{1}{2}}.
\end{equation}

The magnetic moment potential $V_{MM}^{np/pp}$ has been derived from
the one-photon-exchange diagram while $V_{C1}$ and $V_{C2}$ also
include contributions from the two-photon-exchange box and cross-boxed
diagrams \cite{Austen:1983te}. The vacuum polarization 
is a loop correction and of a finite range as opposed to the other
electromagnetic (e.m.) interactions. In spite of its finite (but still very long) range, it can
be important at low energies \cite{Bergervoet:1988zz}. 

The potentials given above only represent the long-range part of the
e.m.~interaction. The short-range
part of those interactions can be represented by contact terms which are of the
same type as the contact interactions already included in the nuclear
potential. Thus, these effects are implicitly incorporated into
the values of the corresponding LECs. 

The partial wave expansion of the amplitude converges only slowly
in the presence of long-range e.m.~interactions. Therefore, the
e.m.~amplitudes are separated from the nuclear amplitude, allowing
fast recalculation of the nuclear amplitude during the fitting process
by summing
nuclear partial waves up to the total angular momentum $j=20$, while the
e.m.~amplitudes need to be calculated only once. The total amplitudes
are the given by 
\begin{align}
	M_{np} &= M_{MM} + M_{N}, \\
	M_{pp} &= M_{C1} + M_{C2} + M_{MM} + M_{VP} + M_{N}.
\end{align}
In the case of np scattering both the amplitude and the nuclear
phase shifts are calculated with respect to Riccati-Bessel functions. The
analytic expression of the magnetic moment amplitude in terms of the Saclay
amplitudes can be found e.g. in \cite{Stoks:1990us,LaFrance:1980},
where the point-particle approximation corresponds to the following
Dirac and Pauli form factors: 
\begin{equation}
  F_1^p = 1, \quad F_1^n = 0, \quad F_2^p = \frac{\kappa_p}{2m_p}, \quad F_2^n = \frac{\kappa_n}{2m_n},
\end{equation}
with the anomalous magnetic moments $\kappa_p = 1.792847$ and $\kappa_n = -1.913043$.

Due to the amplitude separation, the nuclear partial wave $S$-matrix
elements have to be adjusted as
$S_{N}-1\rightarrow\left(S_{MM}\right)^{1/2}\left(S_{N}-1\right)\left(S_{MM}\right)^{1/2}$
when calculating the nuclear amplitude, with $S_{MM}$ being the
partial wave $S$-matrix elements of the magnetic moment interaction. We
again refer to \cite{Stoks:1990us} for explicit expressions and a detailed
discussion. 

For pp scattering, the potentials $V_{C1}$, $V_{C2}$ and $V_{VP}$
are spin-independent and their anti-symmetrized spin-scattering
matrix is given by 
\begin{equation}
  M_{m'_s,m_s}^s = \left[ f_{X}(\theta) + (-1)^s f_{X}(\pi-\theta) \right]\delta_{m'_sm_s},
\end{equation}
where $X$ can be $C1$, $C2$ or $VP$. The amplitude $f_{C1}$ is given by
\begin{equation}
  f_{C1}(\theta) = \frac{1}{2iq} \sum_l (2l+1) \left(
    e^{2i(\sigma_l-\sigma_0)}-1 \right) P_l(\theta) = -\frac{\eta'}{q}
  \frac{e^{i\eta'\ln \left[ \tfrac{1}{2}(1-\cos(\theta))\right] }}{1 - \cos(\theta)},
\end{equation}
with $\eta'=\frac{m_p}{2q}\alpha'$ and the Coulomb phase shifts $\sigma_l = \arg \Gamma(l+1+i\eta)$.
Notice that the amplitude includes an additional phase factor of
$e^{-2i\sigma_0}$. The other pp scattering e.m.~amplitudes are
determined with respect to the $V_{C1}$ interaction in the so-called
Coulomb distorted-wave Born approximation (CDWBA) and include the
$e^{-2i\sigma_0}$ phase as well. 
   
For $f_{C2}$, one has to sum over the partial wave contributions to the amplitude
\begin{equation}
  f_{C2}(\theta) = \frac{1}{2iq}\sum_{l=0}^\infty e^{2i(\sigma_l-\sigma_0)} \left( e^{2i\rho_l}-1 \right) P_l(\theta)
\end{equation}
numerically, employing the corresponding phase-shifts $\rho_l$ given by
\begin{equation}
  \rho_l = \sigma_\lambda - \sigma_l + \frac{(l-\lambda)\pi}{2},
\end{equation}
with $\lambda = \tfrac{1}{2}(-1 + \sqrt{(2l+1)^2 -
  \alpha\alpha'})$. In our calculations, the sum in $f_{C2}$ is
performed up to $l=1000$.

The CDWBA vacuum polarization (VP) amplitude to first order in the VP phase
shifts $\tau_l$ is derived by Durand \cite{Durand:1957zz}: 
\beqa
  \label{vp_amplitude}
  f_{VP}(\theta) &=& \frac{\alpha\eta'}{3\pi q} \left|\Gamma(1+i\eta')\right|^2 \int_{0}^{1} \frac{(1+\tfrac{y}{2})\sqrt{1-y}}{(1-\cos\theta)y+\nu} \left( \frac{2y+\nu}{(1-\cos\theta)y+\nu} \right)^{i\eta'} \nonumber\\
  &\times& \exp\left(-2\eta'\arctan\left(\sqrt{\tfrac{\nu}{2y}}\right)\right) {}_2F_1\left( -i\eta',1+i\eta',1;\frac{y}{y+X} \right) dy,
\eeqa
where $\nu = 2m_e^2/q^2$, $X= \nu/(1-\cos\theta)$ and $_2F_1(a,b,c;z)$
is the hypergeometric function. The integral in
Eq.~\eqref{vp_amplitude} can be evaluated numerically via
quadrature.  Analogously, we also evaluate the expressions for the VP phase shifts
$\tau_{l}$ numerically. 
\begin{align}
	\tau_l =& -\frac{a\eta'}{3\pi} \int_0^1 \left( 1+\frac{y}{2} \right) \frac{\sqrt{1-y}}{y} M_l \left( 1+ \nu y^{-1} \right) dy,\nonumber\\
\end{align}
with
\beqa
	M_l \left( 1+\nu y^{-1} \right) &=& \frac{1}{2} e^{-\pi\eta'} \left( \frac{2y}{\nu+2y} \right)^{l+1} \left( \frac{\sqrt{\nu}-i\sqrt{2y}}{\sqrt{\nu}+i\sqrt{2y}} \right)^{i\eta'} \frac{|\Gamma(l+1+i\eta')|^2}{\Gamma(2l+2)} \nonumber\\
	&\times& {}_2F_1\left( l+1+i\eta', l+1-i\eta', 2l+2; \frac{2y}{\nu+2y} \right).
\eeqa
The magnetic moment interaction for pp scattering has been derived
by Stoks and de Swart \cite{Stoks:1990us}. Like in the case of $f_{C2}$, there
exists no simple analytic expression for the scattering amplitude in CDWBA 
and, instead, the corresponding partial wave $S$-matrix contributions have to be
summed up numerically using Eq.~\eqref{pwd_amplitude}. 
The authors of Ref.~\cite{Stoks:1990us} have found that $M_{10}^1$ and $M_{01}^1$ are not sufficiently
converged when the sum is performed up to $j=1000$\footnote{We have
  verified that this is indeed the case.}, but that the biggest
contribution to these amplitudes, 
\begin{equation}
  \label{eq:pwZLS}
  Z_{LS} \vcentcolon = -\sqrt{2\pi} m_p f_{LS} \sum_{j\;\rm even} e^{2i(\sigma_j-\sigma_{0})} \sqrt{\frac{2j-1}{j(j-1)}} Y_{j1}(\theta,0) 
\end{equation}
for $M_{10}^1$ and $-Z_{LS}$ for $M_{01}^1$, can be summed up
analytically leading to 
\begin{equation}
  \label{eq:ZLS}
  Z_{LS} = \frac{m_pf_{LS}}{\sin\theta\sqrt{2}}\left( e^{-i\eta\ln[\frac{1}{2}(1-\cos\theta)]} + e^{-i\eta\ln[\frac{1}{2}(1+\cos\theta)]} - 1\right).
\end{equation}

It remains to calculate the nuclear amplitude in accordance to the
e.m.~amplitude separation. Like in the case of the
np amplitude, the nuclear partial wave $S$-matrix elements have
to be adjusted with the partial wave matrices of the included
e.m.~interactions. 
\begin{align}
\label{smatrix-em-adjustment}
\left( S_{EM+N} -1\right) \rightarrow & \left(S_{C1}\right)^{1/2}
                                        \left(S_{C1+C2}^{C1}\right)^{1/2}
                                        \left(S_{C1+MM}^{C1}\right)^{1/2}
                                        \left(S_{C1+VP}^{C1}\right)^{1/2}
                                        \left( S_{EM+N}^{EM} -
                                        1\right) \nonumber \\ \times&
                                                                      \left(S_{C1+VP}^{C1}\right)^{1/2} \left(S_{C1+MM}^{C1}\right)^{1/2} \left(S_{C1+C2}^{C1}\right)^{1/2} \left(S_{C1}\right)^{1/2} 
\end{align}
Here we follow the notation of \cite{Bergervoet:1988zz} and denote with
e.g.~$S_{EM+N}^{EM}$ the partial wave $S$-matrix of all
e.m.~interactions plus the nuclear interaction with respect to the
e.m.~wavefunctions. Analogously,  $S_{C1+X}^{C1}$ corresponds to the
CDWBA of an arbitrary interaction $X$. 
The nuclear phase-shifts, however, are calculated with respect to $C1$
alone, employing the method of Vincent and Phatak
\cite{Vincent:1974zz}. Using these phase-shifts in the construction of
the amplitude, i.e. $\delta_{EM+N}^{EM} \approx \delta_{C1+N}^{C1}$,
is sufficiently accurate except for the ${}^1S_0$ partial wave \cite{Stoks:1993tb}. Here
we use the decomposition
\begin{equation}
  \label{1S0pp-corrections}
  \left(\delta_{EM+N}^{EM}\right)_0 = \left(\delta_{C1+N}^{C1}\right)_0 + \tilde{\Delta}_0 - \rho_0 - \tau_0.
\end{equation}
The so-called improved Coulomb-Foldy correction $\tilde{\Delta}_0$
has been calculated for the Nijmegen78 potential and tabulated in the
range of $E_{\rm lab} = 0 - 30$~MeV in Ref.~\cite{Bergervoet:1988zz}. It is
argued in that paper  that the model-dependence of this quantity is small, and
we use the corresponding values from Ref.~\cite{Bergervoet:1988zz} in the
given energy range. 
Above $E_{\rm lab} = 30$~MeV, the corrections to 
$\delta_{C1+N}^{C1}$ in Eq.~\eqref{1S0pp-corrections} are small and can 
be neglected.

\section{Fits to the data}
\def\theequation{\arabic{section}.\arabic{equation}}
\label{sec3}

\subsection{Parameters of the potential}
\label{sec:Parameters}

We are now in the position to specify the values of the nucleon masses
and various parameters entering the long-range part of the potentials. 
For the proton and neutron masses we use the
values of $m_p=938.272$~MeV and $m_n=939.565$~MeV, respectively. 
When calculating the relativistic corrections to the NN interaction,
the average value of the nucleon mass of $m_N = 2 m_p m_n/(m_p + m_n)
= 938.918$~MeV is adopted.  
As explained in section \ref{sec:reg}, we take into account isospin-breaking
corrections to the OPEP due to the different values of the charged and
neutral pion masses, for which the values of $M_{\pi^\pm} =
139.57$~MeV and $M_{\pi^0} =
134.98$~MeV are employed. In the expressions for the TPEP, we use the
pion mass value of $M_\pi = 138.03$~MeV. Following our recent work
\cite{Epelbaum:2014efa,Epelbaum:2014sza}, we adopt the value of $F_\pi = 92.4$~MeV for the pion decay
constant. We set the LECs $d_{18}$, which accounts for the 
Goldberger-Treiman discrepancy, equal to zero and employ the shifted 
value of $g_A = 1.29$ for the nucleon
axial coupling constant, which is  larger than the experimental one
of $g_A \simeq 1.267$. This yields for the  pion-nucleon coupling
constant $g_{\pi N} = g_A m_N/F_\pi$ the value of $g_{\pi N}^2/(4 \pi)
= 13.67$, which is consistent with the recent determination based on
the Goldberger-Miyazawa-Oehme sum rule $g_{\pi N}^2/(4 \pi)
= 13.69 \pm 0.20$ \cite{Baru:2010xn,Baru:2011bw}. 

For the pion-nucleon LECs, we employ the central values from Ref.~\cite{Hoferichter:2015tha},
obtained by matching the solution of the Roy-Steiner equations to
chiral perturbation theory at the subthreshold point as specified in
Table \ref{piNLECs} (set 1). Notice that this paper also quotes
uncertainties related to the accuracy of the subthreshold parameters
which, however, turn out to be rather small.  
\begin{table}[t]
\caption{Values of the $\pi N$ LECs $c_i$, $\bar d_i$ and $\bar e_i$
  in units of GeV$^{-1}$,  GeV$^{-2}$ and GeV$^{-3}$, respectively,
  used in the semilocal momentum-space regularized potentials of this
  work (set 1). Values given in set 2 were employed in the
  potentials of Refs.~\cite{Epelbaum:2014efa,Epelbaum:2014sza} and are
  used in the error analysis to assess the sensitivity 
  of our results to the  $\pi N$ LECs as discussed in section \ref{sec:UncertaintypiN}. 
\label{piNLECs}}
\smallskip
\begin{tabular*}{\textwidth}{@{\extracolsep{\fill}}lrrrrrr}
\hline 
\hline
\noalign{\smallskip} 
& \multicolumn{3}{c}{ ---  set 1
    --- } &
\multicolumn{3}{c}{ --- set 2  --- } 
\\
   &   N$^2$LO    &  N$^3$LO   & N$^4$LO   & 
N$^2$LO    &  N$^3$LO   & N$^4$LO 
\smallskip
 \\
\hline 
\hline
\smallskip
$c_1$ & $-0.74$ &  $-1.07$ &  $-1.10$ &  $-0.81$ &  $-0.81$ &  $-0.75$
  \\
$c_2$ & --- &  $3.20$ &  $3.57$ &  --- &  $3.28$ &  $3.49$
  \\
$c_3$ & $-3.61$ &  $-5.32$ &  $-5.54$ &  $-4.69$ &  $-4.69$ &  $-4.77$
  \\
$c_4$ & $2.44$ &  $3.56$ &  $4.17$ &  $3.40$ &  $3.40$ &  $3.34$
  \\
$\bar d_1 + \bar d_2$ & --- &  $1.04$ &  $6.18$ &  --- &  $3.06$ &  $6.21$
  \\
$\bar d_3$ & --- &  $-0.48$ &  $-8.91$ &  --- &  $-3.27$ &  $-6.83$
  \\
$\bar d_5$ & --- &  $0.14$ &  $0.86$ &  --- &  $0.45$ &  $0.78$
  \\
$\bar d_{14} - \bar d_{15}$ & --- &  $-1.90$ &  $-12.18$ &  --- &  $-5.65$ &  $-12.02$
  \\
$\bar e_{14}$ & --- &  --- &  $1.18$ &  --- &  --- &  $1.52$
  \\
$\bar e_{17}$ & --- &  --- &  $-0.18$ &  --- &  --- &  $-0.37$
  \\
\hline \hline
\end{tabular*}
\end{table}

To estimate the sensitivity of our results to the values of the $\pi
N$ LECs, we will also perform calculations using the values from
set 2, which were employed in our recent studies
\cite{Epelbaum:2014efa,Epelbaum:2014sza},  see section
\ref{sec:UncertaintypiN} for more details. The LECs $c_1$, $c_3$ and
$c_4$ from set 2 used at N$^2$LO and N$^3$LO and listed in the fifth and
sixth columns of Table \ref{piNLECs}, respectively, correspond to the
central values of the
determination of Ref.~\cite{Buettiker:1999ap} from $\pi N$ scattering
inside the Mandelstam triangle.  The LEC $c_2$ could not be reliably extracted in
that analysis, and the quoted value of $c_2 = 3.28$~GeV$^{-1}$
is taken from the order $Q^3$ calculation of $\pi N$ scattering of Ref.~\cite{Fettes:1998ud} in the
framework of heavy-baryon chiral perturbation theory (in the physical
region).  Also the values
of the LECs $\bar d_i$ used at N$^3$LO and listed in the sixth column of Table
\ref{piNLECs} are taken from fit 1 of that paper. Finally, the values
of the LECs used at N$^4$LO and given in the last column of  Table \ref{piNLECs})
correspond to the order-$Q^4$ extraction of Ref.~\cite{Krebs:2012yv}
based on the Karlsruhe-Helsinki partial 
wave analysis of $\pi N$ scattering \cite{Koch:1985bn}. 
We emphasize once again that the $\pi N$ LECs from set 2 are only
used in the uncertainty estimations, while all our results are
obtained with the LECs from the Roy-Steiner equation analysis (set
1). We refer the reader to
Refs.~\cite{Bernard:1995dp,Bernard:2007zu,Hoferichter:2015hva} for
review articles on the 
applications of chiral perturbation theory in the single-nucleon
sector. 

Finally, for the cutoff $\Lambda$, we consider in this study the
values in the range of $\Lambda = 350 - 550$~MeV, namely $\Lambda =
350$, $400$, $450$, $500$ and $550$~MeV.  As will be shown in
section \ref{sec:Data}, choosing $\Lambda = 350$~MeV already leads to 
large regulator artefacts in the description of NN scattering, and for
this reason we do not consider lower values of $\Lambda$.  On the
other hand, increasing $\Lambda$ beyond the highest considered value
of $\Lambda = 550$~MeV leads to highly nonperturbative potentials, see
the discussion in section \ref{sec:WE}, and eventually to the emergence of spurious
deeply-bound states signaling that one enters the highly nonlinear
regime, in which the arguments based on naive dimensional analysis
and assumed in our study become invalid. Notice further that the
cutoff $\Lambda = 550$~MeV in the local regulator $\exp{(- q^2/
  \Lambda^2)}$ corresponds to the distance of $R = 2 \Lambda^{-1}
\simeq 0.7$~fm, which is close to the breakdown distance of the chiral
expansion for the multi-pion exchange potential estimated in
Ref.~\cite{Baru:2012iv}.   We also emphasize that
strongly nonperturbative NN potentials featuring spurious deeply bound
states are of little use for {\it ab initio} few- and many-body
calculations.

\subsection{Database}
\label{sec:data}

For experimental data, we use the SAID database \cite{Gross-db}
containing elastic NN scattering data from 1950 up to 2008. The data
are divided into different experiments, which consist of individual
measurements taken either at a fixed beam energy and different angles
(angle-dependent observables) or at different energies (total cross
sections or observables measured at a fixed angle). 

Not all data are mutually compatible, and the exclusion of outliers is
important in order to obtain good fits. We, therefore, restrict the database
to the mutually compatible data listed in \cite{Perez:2013jpa}, which
is also referred to as the 2013 Granada database. The only data set from
the 2013 Granada database not included in our analysis is that from
Ref.~\cite{Daub:2012qb}, which should be updated by more accurate
data from an improved version of the experiment carried out by the
same group \cite{June}. 
The
database was obtained by iteratively fitting all scattering data in
the range of $E_{\rm lab}=0-350$ MeV to a delta-shell potential (and
including the one-pion exchange and electromagnetic potentials for the
long-range part of the interaction) and then rejecting those
experimental data whose $\chi^2$-residuals lay outside of the 99.46\%
confidence interval of the $\chi^2$-distribution. This data-selection
procedure is also known as $3\sigma$-criterion. The self-consistency
of the selected data was verified in a subsequent publication
\cite{Perez:2014yla}.

In our analysis, the values of $\chi^2$ are calculated up to the 
maximum energy of $E_{\rm lab}=300$ MeV. The resulting  database,
restricted to the energy range of $E_{\rm lab}=0-300$, consists of 2697
np and 2158 pp experimental data. Notice that the actual number of
experimental data used in a fit depends on the energy range which is
chosen differently from order to order, see section \ref{sec:fitting} for more
details.

\subsection{Fitting procedure}
\label{sec:fitting}

Our fitting procedure is divided into two steps: first, we fit phase
shifts and mixing angles to the Nijmegen PWA \cite{Stoks:1993tb} in
the same way as it was done for the EKM
potentials of Ref.~\cite{Epelbaum:2014efa,Epelbaum:2014sza}. The
resulting LECs then serve as a starting point for the second
step, in which the optimization with respect to scattering data is performed. The
fit to phase shifts allows us to decompose the parameter space of the
LECs accompanying the contact interactions into smaller subspaces, 
since each LEC enters only one partial wave. Thus, fits can be
performed on the partial wave basis.  This remains
true if the contact potential is regularized according to
Eq.~\eqref{RegMom}. In this way we avoid the need to perform much more
costly searches in the whole parameter space including parameter values 
far away from the physical minimum. The fitting procedure for
the phase shifts is described in detail in Ref.~\cite{Epelbaum:2014efa}. However,
the removal of the redundant contact interactions renders the
additional constraints on the approximate Wigner SU(4) symmetry and
the deuteron D-state probability employed in
Ref.~\cite{Epelbaum:2014efa} obsolete since we do not observe any
ambiguities in the $^3S_1 - {}^3D_1$ channel without them. 
 
Having determined the initial LECs from phase shifts, we refine them
in a direct fit to scattering data. The comparison
between experimental and calculated scattering observables is done via
a standard $\chi^2$ measure whose value is minimized during the
fit. We employ the same definition of $\chi^2$ as used in
Ref.~\cite{Bergervoet:1988zz,Stoks:1993tb,Gross:2010qm,Perez:2013jpa,Ekstrom:2014dxa}. 
For the sake of completeness, we provide below the definition
of the $\chi^2$ function. 

The experimental data are grouped in experiments as discussed in
Section \ref{sec:data}. The total $\chi^2$ is the sum over all
experiments and the contribution of the j\textsuperscript{th}
experiment $\chi^2_j$ comprising $n_j$ measured observable values is
defined as 
\begin{equation}  
  \label{chi2}
  \chi^2_j = \sum_{i=1}^{n_j} \left(\frac{O_i^{\rm exp}-ZO_i^{\rm
        theo}}{\delta O_i}\right)^2 + \left( \frac{Z-1}{\delta_{\rm
        norm}}\right)^2, 
\end{equation}
 where $O_i^{\rm exp}$ is the measured value of the observable,
 $O_i^{\rm theo}$ is its calculated value while $\delta O_i$ is the
 (statistical) error. Furthermore, $\delta_{\rm norm}$ is the
 normalization error of the data set and $Z$ is the inverse of the estimated normalization. 
 The contribution of the term $\chi^2_{\rm norm} = (Z-1)^2\delta_{\rm norm}^{-2}$ 
 in Eq.~\eqref{chi2} can be thought of 
 as an additional data point for the overall normalization. In the
 case of absolute data ($\delta_{\rm norm}=0$), we have $Z=1$ and 
 $\chi^2_{\rm norm} = 0$. For experiments which give a normalization 
 error $\delta_{\rm norm} \ne 0$, the inverse normalization estimator 
 $Z$ is calculated as 
\begin{equation}
  \label{eq:normalizationestimator}
  Z = \left( \sum_{i=1}^{n}\frac{O_i^{\rm exp}O_i^{\rm theo}}{\delta O_i^2} +
    \frac{1}{\delta_{\rm norm}^2}\right) \left(
    \sum_{i=1}^{n}\frac{\left(O_i^{\rm theo}\right)^2}{\delta O_i^2} +
    \frac{1}{\delta_{\rm norm}^2} \right)^{-1},
\end{equation}
which minimizes $\chi^2_j$ with respect to $Z$. In this way, all $\chi^2_j$
become independent of the norms $Z$ and only the LECs entering the 
NN potential have to be treated as free parameters in the fit. Some
experiments allow the normalization of the data to float thus
providing relative data which have essentially $\delta_{\rm
  norm}=\infty$. Further, if data with  
$\delta_{\rm norm} < \infty$ yield $\chi^2_{\rm norm} > 9$, the estimated 
normalization is deemed incompatible with the given normalization error 
$\delta_{\rm norm}$ and these data are also treated as floating (i.e we 
calculate $Z$ and $\chi^2_j$ with $\delta_{\rm norm} = \infty$). This
corresponds to the application of the $3\sigma$-criterion to the norm $Z$
as a single data point.

One note is in order regarding the $\chi^2$/datum values. We again follow
the convention of Refs.~\cite{Stoks:1993tb,Perez:2013mwa} and divide $\chi^2$ by the total number of data
$N=N_{\rm exp} + N_{\rm norm}$, which consists of the number of measurements
$N_{\rm exp}$ and the number of estimated normalizations $N_{\rm norm}$ for
all experiments which have not been floated. %Compared to the
                                %convention of counting degrees of
                                %freedom, which will be reduced by the
                                %estimation of $Z$, this will lead to
                                %lower $\chi$/datum values. However,
                                %the true number of \textit{dofs}
                                %cannot be reliably estimated for
                                %non-linear fits. \cite{Andrae:2010gh}     
                                
The $\chi^2$-function defined above gets minimized in a simultaneous fit to
both neutron-proton and proton-proton scattering data and, therefore,
all NN contact LECs are varied at once with the exception of the
isospin-breaking $\tilde{C}_{1S0}^{nn}$ contact interaction in the $^1S_0$ phase,
whose determination requires neutron-neutron scattering
data. Its value is fixed afterwards from a fit to the neutron-neutron
scattering length, for which the value of $a_{1S0}^{\rm nn} = -18.90$~fm from
Ref.~\cite{Machleidt:2000ge} is taken. Notice, however, that this
scattering length is not well known experimentally, and the results from different
extractions and experiments yield different values.  

The energy range of included scattering data is chosen differently at
different chiral orders. Starting from N$^4$LO, we found that setting
the maximum fitting energy $E_{\rm lab}^{\rm max} \sim 250$~MeV leads
to stable results for the LECs and phase shifts for all employed
cutoff values in the range of $\Lambda = 350 - 550$~MeV. Given the  
abundance of np spin-observables available at the energies around  $E_{\rm lab} \sim 260$~MeV
and coming from the recent experiments of
\cite{Ahmidouch:1998gf,Arnold:2000vm,Arnold:2000vn},
we decided to set $E_{\rm lab}^{\rm max} = 260$~MeV at this chiral
order. We will address the sensitivity of our results on the choice of 
$E_{\rm lab}^{\rm max}$ in section \ref{sec:Errors}. 
At N$^3$LO, we fit up to $E_{\rm lab}=200$~MeV while the maximum energy of
included scattering data at LO, NLO
and N$^2$LO is chosen to be $E_{\rm lab}=25, 100$ and $125$~MeV. 
Similar energy ranges were used in the fits of the EKM potentials in
Refs.~\cite{Epelbaum:2014efa,Epelbaum:2014sza}. 

In addition to the scattering data, we also include the deuteron
binding energy $B_d = 2.224575(9)$~MeV \cite{DBD} and the world average
for the np coherent scattering length $b_{\rm  np} = -3.7405(9)$~fm
\cite{Schoen:2003my} to additionally stabilize our fits  starting from N$^3$LO. Tuning these two
observables to their experimental values by simply adding the two
corresponding data points to the $\chi^2$ does, however, significantly
affect the conditioning and thus worsens  the convergence of the fits. 
Instead of following this path, we use routines for constrained
optimization and set the deuteron  binding energy and the np coherent
scattering length to their central values as explicit
constraints while maintaining a fast convergence towards the minimum. 
Specifically, we use the Interior Point OPTimizer
(IPOPT) \cite{IPOPT} in our fits, which allows us to incorporate the
above mentioned constraints and second-derivative information. 

Finally, we comment on the fitting procedures at lower orders,
namely at LO, NLO and N$^2$LO. While we consider the potentials
starting from N$^3$LO accurate enough to perform a reliable estimation of the
relative norms $Z$ throughout the whole energy range, we allow neither for 
an estimation of $Z$ nor for floating data sets in case of a too high
value of $\chi^2_{\rm norm}$ based on those lower-order potentials. Instead, we
use this information from the highest order fit at the given cutoff
value and keep the normalizations unchanged during the
fit. We also do not include the deuteron binding energy and the np
coherent scattering length in the fits at these chiral orders.

\section{Results for the semilocal momentum-space regularized potentials}
\def\theequation{\arabic{section}.\arabic{equation}}
\label{sec4}

In the following sections, we provide a detailed description of our
results for the new family of semilocal momentum-space regularized chiral potentials. 

\subsection{Convergence of the chiral expansion}

\begin{figure}[tb]
\vskip 1 true cm
\begin{center}
\includegraphics[width=1.0\textwidth,keepaspectratio,angle=0,clip]{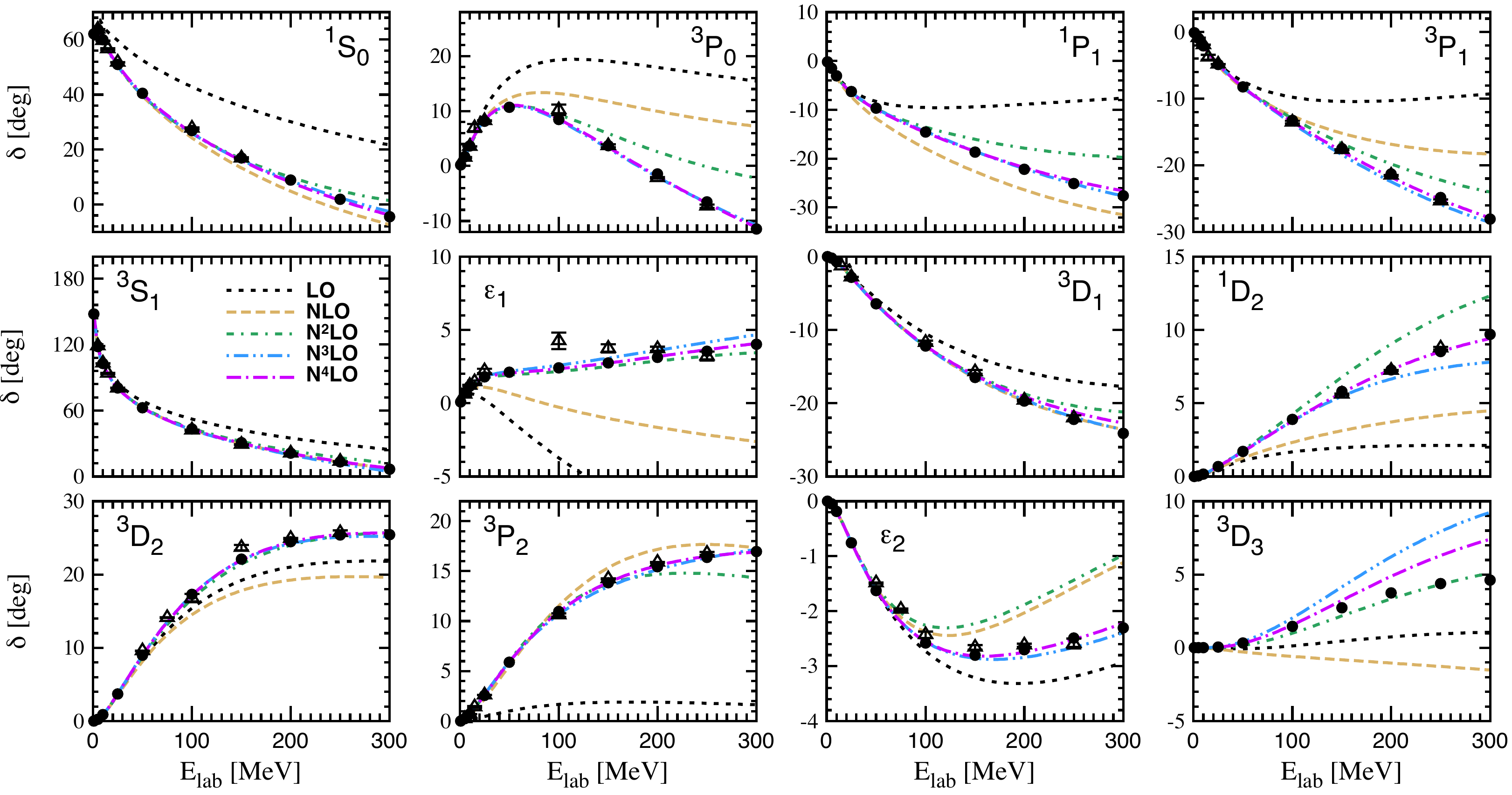}
\end{center}
    \caption{(Color online) Chiral expansion of the np phase shifts in
      comparison with the Nijmegen \cite{Stoks:1993tb} (solid dots)
      and the GWU \cite{Arndt:1994br}
      (open triangles) np partial wave analysis. Black dotted, orange
      dashed, green short-dashed-dotted, blue dashed-double-dotted and
      violet long-dashed-dotted lines  show the
      results at LO, NLO, N$^2$LO, N$^3$LO and N$^4$LO, respectively,
      calculated using the cutoff $\Lambda=450\,$MeV.  Only those
      partial waves are shown which  
     involve contact interactions at N$^4$LO.    
\label{fig:SPD}
}
\end{figure}

Having specified the fitting procedure, we are now in the position to
present our results. In Fig.~\ref{fig:SPD}, we show the resulting np phase
shifts and mixing angles in the fitted channels, namely~in the S-, P-
and D-waves and the mixing angles $\epsilon_1$ and $\epsilon_2$, in
comparison with the Nijmegen multi-energy \cite{Stoks:1993tb} and the GWU
single-energy \cite{Arndt:1994br}  partial wave analyses.  
While we restrict ourselves to the case of the
intermediate cutoff of $\Lambda = 450$~MeV throughout this section,
the results for other cutoff values are qualitatively
similar. Notice further that the P- and D-wave phase shifts and the mixing
angles $\epsilon_1$ and $\epsilon_2$ (the D-wave phase shifts and the mixing angle
$\epsilon_2$) are predicted in a parameter-free way at LO (up to
N$^2$LO). As expected, our results are similar to the ones
of Refs.~\cite{Epelbaum:2014efa,Epelbaum:2014sza}, where a coordinate-space version of
the local regulator was employed and the fits were performed to
the NPWA. In most of the channels shown, one observes a good
convergence of the chiral expansion with the results showing little
changes between N$^3$LO and N$^4$LO. 

\begin{figure}[tb]
\vskip 1 true cm
\begin{center}
\includegraphics[width=1.0\textwidth,keepaspectratio,angle=0,clip]{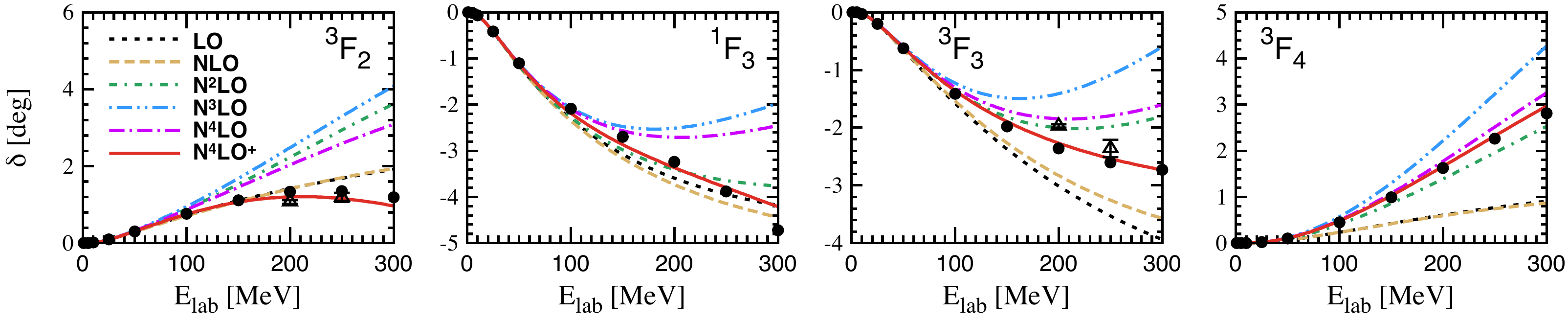}
\end{center}
    \caption{(Color online) Chiral expansion of the np F-wave phase shifts in
      comparison with the Nijmegen \cite{Stoks:1993tb} (solid dots)
      and the GWU \cite{Arndt:1994br}
      (open triangles) np partial wave analysis. Black dotted, orange
      dashed, green short-dashed-dotted, blue dashed-double-dotted, 
      violet long-dashed-dotted and red solid lines  show the
      results at LO, NLO, N$^2$LO, N$^3$LO, N$^4$LO and N$^4$LO$^+$, respectively,
      calculated using the cutoff $\Lambda=450\,$MeV.  
\label{fig:F}
}
\end{figure}

The situation is, however,
different for F-waves where the results are still not converged at the
level of N$^4$LO as shown in Fig.~\ref{fig:F}. 
Here, the differences between the N$^3$LO and N$^4$LO predictions are
clearly visible, and the empirical phase shifts are still not
reproduced at N$^4$LO.  To further elaborate on this issue, we
performed fits based on the N$^4$LO chiral potential and including
the leading F-wave contact interactions which apear at N$^5$LO
and are given in Eq.~(\ref{LECsF}).
Here and in what follows, the resulting partial
N$^5$LO potential is referred to as N$^4$LO$^+$. As will be discussed in
section \ref{sec:LECs}, the resulting N$^5$LO contact interactions appear to be of
a natural size. The difference between the N$^4$LO and N$^4$LO$^+$
results can thus be regarded as a lower bound of the N$^4$LO theoretical
uncertainty. In the resonance-saturation picture, it can be traced
back to the short-range contributions due to heavy-meson exchanges
which are not accounted for at the level of N$^4$LO. The fact that the
LECs $E_i$ come out of a natural size suggests that the poor
convergence pattern for F-waves shown in Fig.~\ref{fig:F} does not reflect
any failure of the chiral EFT. Rather, the $^3$F$_2$, $^1$F$_3$ and
$^3$F$_3$ partial waves simply do not provide a suitable testing ground for
the chiral pion exchange potential as 
originally suggested in Ref.~\cite{Kaiser:1997mw} due to the large short-range
contributions to the corresponding phase shifts at energies $E_{\rm
  lab} \gtrsim 150$~MeV. Notice that the impact of short-range
operators decreases rapidly with increasing values of the orbital angular momentum and
becomes small for G- and higher partial waves.  
Finally, we emphasize that the inclusion of the leading F-wave
contact interactions appears to be necessary in order to describe
some of the very precisely measured proton-proton scattering data, see
the discussion in section \ref{sec:Data}.

\subsection{Low-energy constants of the contact interactions}
\label{sec:LECs}

In Table \ref{tab_LEC}, we show the extracted values of the LECs at
\begin{table}[t]
\caption{The determined LECs of the NN contact interactions at
  N$^4$LO$^+$ along with the statistical uncertainties (except for
  $\tilde C_{1S0}^{\rm nn}$ which is fixed from the value of
  the nn scattering length of $a_{\rm nn} = -18.9$~fm). The values
  of the order-$Q^{2n}$ LECs are given in units of $10^4$
  GeV$^{-2-2n}$.
\label{tab_LEC}}
\smallskip
\begin{tabular*}{\textwidth}{@{\extracolsep{\fill}}lrrrrr}
\hline 
\hline
\noalign{\smallskip}
LEC &  $\Lambda=350\,$MeV &  $\Lambda =400\,$MeV &  $\Lambda
                                                   =450\,$MeV &
                                                                $\Lambda
                                                                =500\,$MeV &  $\Lambda=550\,$MeV 
\smallskip
 \\
\hline 
\hline
%\smallskip
\multicolumn{6}{l}{\bf Low-energy constants at order $Q^0$} \\
$\tilde C_{1S0}^{\rm pp}$   &  $-0.061 \pm 0.000$   & $-0.029 \pm  0.000$  & $0.012  \pm 0.000$   & $0.058 \pm 0.000$  & $0.106 \pm 0.000$  \\
$\tilde C_{1S0}^{\rm nn}$   & $-0.061$  \hspace{1.03cm}        & $-0.029$ \hspace{1.03cm}      & $0.012$ \hspace{1.03cm}        & $0.058$ \hspace{1.03cm}                 &  $0.106$  \hspace{1.03cm}   \\
$\tilde C_{1S0}^{\rm np}$   &  $-0.067 \pm 0.001$   & $-0.031 \pm 0.000$   & $0.010 \pm 0.000$    & $0.056 \pm 0.000$  &  $0.103 \pm 0.001$ \\
$\tilde C_{3S1}$           &  $-0.079 \pm 0.002$   &  $-0.037 \pm 0.002$  &    $0.011 \pm 0.002$ & $0.067 \pm 0.002$  &  $0.124 \pm 0.002$ \\
\hline
\multicolumn{6}{l}{\bf Low-energy constants at order $Q^2$} \\
$C_{1S0}^{\rm pp, nn}$   &  $-0.014 \pm 0.008$    & $0.099 \pm 0.006$    & $0.133 \pm 0.005$    & $0.131 \pm 0.003$  &  $0.124 \pm 0.003$ \\
$C_{1S0}^{\rm np}$        &  $0.074 \pm 0.017$   & $0.106 \pm 0.014$    & $0.121 \pm 0.013$    & $0.120 \pm 0.012$  &  $0.119 \pm 0.012$ \\
$C_{3P0}$                  &  $0.040 \pm 0.008$    & $0.350 \pm 0.006$    &    $0.591 \pm 0.005$ & $0.756 \pm 0.004$  &  $0.833 \pm 0.004$ \\
$C_{1P1}$                  &  $0.570 \pm 0.019$    & $0.692 \pm 0.015$    &    $0.754 \pm 0.013$ & $0.794 \pm 0.012$  &  $0.819 \pm 0.012$ \\
$C_{3P1}$                  &  $0.705 \pm 0.005$    & $0.763 \pm 0.004$    &    $0.828 \pm 0.003$ & $0.856 \pm 0.004$  &  $0.844 \pm 0.004$ \\
$C_{3S1}$                  &  $-0.191 \pm 0.015$   & $-0.063 \pm  0.015$  &  $-0.025 \pm 0.017$  & $-0.045 \pm 0.021$ &   $-0.084 \pm 0.030$\\
$C_{\epsilon 1}$          &  $0.750 \pm 0.022$   & $0.550 \pm 0.017$     &   $0.410 \pm  0.014$ & $0.312 \pm 0.011$  &  $0.233 \pm 0.010$ \\
$C_{3P2}$                  &  $-0.119 \pm 0.002$   & $0.008 \pm   0.001$   &    $0.127 \pm 0.001$ & $0.224 \pm 0.001$  &  $0.288 \pm 0.001$ \\
\hline
\multicolumn{6}{l}{\bf Low-energy constants at order $Q^4$} \\
$D_{1S0}$                  &  $6.531 \pm 0.190$    & $4.043 \pm  0.155$   & $2.579 \pm 0.151$    & $1.722 \pm 0.159$   &  $1.182 \pm 0.178$  \\
$D_{3P0}$                  &  $3.161 \pm 0.067$    & $1.829 \pm 0.055$    &    $0.813 \pm 0.053$ & $0.010 \pm 0.052$ &  $-0.626 \pm 0.046$ \\
$D_{1P1}$                  &  $5.393 \pm 0.180$    & $1.474 \pm 0.137$    &    $-0.056 \pm 0.118$ & $-0.733 \pm 0.104$ &  $-1.091 \pm 0.086$ \\
$D_{3P1}$                  &  $2.528 \pm 0.067$    & $1.111 \pm 0.055$    &    $0.216 \pm 0.053$ & $-0.383 \pm 0.050$ &  $-0.771 \pm 0.043$ \\
$D_{3S1}$                  &  $4.698 \pm 0.247$    & $3.531 \pm 0.187$    & $2.670 \pm  0.176$   & $2.351 \pm 0.189$  & $2.473 \pm 0.243 $  \\
$D_{\epsilon 1}$           &  $0.286 \pm 0.282$   & $-0.346 \pm 0.216$    &  $-0.331 \pm  0.200$ & $-0.273 \pm 0.205$ &  $-0.186 \pm 0.233$ \\
$D_{3D1}$                  &  $3.617 \pm 0.180$    & $2.171 \pm 0.133$     &  $1.524 \pm  0.119$  & $1.139 \pm 0.112$  &  $0.902 \pm 0.108$ \\
$D_{1D2}$                  &  $-0.950 \pm 0.016$   & $-0.153 \pm 0.010$     &    $0.218 \pm 0.008$ & $0.377 \pm 0.006$  &  $0.410 \pm 0.006$ \\
$D_{3D2}$                  &  $-3.178 \pm 0.050$   & $-0.857 \pm  0.030$   &  $0.187 \pm 0.022$   & $0.680 \pm 0.018$  &  $0.941 \pm 0.017$ \\
$D_{3P2}$                  &  $-1.402 \pm 0.016$   & $-0.683 \pm 0.010$    & $-0.448 \pm 0.008$   & $-0.386 \pm 0.007$ &  $-0.373 \pm 0.005$ \\
$D_{\epsilon 2}$           &  $-0.909 \pm 0.014$   & $-0.424 \pm 0.009$    & $-0.219 \pm 0.007$   & $-0.118 \pm 0.006$ &  $-0.067 \pm 0.006$ \\
$D_{3D3}$                  & $-0.482 \pm 0.067$    & $-0.080 \pm 0.041$    & $0.141 \pm 0.030$    & $0.257 \pm 0.025$  &  $0.283 \pm 0.022$  \\
\hline
\multicolumn{6}{l}{\bf Low-energy constants at order $Q^6$} \\
$E_{3F2}$                  & $2.297 \pm 0.254$    & $2.488 \pm 0.162$    & $2.289 \pm 0.130$    & $2.138 \pm 0.119$  &  $2.018 \pm 0.122$  \\
$E_{1F3}$                  & $-0.633 \pm 0.470$    & $1.523 \pm 0.327$    & $2.162 \pm 0.275$    & $2.348 \pm 0.261$  &  $2.483 \pm 0.286$  \\
$E_{3F3}$                  & $2.925 \pm 0.339$    & $1.769 \pm 0.218$    & $1.319 \pm 0.172$    & $1.022 \pm 0.149$  &  $0.836 \pm 0.138$  \\
$E_{3F4}$                  & $-1.817 \pm 0.138$    & $-0.221 \pm 0.088$    & $0.277 \pm 0.067$    & $0.427 \pm 0.056$  &  $0.425 \pm0.049$  \\[4pt]
\hline \hline
\end{tabular*}
\end{table}
N$^4$LO$^+$ for all employed cutoff values along with the
corresponding statistical uncertainties.
We remind the reader that 
$\tilde C_i (\Lambda )$, $C_i (\Lambda )$, $D_i (\Lambda )$ and $E_i
(\Lambda )$ are {\it bare} parameters and thus have to be refitted
at every chiral order. 
As discussed in Ref.~\cite{Epelbaum:2014efa}, the natural size of the
LECs in the spectroscopic notation can be estimated via 
\beq
| \tilde C_i | \sim \frac{4 \pi}{F_\pi^2}, \quad \quad
| C_i | \sim \frac{4 \pi}{F_\pi^2 \Lambda_{\rm b}^2}, \quad \quad
| D_i | \sim \frac{4 \pi}{F_\pi^2 \Lambda_{\rm b}^4}, \quad \quad
| E_i | \sim \frac{4 \pi}{F_\pi^2 \Lambda_{\rm b}^6},
\eeq 
with the factor of $4 \pi$ emerging from the angular integration and
$\Lambda_{\rm b}$ denoting the breakdown scale of the chiral
expansion. Using the estimated value of $\Lambda_{\rm b} = 600$~MeV proposed in
Ref.~\cite{Epelbaum:2014efa} and verified in  Ref.~\cite{Furnstahl:2015rha}, the natural size of the
LECs is expected to be of the order of 
\beq
| \tilde C_i | \sim 0.15, \quad \quad
| C_i | \sim 0.4, \quad \quad
| D_i | \sim 1.1 , \quad \quad
| E_i | \sim 3,
\eeq
in units used in Table \ref{tab_LEC}. We observe that all values of
the LECs at N$^4$LO$^+$ are of a natural size for the cutoff values in the range
of $\Lambda = 450 \ldots 550$~MeV. For softer cutoff values and
especially for the lowest value of $\Lambda = 350$~MeV, the magnitude
of some of the LECs increases considerably. This pattern is expected
and points towards significant cutoff artefacts, which effectively
lower the breakdown scale $\Lambda_{\rm b}$ for too small values of
$\Lambda$.  As already mentioned in the previous section, all F-wave
LECs $E_i$ appear to be of a natural size. 
Notice further that the apparently rather smooth
$\Lambda$-dependence of the various LECs does not extend to higher
cutoff values. In particular, in channels where the dominant
short-distance singularity (in the infinite-cutoff limit) of the long-range
interaction is attractive, $\Lambda$-dependence of the LECs accompanying
contact terms has to be discontinuous in order to accommodate
for the appearance of deeply-bound states without, at the same time,  affecting the
low-energy behavior of the amplitude. Such a discontinuous behavior of
the LECs signals a highly nonperturbative regime, in which naive
dimensional analysis and the above-mentioned naturalness estimations
are not applicable anymore. The values of $\Lambda$ at which the first
deeply bound states are generated depend on several factors such as
the chiral order, the employed form of the regulator and the values of
the $\pi N$ coupling constants. While all LECs at N$^4$LO$^+$  
stay natural  for cutoff values $\Lambda \leq
550$~MeV\footnote{The values of $\tilde C_i$, $C_i$ and $D_i$
  at N$^4$LO are similar to the ones at N$^4$LO$^+$ for
  the same cutoff value, but the differences are, in certain cases,
  larger than the statistical uncertainties given in Table
  \ref{tab_LEC}.}, we do observe at N$^3$LO  the LEC $D_{3S1}$
becoming unnaturally large for $\Lambda =
550$~MeV. 
As will be shown in section \ref{sec:WE}, this also manifests itself
in the appearance of a very
large repulsive Weinberg eigenvalue in the $^3$S$_1$-$^3$D$_1$
channel, indicating that the corresponding potential is highly
nonperturbative. The larger available cutoff range at N$^4$LO as compared
to N$^3$LO can presumably be traced back to the more accurate representation of the
two-pion exchange potential at this order in the chiral expansion.  
 
It is instructive to compare the values of the LECs $D_i$ in the
$^1$S$_0$, $^3$S$_1$ and $^3$S$_1$-$^3$D$_1$ channels, which are affected by
the short-range UTs defined in Eq.~(\ref{ShortRangeUT}), with
the results of earlier studies. In particular, we find that our
choice of the off-shell behavior of the contact 
interactions specified in Eqs.~(\ref{ConventionOffShell1}),
(\ref{ConventionOffShell2}) leads to the LECs $D_{1S0}$,
$D_{3S1}$ and $D_{\epsilon 1}$ which are  all of a natural size. This is in
a strong contrast with the large-in-magnitude values of $D_{1S0}^{1}$,
$D_{1S0}^{2}$,    $D_{3S1}^{1}$, $D_{3S1}^{2}$,    $D_{\epsilon
  1}^{1}$ and $D_{\epsilon 1}^{2}$ found in
Ref.~\cite{Epelbaum:2014efa} at N$^3$LO.\footnote{The same statement
  applies to the N$^4$LO EKM
  potentials of Ref.~\cite{Epelbaum:2014sza} as well.} Our findings suggest that
these large values appeared as a result of overfitting in the presence
of the redundant terms as already emphasized in section \ref{Sec:RedundantFits} for the
case of semilocal coordinate-space regularized potentials introduced
in Refs.~\cite{Epelbaum:2014efa, Epelbaum:2014sza}. 
In section \ref{sec:Redundant}, we will further elaborate
on this point by explicitly demonstrating that our fits
to scattering data are, to a very good approximation, independent on
the choice of the redundant parameters $D_{1S0}^{\rm off}$,
$D_{3S1}^{\rm off}$ and $D_{\epsilon 1}^{\rm off}$. 

Next, we comment on the statistical uncertainties of the various LECs listed in Table
\ref{tab_LEC}. First, one observes that the statistical uncertainties
of the LECs, taken relative to their expected natural values, tend to increase
from $\tilde C_i$ to $C_i$ to $D_i$ to $E_i$. This pattern is
consistent with the power counting, which suggests that the
contributions of contact interactions to the scattering
amplitude decrease with the chiral order. We also observe
considerably smaller statistical uncertainties in the isovector
channels as compared to the isoscalar ones, which are extracted from
the combined np and pp data and thus benefit from the larger amount of
experimental data and the generally higher precision of pp data. Last
but not least, the statistical uncertainties tend to increase for 
soft choices of the regulator as one would expect as a consequence of the less
accurate description of the data, which is especially true for the 
case of $\Lambda = 350$~MeV, see section \ref{sec:Data} for more
details, but show little cutoff dependence 
otherwise. 

\begin{figure}[tb]
\vskip 1 true cm
\begin{center}
\includegraphics[width=\textwidth,keepaspectratio,angle=0,clip]{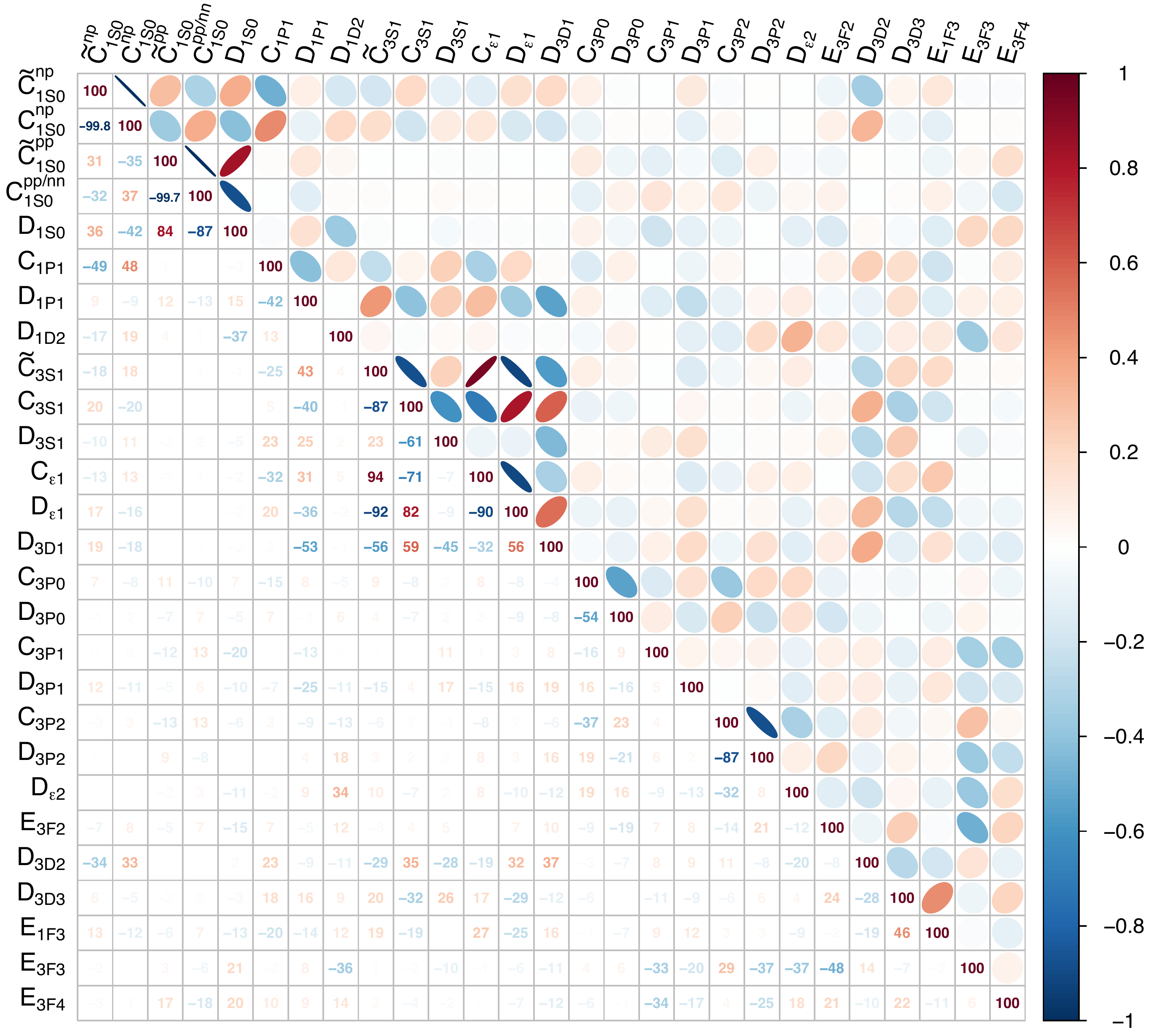}
\end{center}
    \caption{Correlations between the various LECs for the case of the
      N$^4$LO$^+$ potential with $\Lambda = 450$~MeV. The lower
      triangle gives the correlation coefficients in percent.
\label{fig:Correlations}
}
\end{figure}

Finally, we have also looked at the
correlations between the various LECs as visualized in
Fig.~\ref{fig:Correlations} for the case of the intermediate cutoff
$\Lambda = 450$~MeV.  
As one may expect, the strongest correlations and/or anti-correlations
appear between the LECs in the channels with the largest number of
parameters, namely $^1$S$_0$ and in $^3$S$_1$-$^3$D$_1$. This
is reflected in the statistical uncertainties of the
corresponding LECs being larger than the uncertainties of the LECs
$C_i$ and $D_i$ in most of the other channels. Still, in spite of
the strong correlations/anticorrelations present, most of the LECs are
determined by the scattering data with good accuracy. 
The large statistical uncertainty for $E_{1F3}$ reflects the fact 
that the corresponding phase shifts provide only a minor contribution
to the NN scattering amplitude in the considered energy range. 
The impact of the statistical uncertainties of the LECs on the calculated
observables in the NN system will be quantified in section
\ref{sec:StatUncertainty}.

\subsection{Description of NN scattering data}
\label{sec:Data}

We now turn to the description of the scattering data. In Table
\ref{tab_chi2_AllOrders}, we give the $\chi^2/{\rm datum}$ for the description of the
  neutron-proton and proton-proton scattering data at various chiral
  orders for the case of the intermediate cutoff $\Lambda = 450$~MeV.  
Our results agree qualitatively with the ones reported in
Refs.~\cite{Epelbaum:2014efa, Epelbaum:2014sza}, where fits to the
Nijmegen PWA have been performed. As expected, the value of
$\chi^2/{\rm datum}$ decreases with increasing chiral order. It is
particularly comforting to observe a significant improvement in the
description of the experimental data at N$^2$LO and N$^4$LO as
compared to NLO and N$^3$LO, which can
be traced back to the corresponding parameter-free contributions to the
two-pion exchange potential.\footnote{At N$^4$LO, we take into account
  an additional isospin-breaking contact interaction in the $^1$S$_0$
  channel as compared to
  N$^3$LO, which gives the difference between $C_{1S0}^{\rm pp, nn}$
  and $C_{1S0}^{\rm np}$, see Ref.~\cite{Epelbaum:2014sza} for more details.} This provides a clear evidence of the
chiral two-pion exchange, a feature which was already reported in
Refs.~\cite{Epelbaum:2014efa,Epelbaum:2014sza} based on the Nijmegen
PWA, see also Refs.~\cite{Rentmeester:1999vw,Birse:2003nz} for related earlier studies.  

\begin{table}[t]
\caption{$\chi^2/{\rm datum}$ for the description of the
  neutron-proton and proton-proton scattering data at various orders
  in the chiral expansion for $\Lambda = 450$~MeV.
The values in the brackets give the $\chi^2/{\rm datum}$ without
including the ``outlier'' proton-proton scattering data set CO(67) from
      Ref.~\cite{Cox:1968jxz} as described in
the text. 
\label{tab_chi2_AllOrders}}
\smallskip
\begin{tabular*}{\textwidth}{@{\extracolsep{\fill}}lllllll}
\hline 
\hline 
\noalign{\smallskip}
 $E_{\rm lab}$ bin &  LO   &  NLO   &  N$^2$LO   &  N$^3$LO  &  N$^4$LO  & N$^4$LO$^+$  
\smallskip
 \\
\hline 
\hline 
%\smallskip
\multicolumn{5}{l}{neutron-proton scattering data} \\ 
0--100 & 73 & 2.2 & 1.2 & 1.08 & 1.07 & 1.08\\ 
0--200 & 62 & 5.4 & 1.7 & 1.10 & 1.08 & 1.07\\
0--300 & 75 & 14  & 4.1 & 2.01 & 1.16 & 1.06\\ [4pt]
%\smallskip
\hline 
% \smallskip
\multicolumn{5}{l}{proton-proton scattering data} \\ 
0--100 & 2290 & 10 & 2.2 & 0.90 & 0.88 & 0.86\\ 
0--200 & 1770 & 90 & 37  & 1.99 (1.68) & 1.42 (1.16) & 0.95\\ 
0--300 & 1380 & 91 & 41  & 3.43 (3.23) & 1.67 (1.50)  & 1.00\\ [4pt]
\hline 
\hline 
\end{tabular*}
\end{table}

\begin{figure}[tb]
\vskip 1 true cm
\begin{center}
\includegraphics[width=1.0\textwidth,keepaspectratio,angle=0,clip]{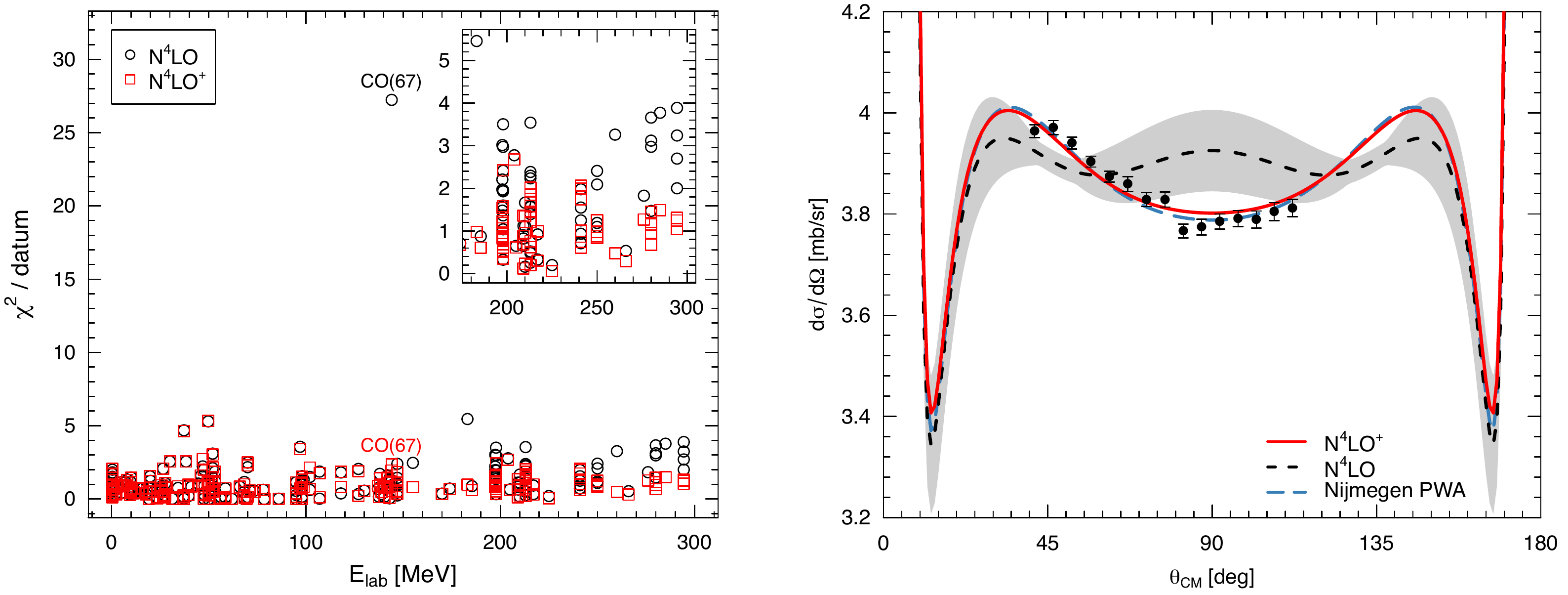}
\end{center}
    \caption{(Color online) Left panel: contributions of various pp data
      sets to the $\chi^2/{\rm datum}$. Open black circles and red squares
      correspond to the results at N$^4$LO and N$^4$LO$^+$,
      respectively. Right panel: pp differential
      cross section at $E_{\rm lab} = 144$~MeV as function of the
      scattering angle calculated at N$^4$LO (black short-dashed line) and
      N$^4$LO$^+$ (red solid line) in comparison with the
      experimental data of the CO(67) data set of
      Ref.~\cite{Cox:1968jxz}, which have been divided by the inverse
      norm of $Z=1.012$ estimated in the N$^4$LO$^+$
      fit.
      The grey shaded band shows the theoretical uncertainty
      at
      N$^4$LO from the truncation of the chiral expansion estimated
      using the approach formulated in
      Refs.~\cite{Epelbaum:2014efa,Epelbaum:2014sza} (without taking
      into account the information at order N$^4$LO$^+$).  Long-dashed
      blue line shows the result of the Nijmegen PWA
      \cite{Stoks:1993tb}. In both panels, the chiral potentials
      correspond to the intermediate cutoff of $\Lambda = 450$~MeV.   
\label{fig:Chi2pp}
}
\end{figure}

While the description of the np scattering data up to $E_{\rm lab} =
200$~MeV and pp data up to $E_{\rm lab} =
100$~MeV is found to be very good starting from N$^3$LO, the $\chi^2/{\rm datum}$
is significantly larger than $1$ for pp data at intermediate
energy. To better understand the origin of this behavior, we show
in the left panel  of Fig.~\ref{fig:Chi2pp} the
contributions of the individual  data sets to the  $\chi^2$.
One observes that the large value of the $\chi^2/{\rm datum}$ for the
energy bin of $E_{\rm lab} = 0 \ldots 200$~MeV at
N$^4$LO is, to a considerable extent, caused by the CO(67) \cite{Cox:1968jxz} 
set of differential cross sections, which generates a large
contribution of $[\chi^2/{\rm datum}]_{\rm CO(67)} = 27.88$. This can be traced back to
the very high precision of these experimental data, which exceeds
substantially the accuracy of our N$^4$LO results. We emphasize that  this
does not indicate any failure of the chiral EFT: indeed, as shown in the
right panel of Fig.~\ref{fig:Chi2pp}, the CO(67)
experimental data are actually fairly well described at N$^4$LO if
one takes into account the theoretical uncertainties at this order
estimated using the approach of
Ref.~\cite{Epelbaum:2014efa}.\footnote{Notice that the leftmost point
  at $\theta_{\rm 
            CM}=41.4^\circ$ is declared as an outlier in the SAID
          database and is, therefore, not included in the fits.}  
Still, the shape of
the CO(67) cross section data is not properly described at
N$^4$LO, see also Ref.~\cite{Ekstrom:2013kea} for a related discussion. 
We found that this can be traced back to the still poor description
of F-waves at N$^4$LO as visualized in Fig.~\ref{fig:F}. To further
elaborate on this point, we have redone the N$^4$LO fits including
the leading contact interactions in the F-waves defined in
Eq.~(\ref{LECsF}). With 4 additional parameters, we indeed observe that the
shape of the cross section data is well described, see the right
panel of Fig.~\ref{fig:Chi2pp}. We also show in the left panel of
Fig.~\ref{fig:Chi2pp} that the $\chi^2/{\rm datum}$ for the description of
pp data at higher energies decreases significantly upon including 
these contact interactions in F-waves. We are then able to achieve a nearly
perfect description of the np and pp scattering data up to $E_{\rm
  lab} = 300$~MeV as shown in the last column of Table
\ref{tab_chi2_AllOrders}. 
We refer to the resulting potentials as
N$^4$LO$^+$ and expect them to be particularly useful for a reliable
quantification of the theoretical uncertainties at N$^4$LO. They also
allow for a more precise extraction of the LECs $\tilde C_i$, $C_i$
and $D_i$ at N$^4$LO. 

Finally, in Table \ref{tab_chi2_AllCutoffs}, we show the values
of the $\chi^2/{\rm datum}$ for the description of the np and pp
scattering data at N$^4$LO$^+$ for all considered cutoff values. 
\begin{table}[t]
\caption{$\chi^2/{\rm datum}$ for the description of the
  neutron-proton and proton-proton scattering data at N$^4$LO$^+$ for
  all considered cutoff values.
\label{tab_chi2_AllCutoffs}}
\smallskip
\begin{tabular*}{\textwidth}{@{\extracolsep{\fill}}cccccc}
\hline 
\hline 
\noalign{\smallskip}
 $E_{\rm lab}$ bin &  $\Lambda = 350$~MeV   &  $\Lambda = 400$~MeV   &  $\Lambda = 450$~MeV   &  $\Lambda = 500$~MeV  &  $\Lambda = 550$~MeV  
\smallskip
 \\
\hline 
\hline 
%\smallskip
\multicolumn{5}{l}{neutron-proton scattering data} \\ 
0--100 & 1.13 & 1.08 & 1.08 & 1.07 & 1.07 \\ 
0--200 & 1.17 & 1.07 & 1.07 & 1.07 & 1.07 \\
0--300 & 1.84 & 1.17 & 1.06 & 1.06 & 1.07 \\ [4pt]
%\smallskip
\hline 
% \smallskip
\multicolumn{5}{l}{proton-proton scattering data} \\ 
0--100 & 0.98 & 0.88 & 0.86 & 0.86 & 0.87 \\ 
0--200 & 1.10 & 0.96 & 0.95 & 0.97 & 0.99 \\ 
0--300 & 1.30 & 1.02 & 1.00 & 1.04 & 1.11  \\ [4pt]
\hline 
\hline 
\end{tabular*}
\end{table}
As expected, the softest version of the potential with $\Lambda =
350$~MeV leads to significant cutoff artefacts, which manifest
themselves in the rather large values of the  $\chi^2/{\rm datum}$ at
higher energies. The description of the data improves with increasing
the cutoff $\Lambda$ and reaches its optimum for $\Lambda = 450$~MeV. 
No improvement is observed when increasing the cutoff $\Lambda$
beyond this value. A similar picture emerges at lower orders in the
chiral expansion. This pattern is qualitatively similar to the one
observed for the coordinate-space regularized potentials of
Refs.~\cite{Epelbaum:2014efa, Epelbaum:2014sza}.

\subsection{Weinberg eigenvalue analysis}
\label{sec:WE}

\begin{figure}[h]
\vskip 1 true cm
\begin{center}
\includegraphics[width=\textwidth,keepaspectratio,angle=0,clip]{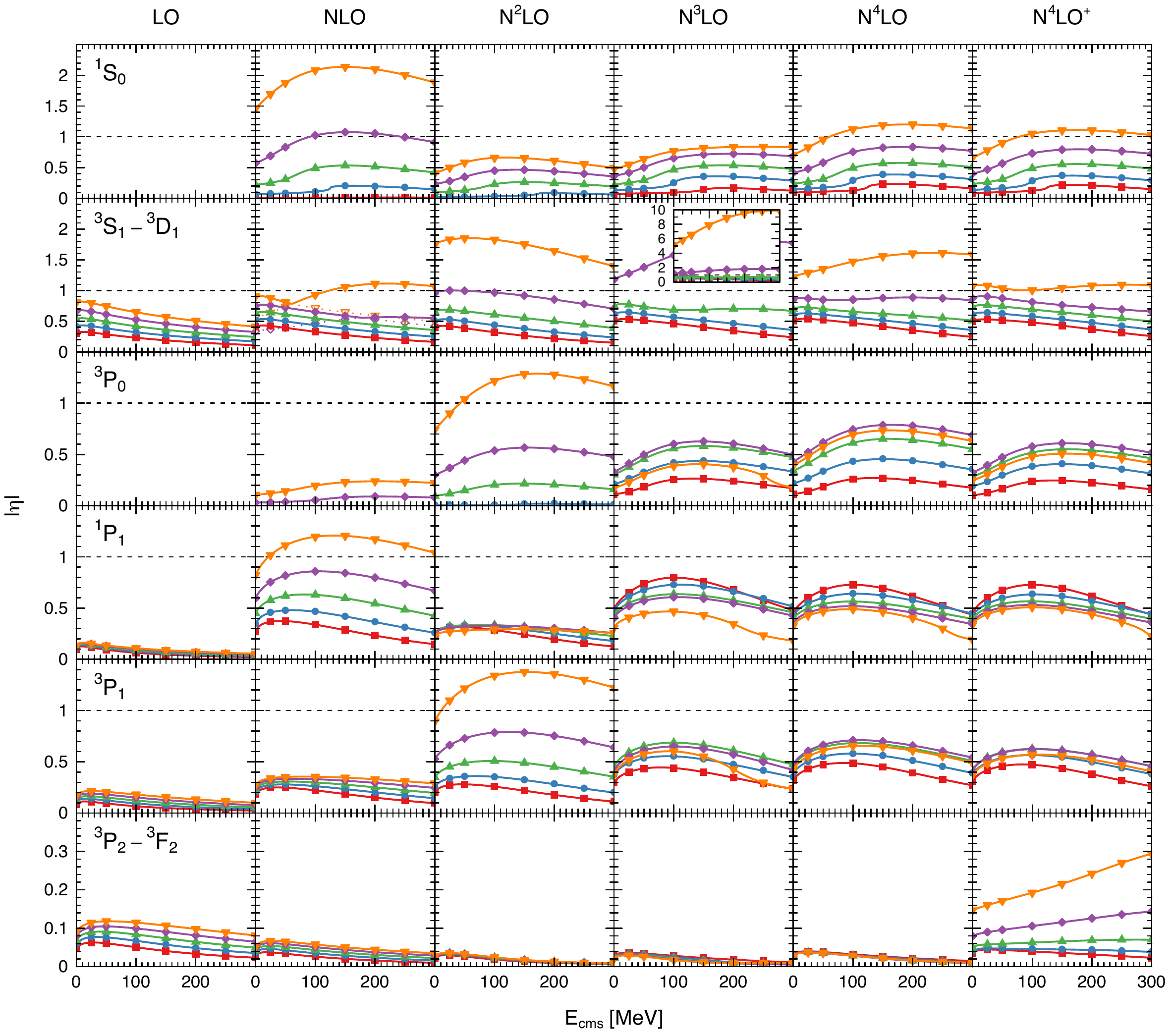}
\end{center}
    \caption{(Color online) The magnitude of the largest repulsive
      Weinberg eigenvalues $|\eta |$ in the $^1$S$_0$, $^3$S$_1$-$^3$D$_1$,
      $^3$P$_0$, $^1$P$_1$,  $^3$P$_1$ and  $^3$P$_2$-$^3$F$_2$
      channels at different orders in the chiral expansion. Red
      squares, blue dots, green up triangles,
      violet diamonds and orange down triangles correspond to $\Lambda =
      350$, $400$,   $450$,   $500$ and $550$~MeV, respectively.       
\label{fig:WeinbergEVnew}
}
\end{figure}
\begin{figure}[h]
\vskip 1 true cm
\begin{center}
\includegraphics[width=\textwidth,keepaspectratio,angle=0,clip]{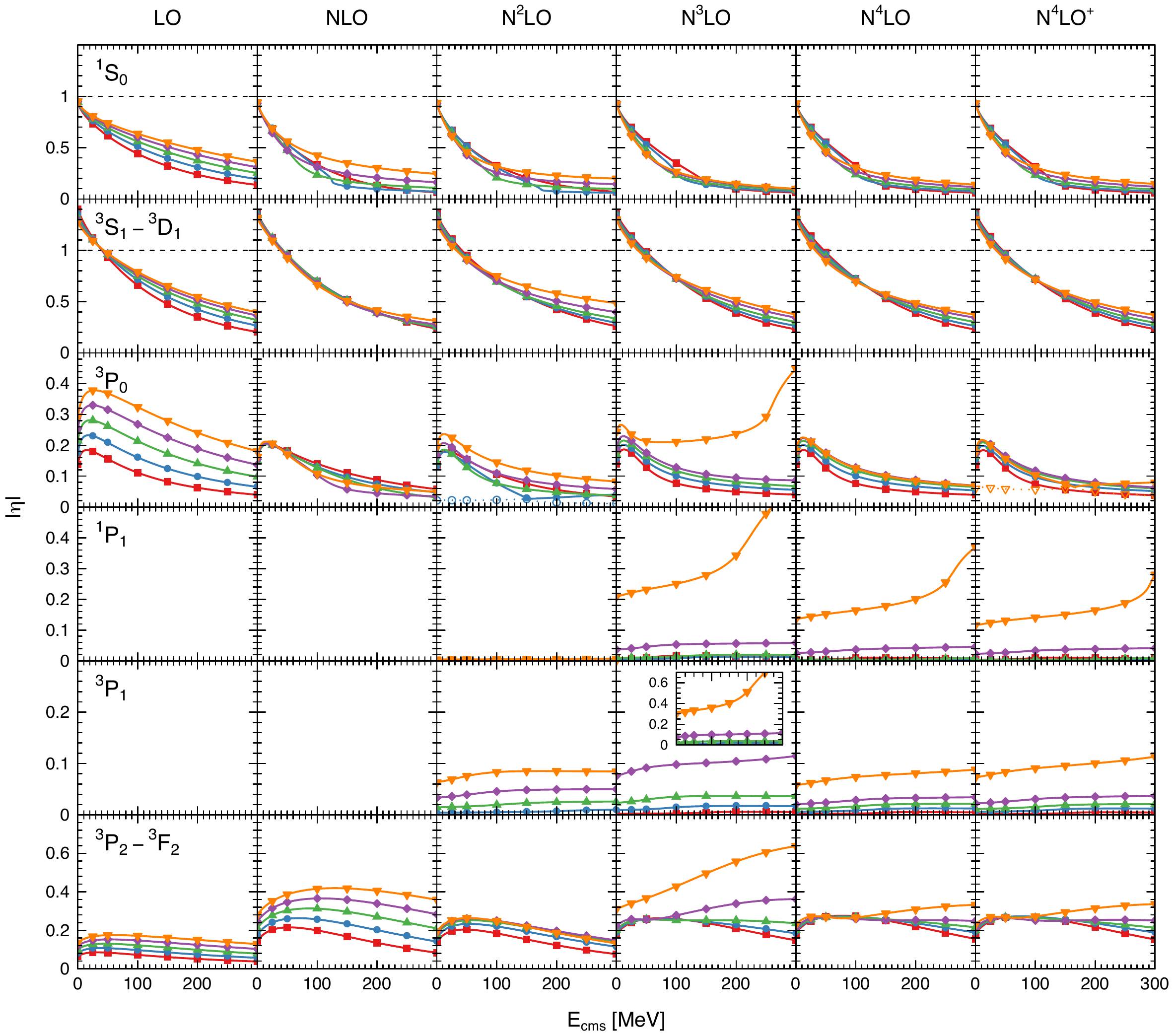}
\end{center}
    \caption{(Color online) The magnitude of the largest attractive 
      Weinberg eigenvalues $|\eta |$ in the $^1$S$_0$, $^3$S$_1$-$^3$D$_1$,
      $^3$P$_0$, $^1$P$_1$,  $^3$P$_1$ and  $^3$P$_2$-$^3$F$_2$
      channels at different orders in the chiral expansion. For
      notation see Fig.~\ref{fig:WeinbergEVnew}.  
\label{fig:WeinbergEVnewAttr}
}
\end{figure}

To assess perturbativeness of the new SMS potentials we have
performed a Weinberg eigenvalue analysis as outlined in section \ref{Sec:RedundantFits}. In
Figs.~\ref{fig:WeinbergEVnew} and \ref{fig:WeinbergEVnewAttr}, we show the magnitude of the largest
repulsive and attractive eigenvalues as function of the cms 
energy at all orders and for all considered cutoff values.  
In most cases, the largest in magnitude eigenvalues are given by the
trajectory of a single eigenvalue as function of the energy. 
In a few exceptions, however, the trajectories corresponding to the two
largest in magnitude 
eigenvalues cross each other within the considered energy range. 
In such cases, we show the part of the eigenvalue trajectory
corresponding to the second-largest in
magnitude eigenvalue by open symbols connected with dotted lines. 
The remaining kinks do not reflect any
discontinuous behavior and emerge when the corresponding
eigenvalues change primarily in the
radial direction in the complex plane when increasing the energy $E_{\rm cms}$.  

As expected, one observes a clear tendency towards increasingly
nonperturbative potentials for larger cutoff values. Except for the
$^3$S$_1$-$^3$D$_1$ channel, all potentials are perturbative
in the considered energy range for the cutoff values of $\Lambda = 350
\ldots 450$~MeV. In this perturbative regime, the eigenvalues do change
strongly at lower orders but appear to reach convergence starting from
N$^3$LO. The only exception is the $^3$P$_2$-$^3$F$_2$ channel, where 
the repulsive eigenvalues are driven by the leading contact
interaction first taken into account at N$^4$LO$^+$.  The largest repulsive
eigenvalues increase significantly in S- and P-waves at
NLO and N$^2$LO and in S-waves at N$^3$LO starting from the cutoff
$\Lambda = 500$~MeV, but the corresponding potentials still remain
perturbative in most cases.  This trend becomes much  more pronounced
for the largest considered cutoff value of $\Lambda =
550$~MeV. Here, by far the largest in magnitude repulsive eigenvalue
emerges at N$^3$LO in the  $^3$S$_1$-$^3$D$_1$ channel, and the
corresponding potential is strongly non-perturbative. This pattern
is presumably related to the unnaturally large value of the order-$Q^4$
LEC  $D_{3S1}$ we find in this case, namely $D_{3S1} = 22.7 \times
10^4$~GeV$^{-6}$. Notice that this is the only case we observe
an unnaturally large LEC.  Interestingly, the potentials at N$^4$LO
and N$^4$LO$^+$ are considerably more perturbative for the
largest considered cutoff value than that at N$^3$LO.

\subsection{Error analysis}
\label{sec:Errors}

We now turn to the discussion of the various sources of uncertainties
of our calculations. Specifically, we address below the statistical
uncertainty, truncation error of the chiral expansion, uncertainties
in the pion-nucleon LECs $c_i$, $\bar d_i$ and $\bar e_i$ as well as the uncertainty associated with
the choice of the energy range in the fits. We emphasize that the
discussed uncertainties are not completely independent of each other. For
this reason we  refrain from combining them into a single theoretical error
and prefer to specify them separately when giving results for
observables.

\subsubsection{Statistical uncertainty}
\label{sec:StatUncertainty}

We begin with the discussion of the statistical uncertainty of the LECs
accompanying NN contact interactions, which we collectively denote as $c$. 
In the context of parameters estimation, the statistical errors of the
LECs quantify the accuracy of their determination by the scattering
data used in the fit. For the case at hand, we regard the
$\chi^2$-function as a measure of the goodness of the fits
with the optimal values of LECs $c = c_{\rm min}$ corresponding to the minimum of
$\chi^2$. The statistical uncertainty of the determined LECs can be
naturally assessed via corresponding variations of
$\chi^2$. Specifically, we define the
uncertainty of the LECs to comprise all values $c$
compatible with the maximum allowed variation $T$  of $\chi^2$ 
\be
	\label{chi2-variation}
	\chi^2(c) \le \chi^2(c_{\rm min}) + T.
\ee
Different choices of $T$ lead to different confidence levels for
the obtained parameter uncertainties. While the exact connection 
between $T$ and the confidence level is
not immediately clear in a most general case, we can be guided by the
more specific situation described below, which can often be regarded
as a good approximation. Namely, for the case of normally distributed
residuals, which are further assumed to be approximately linear
functions of the parameters in the vicinity of the minimum, the
increase of $\chi^2$ can be 
related to a confidence level via the $\chi^2$ distribution. For the
uncertainty of a single parameter, i.e.~while other parameters are allowed
to vary, $T=1$ leads to a $1\sigma$ or $68\%$ confidence
level. Under the aforementioned assumptions, one would also expect
$\chi^2_{\rm min}$ to follow the $\chi^2$ distribution with $N_{\rm dof}
= N_{\rm exp} - N_{\rm float} - N_c$ degrees of freedom, where $N_{\rm
  exp}$ is the number of experimental data (not including the
norms),  $N_{\rm float}$ is the number of floated data while $N_c$ is
the number of LECs entering the fit. For a proper description of the
data, $\chi^2_{\rm min}$ is expected to have the value of $N_{\rm dof}
\pm \sqrt{2N_{\rm dof}}$. However, in practice, one often finds 
higher values of $\chi_{\rm min}^2$. To account for this situation, we
follow a commonly used approach, see
e.g.~Ref.~\cite{Carlsson:2015vda},  and normalize the residuals
with the  Birge-factor \cite{Birge} 
according to $\chi^2 \rightarrow \chi^2  N_{\rm dof}/\chi^2_{\rm min}
$, so that the expectation value $\chi^2_{\rm min} = N_{\rm dof}$ is
trivially fulfilled. We then get $T=\chi^2_{\rm min}/N_{\rm dof}$, which
increases the uncertainties to account for the non-optimal description
of the data. 

While it is certainly important to quantify the uncertainty of the LECs, our
main goal is to estimate the statistical errors of the calculated
observables. In a complete analogy with the above considerations for 
the LECs,  the uncertainty regions $\delta O_\pm$ for an observable of
interest $O$ are defined in terms of the allowed maximum variation 
of the $\chi^2$-function as 
\beq
	\label{general-statistical-errors}
	\delta O_-= -\min \left[O(c) - O(c_{\rm min}) \right]\,,  \quad
        \delta O_+ = \max \left[O(c) - O(c_{\rm min})\right] \,,
\eeq
subject to the constraint in Eq.~(\ref{chi2-variation}), see also
Ref.~\cite{Carlsson:2016saa} for a similar approach. 
For the sake of generality, we have separately defined the lower and
upper errors $\delta O_-$ and $\delta O_+$, which, of course,
coincide if the underlying probability distribution is symmetric as it
is e.g.~the case for normally-distributed observables. 
This definition can also be straightforwardly extended to
include other constraints in addition to Eq.~\eqref{chi2-variation}. For
the case at hand, the allowed variation of the LECs $c$ is further restricted
to exactly reproduce the deuteron binding energy and
the np coherent scattering length starting from N$^3$LO. 
In our analysis, we have
directly implemented Eq.~\eqref{general-statistical-errors} as a
constrained optimization problem and applied it to calculate the statistical uncertainty for
selected observables. Notice that in such an optimization procedure,
which explores variations of $\chi^2$ away from the optimal solution,
the set of data sets with floated norm must be fixed according to
its determination at the $\chi^2$ minimum. While it, by itself, makes little sense
to apply the decision criterion for floating the data based on a non-optimal
solution, additional floating would also introduce discontinuities in the
$\chi^2$ hypersurface and can lead to $\chi^2$ values lower than 
the previously determined minimum, a feature we certainly  want
to avoid in our error analysis. 

While the approach for estimating the statistical uncertainty outlined
above is quite general, 
it requires performing additional fits for every quantity of
interest. Consequently, it is usually too expensive to be employed in practical
calculations, especially when carrying out error analysis beyond the
two-nucleon system. 
We will, therefore, use  a more convenient approach to
propagating the statistical uncertainties based on the covariance
matrix throughout our analysis except for the few cases, where this
approach is expected to be inaccurate and where we will resort to the
general method described at the beginning of this section. 
This also allows us to easily assess the amount of linear correlations between the LECs. 
We emphasize that many of the results discussed below are, strictly
speaking, only valid for model functions (in our case
observables), which depend linearly on the parameters. For nonlinear
fits like the ones we perform, several assumptions have to be made. First, we
assume that the $\chi^2$ function can be approximated  around the
minimum $c = c_{\rm min}$
via 
\beq
\label{quadratic-approximation}
\chi^2(c) \approx \chi^2_{\rm min} + \frac{1}{2} (c-c_{\rm min})^T H (c-c_{\rm min})\,,
\eeq
where $c$ is the vector of the LECs while $H$ denotes the Hessian of
$\chi^2$ at the minimum, i.e.~$H_{ij} = \frac{\partial^2
  \chi^2}{\partial c_i \partial c_j}\big|_{c=c_{\rm min}}$. The
parameters $c$ are assumed to follow a multivariate Gaussian
probability distribution 
\beq
	\label{multivariate-gaussian}
	p(c,c_{\rm min},\Sigma) = \frac{1}{\sqrt{(2\pi)^n
            \det(\Sigma)}} \exp \left[-\frac{1}{2}\left( c-c_{\rm min}
          \right)^T \Sigma^{-1} \left( c - c_{\rm min} \right)
        \right]
\eeq
in the vicinity of the minimum, where the covariance matrix $\Sigma$ is given by:
\beq
\label{covarianceMatrix}
\Sigma = 2\frac{\chi^2}{N_{\rm dof}} H^{-1}.
\eeq
This relation (modulo the $\chi^2/N_{\rm dof}$ factor) can be easily 
obtained by inserting the quadratic approximation 
\eqref{quadratic-approximation} into the Likelihood function 
$\mathcal{L} \propto \exp\left(-\chi^2/2\right)$ and by matching the
resulting expression to  
Eq.~\eqref{multivariate-gaussian}. The additional factor of $\chi^2/N_{\rm dof}$ 
is related to a rescaling of $\chi^2$ with the Birge-factor as discussed above.
In our fits, the deuteron binding energy and the value of the
coherent np scattering length are adjusted via
constrained optimization techniques, which cannot be included in the
estimation of the covariance matrix in a straightforward way. Instead,
we employ an augmented $\chi^2$-function, where we have added two 
quadratic penalty terms for the deuteron binding energy $B_d$ and the
np coherent scattering length $b_{\rm np}$
\be
	\label{aug-chi2}
	\chi^2_{\rm aug} = \chi^2_{\rm data} + \left(
          \frac{B_d-B_d^{\rm exp}}{\Delta B_d} \right)^2 + \left(
          \frac{b_{\rm np}-b_{\rm np}^{\rm exp}}{\Delta b_{\rm np}}
        \right)^2, 
\ee
with $\Delta B_d = 5\times 10^{-5}$ MeV and $\Delta b_{\rm np} =
9\times  10^{-4}$ 
fm. Hence, we have relaxed the constraints to the additional data points. The 
augmented $\chi^2$ in Eq.~\eqref{aug-chi2} does not have exactly the
same location of the minimum 
as the $\chi^2$ in the constrained problem, and, consequently, we have to
readjust the LECs to be at the minimum
of $\chi^2_{\rm aug}$ before we can estimate the covariance matrix from its Hessian.
The changes in the LECs and observables compared to our constrained fits are very 
small and well within the statistical uncertainties, but the additional fits 
of $\chi^2_{\rm aug}$ are, nonetheless, rather time-consuming due to
the slow convergence as 
already mentioned in Sec.~\ref{sec:fitting}. 
The resulting values of the LECs are only used for the calculation of the covariance matrix.
The statistical errors $\sigma_i = \sqrt{\Sigma_{ii}}$ and correlation coefficients
$\text{Corr}(c_i,c_j) = \Sigma_{ij}/ \sqrt{\Sigma_{ii}\Sigma_{jj}}$
of the LECs given in Sec.~\ref{sec:LECs} are then easily obtained from
the covariance matrix. 

When calculating observables, propagation of statistical uncertainties
can be carried out in a general way by Monte Carlo
sampling of the assumed multivariate Gaussian distribution of the LECs
in the vicinity of the minimum. We avoid the costly sampling by
expanding an observable of interest $O$ in powers of $c-c_{\rm min}$. 
While it is common to use just a linear expansion, it was
argued in \cite{Carlsson:2015vda} that some observables need to be
expanded quadratically in $c-c_{\rm min}$, 
\beq
\label{observable-expansion}
O(c) = O(c_{\rm min}) + J_O(c-c_{\rm min}) + \frac{1}{2} (c-c_{\rm min})^T H_O (c-c_{\rm min}),
\eeq
in order to have an accurate estimation of uncertainties. Here, $J_O$
and $H_O$ denote the corresponding Jacobian and Hessian,
respectively. Here and in what follows, we assume the validity of this quadratic
approximation. 
Using Eq.~\eqref{observable-expansion} it is easy to
obtain the  
mean or expectation value $\mu$ of $O(c)$
\beq
\label{observable_mean}
\mu(O) \equiv \langle O \rangle = O + \frac{1}{2} \Tr \left( H_O \Sigma \right)
\eeq
as well as its variance
\beq
{\rm Var}(O) \equiv \langle O^2\rangle - \langle O \rangle^2 =
J_O^T\Sigma J_O + \frac{1}{2} \Tr \left( \left( H_O \Sigma \right)^2
\right) \,,
\eeq
where $\langle X\rangle$ denotes the expectation value of a random
variable $X$ and it was again assumed that the parameters $c_i$ follow
the multivariate Gaussian distribution \eqref{multivariate-gaussian}
in order to evaluate their moments. We then define the confidence
interval for $O$ as $\left[ \mu(O) - \sqrt{{\rm Var}(O)}, \mu(O) +
  \sqrt{{\rm Var}(O)} \right]$. Notice that, due to the second-order
term in Eq.~\eqref{observable_mean}, $\mu(O)$ does not necessarily
coincide with $O(c_{\rm min})$, making the error bars for $O(c_{\rm
  min})$ asymmetric. We further emphasize that given that $O(c)$ do
not necessarily  follow a Gaussian distribution, the interpretation of
the quoted statistical uncertainties in terms of confidence intervals
should be taken with care.

\subsubsection{Truncation error}
\label{sec:ErrorTrunc}

To estimate the uncertainties from the truncation of
the chiral expansion we
employ the approach formulated in
\cite{Epelbaum:2014efa,Epelbaum:2014sza}. Specifically, let $X (p)$ be
a NN observable calculated using the chiral nuclear forces in terms of
the expansion 
\beq
X^{(\nu)} (p) = \sum_{i=0}^\nu a_i \, Q^i\,,
\eeq
where $p$ is the characteristic cms momentum scale, $Q$ denotes the
expansion parameter of the chiral EFT and $a_i$ are the 
corresponding coefficients with $a_1 =0 $ in accordance with the
vanishing contribution to the nuclear Hamiltonian at that order.
Following Ref.~\cite{Epelbaum:2014efa}, the expansion parameter $Q$ is 
defined as 
\beq
Q= \max\left( \frac{M_\pi}{\Lambda_{\rm b}}, \; \frac{p}{\Lambda_{\rm b}} \right)\,,
\eeq
and we use for the breakdown scale $\Lambda_{\rm b}$ the value of
$\Lambda_{\rm b} = 600$~MeV as suggested in that paper except for the
results in section \ref{PWD} and appendix \ref{ResPWA}, where a
slightly larger value of $\Lambda_{\rm b} = 650$~MeV is adopted. Notice that
due to the appearance of significant cutoff
artefacts for the softest choice of the regulator $\Lambda = 350$~MeV,
we do not expect this assignment to provide a realistic estimation  
of the breakdown scale in that case, see also the related discussions in 
Refs.~\cite{Epelbaum:2014efa,Melendez:2017phj}. Following
Refs.~\cite{Epelbaum:2014efa,Epelbaum:2014sza,Binder:2015mbz}, we use 
$Q = M_\pi / \Lambda_{\rm b}$ when estimating the uncertainties
of various deuteron properties. 

To calculate the expansion coefficients $a_i$ we exploit the knowledge
of the available results for $X^{(i)}$ at various chiral orders. Specifically, we build the 
differences
\begin{align}
\Delta X^{(2)} &= X^{(2)}-X^{(0)}, \nonumber\\
\Delta X^{(i)} &= X^{(i)}-X^{(i-1)} \;, \quad i\ge 3,
\end{align}
and identify $a_0 \equiv X^{(0)}$ and $a_iQ^i \equiv \Delta X^{(i)}$
for $i \ge 2$. As the expansion of $X^{(\nu )}$ is assumed to be
convergent for $\nu \rightarrow \infty$ and, therefore, the
contributions are expected to become smaller at higher orders, it
is reasonable to assume the truncation error to be dominated 
by the first neglected chiral order, i.e.~we assume that the error scales with $Q^{\nu+1}$. 
Specifically, we  define the truncation error for an observable at
chiral order $\nu$ to be  \cite{Epelbaum:2014efa}
\begin{align}
	\delta X^{(0)} &= Q^2|X^{(0)}|\,, \nonumber \\
	\delta X^{(\nu)} &= \max_{2 \le i \le \nu}\left( Q^{\nu+1}
                           \big|X^{(0)}\big|, \; Q^{\nu+1-i}\big|\Delta X^{(i)}\big|
                           \right)  \,, \quad \nu \ge 2,
\end{align}
subject to the additional constraint 
\beq
\delta X^{(\nu)} \ge  \max_{i, j} \left( \big| X^{(i \ge \nu)} -
  X^{(j \ge \nu)} \big| \right)\,.
\eeq
Notice that the Bayesian analysis of the results of Refs.~\cite{Epelbaum:2014efa,Epelbaum:2014sza} 
for various NN observables
%np total cross section 
carried out in Refs.~\cite{Furnstahl:2015rha,Melendez:2017phj} has revealed that
the above algorithm corresponds to a particular form of the prior
probability distribution, and that the estimated truncation
errors are (on average) consistent with the $68\%$ degree-of-belief
intervals for not too soft cutoff choices. Those studies have also
confirmed the estimation of the breakdown scale $\Lambda_{\rm b}
\sim 600$~MeV of  Refs.~\cite{Epelbaum:2014efa}. For further
discussion of this approach to quantifying the truncation error and of
its limitations the reader is refereed to Ref.~\cite{Epelbaum:2015pfa}.

\subsubsection{Uncertainty in the $\pi N$ LECs}
\label{sec:UncertaintypiN}

Another source of uncertainty is related to the employed values of
the $\pi N$ LECs $c_i$, $\bar d_i$ and $\bar e_i$, which are taken from matching
chiral perturbation theory to the recent
Roy-Steiner-equation analysis of $\pi N$ scattering in the
subthreshold region \cite{Hoferichter:2015tha}. This paper also quotes
uncertainties emerging from the known accuracy of  the
subthreshold coefficients used as input in the determination of the
$\pi N$ LECs, which can, in principle, be propagated when calculating
NN observables.  Such an error analysis would, however,
not reflect the systematic uncertainty in the $\pi N$ LECs related to
the truncation error of the chiral expansion and to the employed 
method and/or kinematics chosen to determine the LECs. In fact, by far
the dominant source of uncertainty in these  LECs is
known to reside in the chiral expansion of the $\pi N$ scattering
amplitude as reflected by the significant shifts in the LECs
extracted at different orders, see Table \ref{piNLECs}. 
Thus, we follow here a different procedure and estimate the sensitivity of our
predictions to the employed values of these parameters by calculating 
NN observables using an alternative set of the  $\pi N$  LECs as
specified in table \ref{piNLECs} (set 2). 
This allows
us to probe also some of the systematic uncertainties mentioned
above since the $\pi N$ LECs employed at N$^4$LO/N$^4$LO$^+$ are
determined from
different kinematical regions of $\pi N$ scattering, namely from 
the unphysical region around the subthreshold point for set 1 and  
from the physical region for set 2, see section \ref{sec:Parameters} for more
details. 

To this aim, we have redone the fits of the NN LECs using
the same protocol at all orders starting from N$^2$LO and for all
cutoff values using the $\pi N$ LECs from set 2. The resulting
description of the data is generally found to be 
comparable to the one based on the LECs from the Roy-Steiner equation
analysis (set 1) as given in Tables \ref{tab_chi2_AllOrders} and \ref{tab_chi2_AllCutoffs}.
More precisely, the $\pi N$ LECs from set 2 typically yield a
better description of the NN data at N$^3$LO and N$^4$LO (except for
low-energy data). On the other hand, the results at N$^4$LO$^+$ appear
to be
somewhat worse than those based on  the $\pi N$ LECs from set 1. For example, 
for the intermediate cutoff $\Lambda = 450$~MeV, we obtain at N$^4$LO$^+$ the values of 
$\chi^2/{\rm datum}$ of $1.10$ ($0.93$), $1.09$ ($0.98$) and $1.09$ ($1.05$)
for np (pp) data in the energy ranges of $E_{\rm lab} = 0-100$~MeV,
$E_{\rm lab} = 0-200$~MeV and $E_{\rm lab} = 0-300$~MeV, 
respectively, which have to be compared with the ones for $\pi N$ LECs
from set 1 given in the
last column of Table \ref{tab_chi2_AllOrders}. The worse description of the low-energy pp
data can be traced back to deviations in the $^1$D$_2$ partial wave. 
Interestingly, we find that  for hard cutoff choices, the potentials based on the $\pi N$ LECs
from set 2 tend to be more perturbative
than the ones based on the LECs
from the Roy-Steiner equation analysis. This  is especially
pronounced at N$^3$LO for  $\Lambda = 550$~MeV. 

\subsubsection{Systematic uncertainty due to the choice of the energy
  range in the fits}

Last but not least, we address the uncertainty related to the choice
of the maximum energy when performing fits to NN data. Here and in
what follows, we restrict ourselves to the highest chiral order and
thus consider only N$^4$LO$^+$ potentials. As will be shown in the
following sections, the resulting uncertainty typically appears to be small
compared with the other error sources. When lowering the chiral order,
the truncation error increases rapidly, and the systematic 
uncertainty due to the choice of $E_{\rm lab}^{\rm max}$ in the fits may
be expected to become even less relevant.  

\begin{figure}[tb]
\vskip 1 true cm
\begin{center}
\includegraphics[width=0.6\textwidth,keepaspectratio,angle=0,clip]{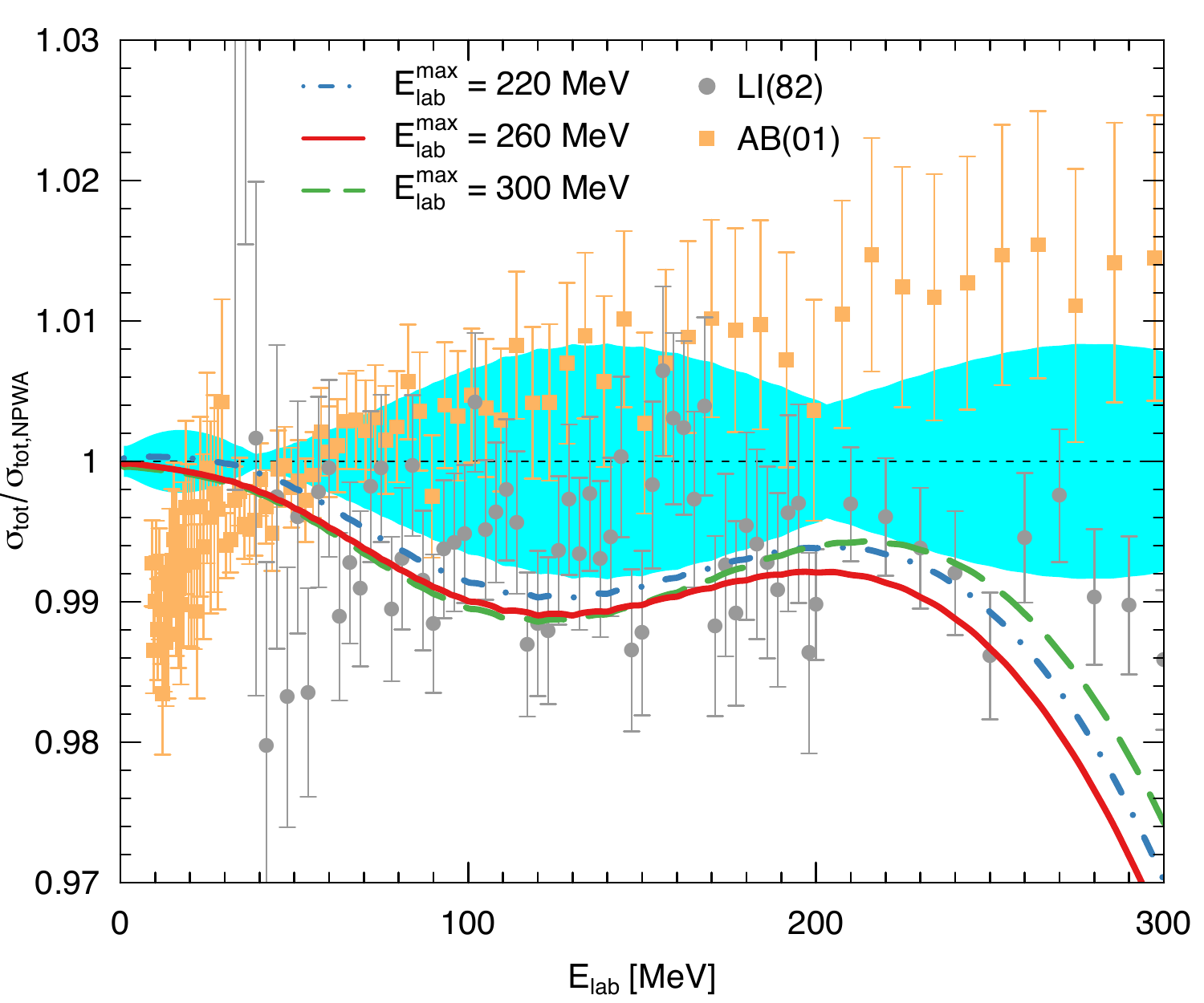}
\end{center}
    \caption{(Color online) The total cross section
      $\sigma_{\rm tot}$, normalized to the values from the Nijmegen
      PWA, as obtained at N$^4$LO$^+$ using $\Lambda = 450$~MeV. Blue dash-dotted, red
      solid and green dashed lines correspond to the fits up to $E_{\rm
        lab}^{\rm max} = 220$, $260$ and $300$~MeV, respectively. The
      cyan band shows the results based on  the 
      Nijm I, II and Reid93 potentials of Ref.~\cite{Stoks:1994wp}. The LI(82) and AB(01)
      data are taken from Refs.~\cite{Lisowski:1982rm}  and
      \cite{Abfalterer:2001gw}, respectively. 
\label{fig:SGT}
}
\end{figure}

As described in section \ref{sec:fitting}, the fits of the N$^4$LO$^+$ potentials
are performed to scattering 
data with the laboratory energy of up to $E_{\rm lab}^{\rm max}=260$ MeV. To assess
the sensitivity to this choice, we performed two additional fits by
varying the maximum energy by $\pm 40$~MeV, i.e.~by  
including the data up to $E_{\rm lab}^{\rm max}=220$ and $300$~MeV, and using the
same fitting protocol. Notice that further reducing $E_{\rm lab}^{\rm max}$ makes our
fits unstable simply because the data do not contain sufficient 
information to accurately constrain the values of the NN LECs. The
resulting values of the $\chi^2/{\rm datum}$ for three different
fits are collected in Table \ref{tab_chi2_energyrange} for the
intermediate cutoff of $\Lambda = 450$ MeV. 
\begin{table}[tb]
	\caption{$\chi^2/{\rm datum}$ for the description of the
		neutron-proton and proton-proton scattering data at N$^4$LO$^+$ for  $\Lambda = 450$~MeV as a function of the
		maximum fitting energy $E_{\rm lab}^{\rm max}$. 
		\label{tab_chi2_energyrange}}
	\smallskip
\begin{tabular*}{\textwidth}{@{\extracolsep{\fill}}llll}
	\hline 
	\hline 
	\noalign{\smallskip}
	$E_{\rm lab}$ bin &  $E_{\rm lab}^{\rm max} = 220$ MeV   &  $E_{\rm lab}^{\rm max} = 260$ MeV   &  $E_{\rm lab}^{\rm max} = 300$ MeV 
	\smallskip
	\\
	\hline 
	\hline 
	%\smallskip
	\multicolumn{4}{l}{neutron-proton scattering data} \\ 
	0--100 & 1.07 & 1.08 & 1.08 \\ 
	0--200 & 1.06 & 1.07 & 1.07 \\
	0--300 & 1.10 & 1.06  & 1.06 \\ [4pt]
	%\smallskip
	\hline 
	% \smallskip
	\multicolumn{4}{l}{proton-proton scattering data} \\ 
	0--100 & 0.86 & 0.86 & 0.87 \\ 
	0--200 & 0.95 & 0.95 & 0.96 \\ 
	0--300 & 1.00 & 1.00 & 0.98 \\ [4pt]
	\hline 
	\hline 
\end{tabular*}
\end{table}
The description of pp data in terms of the $\chi^2/{\rm datum}$
turns out to be very stable with
respect to the included energy range. We do, however, observe minor improvement
for high-energy data upon including them 
in the fits as may be expected. Quite remarkably, there is almost no
worsening in the description of pp data when lowering $E_{\rm lab}^{\rm max}$
down to $220$~MeV.   The stability of
the pp results is also reflected in the very small changes of the
corresponding NN LECs and phase shifts. For np data, the dependence of the 
$\chi^2/{\rm datum}$ on $E_{\rm lab}^{\rm max}$ appears to be somewhat larger.  
This is in line with the discussion in
section \ref{sec:StatUncertainty} and presumably can be traced back to
larger experimental errors of np data as compared to pp ones, 
along with the larger number of parameters which need to be determined
for np  observables. The resulting variations in the values of LECs
turn out to be roughly of the same size as the corresponding statistical uncertainties except
for $E_{1F3}$, which does change strongly when performing fits up to 
$E_{\rm lab}^{\rm max} = 220$~MeV. Again, this is a consequence of the
fact that the
scattering data below this energy do not allow for an accurate
determination of this LEC. Still, even for np results, it is still
fair to say that the sensitivity to the variation of $E_{\rm lab}^{\rm
  max}$ is very small except for the softest cutoff choice of $\Lambda
= 350$~MeV. As a representative example, we show in Fig.~\ref{fig:SGT} 
the
dependence of the total cross section for np scattering as function of the
laboratory energy, normalized to the values from the Nijmegen partial
wave analysis, on the choice of $E_{\rm lab}^{\rm
  max}$ for N$^4$LO$^+$ using $\Lambda = 450$~MeV. Notice that the
data by Abfalterer {\it et al.}, AB(01), are not included in the 2013 Granada
database. 
It is clear that the size of the variations is small compared with the
other uncertainties such as especially the truncation error, which
e.g.~amounts to $\delta (\sigma_{\rm tot}/\sigma_{\rm tot, NPWA}) \sim 1.5\%$
at the energy of $E_{\rm lab} = 200$~MeV. 

\subsection{Effective range parameters}
\label{ERE}

Having specified our approach to uncertainty quantification, 
we are now in the position to present results for the S-wave 
scattering lengths and effective range parameters. Our N$^4$LO$^+$
predictions for the various quantities are collected in Table \ref{tab_ere} for
the cutoff values of $\Lambda = 400 - 550$~MeV. 
\begin{table}[t]
\caption{S-wave scattering lengths and effective
  range parameters as obtained at N$^4$LO$^+$ for the cutoff values of $\Lambda =
  400$, $450$, $500$ and $550$~MeV. The first uncertainty is
  statistical, the second one corresponds to the truncation error at
  N$^4$LO, the third one estimates the sensitivity to the $\pi
  N$ LECs while the last uncertainty reflects the sensitivity to the
  choice of the maximum energy in the fits as explained in the
  text. The statistical uncertainty of the np scattering lengths are
  calculated from the corresponding allowed variation of the
  $\chi^2/{\rm datum}$ as described in section \ref{sec:StatUncertainty}. For all other
  observables, the estimations are based on the quadratic approximation
  in Eq.~(\ref{observable-expansion}) along with  the corresponding
  covariance matrices. 
The np (pp) phase shifts
are calculated with respect to Riccati-Bessel (Coulomb) functions. 
\label{tab_ere}}
\smallskip
\begin{tabular*}{\textwidth}{@{\extracolsep{\fill}}lrrrrr}
\hline 
\hline
\noalign{\smallskip }
 &$\Lambda = 400$ MeV  &$\Lambda = 450$ MeV  &$\Lambda = 500$ MeV
  &$\Lambda = 550$ MeV  & Empirical
\smallskip 
 \\
\hline \hline
\noalign{\smallskip} 
$a_{1S0}^{\rm np}$ (fm)  &  $-23.739_{(-4)}^{(+4)} (4)(4)(5)$  &
                                                            $-23.739_{(-4)}^{(+4)}
                                                                  (4)(6)(2)$
                                             & $-23.742_{(-4)}^{(+4)}
                                               (4)(8)(0)$  &
                                                             $-23.744_{(-4)}^{(+4)} (4)(9)(1)$ & $-23.748(10)$ \cite{Dumbrajs:1983jd}\\ [0pt]
&&&&&$-23.740(20)$ \cite{Machleidt:2000ge}\\ [7pt]
$r_{1S0}^{\rm np}$ (fm)  &  $2.706_{(-8)}^{(+8)} (19)(20)(30)$  &
                                                            $2.697_{(-8)}^{(+7)}
                                                                (16)(22)(17)$
                                             & $2.698_{(-7)}^{(+6)}
                                               (12)(24)(15)$  &
                                                              $2.702_{(-7)}^{(+6)} (11)(25)(14)$ & $2.75(5)$\cite{Dumbrajs:1983jd}\\ [0pt]
&&&&&$2.77(5)$ \cite{Machleidt:2000ge}\\ [7pt]
$a_{3S1}$ (fm)  &  $5.420_{(-1)}^{(+1)} (1)(1)(2)$  &
                                                            $5.420_{(-1)}^{(+1)}
                                                      (1)(2)(1)$  &
                                                                    $5.421_{(-1)}^{(+1)}
                                                                    (1)(3)(0)$
  & $5.422_{(-1)}^{(+1)} (1)(3)(0)$ & $5.424(4)$ \cite{Dumbrajs:1983jd}\\ [0pt]
&&&&&$5.4194(20)^\star$ \cite{deSwart:1995ui}\\ [7pt]
$r_{3S1}$ (fm)  &  $1.754_{(-2)}^{(+2)} (2)(1)(2)$  &
                                                            $1.754_{(-2)}^{(+2)} (1)(2)(1)$  & $1.754_{(-2)}^{(+2)} (1)(3)(0)$  & $1.755_{(-2)}^{(+2)} (1)(3)(0)$ & $1.759(5)$\cite{Dumbrajs:1983jd}\\ [0pt]
&&&&&$1.7536(25)^\star$ \cite{deSwart:1995ui}\\ [7pt]
$a_{1S0}^{\rm pp}$ (fm)  &  $-7.814_{(-1)}^{(+1)} (1)(2)(0)$  &
                                                            $-7.815_{(-1)}^{(+1)}
                                                                (1)(2)(0)$
                                             & $-7.816_{(-2)}^{(+1)}
                                               (1)(2)(0)$  &
                                                             $-7.816_{(-3)}^{(+1)} (2)(3)(0)$ & $-7.817(4)^\dagger$ \cite{Bergervoet:1988zz}\\ [7pt]
$r_{1S0}^{\rm pp}$ (fm)  &  $2.766_{(-2)}^{(+2)} (4)(7)(1)$  &
                                                            $2.770_{(-2)}^{(+2)}
                                                               (3)(9)(1)$
                                             & $2.772_{(-2)}^{(+2)}
                                               (6)(10)(0)$  &
                                                              $2.773_{(-2)}^{(+2)} (8)(10)(0)$ & $2.780(20)^\dagger$ \cite{Bergervoet:1988zz}\\ [7pt]
\hline \hline
\multicolumn{6}{l}{$^\star$Recommended values; the errors are claimed
  to be 
  ``educated guesses''.} \\[-2pt]
\multicolumn{6}{l}{$^\dagger$Recommended values for the phase shifts
  $\delta_C$ with respect to Coulomb functions.} 
\end{tabular*}
\end{table}
A
significant amount of cutoff artefacts for the case of  
$\Lambda = 350$~MeV makes the uncertainty
analysis in that case less straightforward, and for this reason we refrain from
showing the corresponding results. We have, however, verified that 
the values for all considered quantities based on the softest cutoff
choice are consistent with the 
ones given in  Table \ref{tab_ere}.
Finally, we remind the
reader that given the essentially perfect description of the data at
N$^4$LO$^+$, the results of our partial wave analysis at this order
can be regarded as reference values for these quantities and include a
comprehensive analysis of the various sources of uncertainty. The
quoted truncation errors correspond to N$^4$LO. We
regard the results for the cutoff choice of $\Lambda = 450$~MeV, which leads
to the best description of the experimental data, as our final
predictions. 

For np scattering,
our results correspond to the standard phase shifts calculated with respect to
Riccati-Bessel functions. For pp $^1$S$_0$ phase shift, the values
shown in  Table \ref{tab_ere} correspond to the phase shifts
$\delta_{C1 + N}^{C1}$
calculated with respect to Coulomb wave functions by means of the
well-known modified effective range expansion, see
e.g.~Ref.~\cite{vanHaeringen:1981pb} and references therein. We emphasize that
due to the presence of the long-range electromagnetic interactions
beyond the Coulomb one, the definition of the corresponding phase
shift is, strictly speaking, model dependent, see Ref.~\cite{Bergervoet:1988zz} for more
details. We, however, expect the emerging model dependence to be
small at the level of accuracy of our calculations. The quoted
empirical values for the np $^1$S$_0$ parameters are the ones which
are most frequently used in the literature, see
Refs.~\cite{Babenko:2007ss,Machleidt:2000ge} and references therein for more 
details.  For the pp $^1$S$_0$ parameters, we show the recommended
values from the partial wave analysis of
Ref.~\cite{Bergervoet:1988zz}. More precisely, we have
used the results from Table VI of that paper to 
adjust their recommended values of $a_{1S0, \; \rm EM}^{\rm pp} =
-7.804 \pm 0.004$~fm,  $r_{1S0, \; \rm EM}^{\rm pp} =
2.784 \pm 0.020$~fm, calculated from the phase shift $\delta_{\rm EM +
N}^{EM}$ with respect to the full
electromagnetic interaction and employing a more complicated definition of the 
effective range function, to the ones corresponding to
$\delta_{C1+N}^{C1}$. Finally, for the $^3$S$_1$ parameters, we provide, in
addition to the compilation of experimental values of Ref.~\cite{Dumbrajs:1983jd},
also the results from the Nijmegen partial wave analysis \cite{deSwart:1995ui}. 

Our results for the scattering lengths and effective range parameters
are in an excellent agreement with the empirical numbers. The largest
deviations are observed for $r_{1S0}^{\rm np}$, but the discrepancy is
still only at the level of $\sim 1.5\, \sigma$. We further emphasize that our
predictions are, in many cases, more precise than the quoted
experimental and/or empirical values. The results collected in Table
\ref{tab_ere} also illustrate the usefulness of the approach for quantifying
truncation errors in chiral EFT advocated in Ref.~\cite{Epelbaum:2014efa}, which
allows one to perform independent error analysis for each cutoff value. 
While the results corresponding to different cutoffs
are correlated with each other and, therefore, cannot be combined together
to decrease the uncertainty, they do provide a useful consistency
check of the calculations and error analysis. In particular, it is
comforting to see that the analyses carried out at different values of
$\Lambda$ yield consistent results for all extracted quantities.  

We find that the np $^1$S$_0$ effective range shows by far the
largest uncertainty. This can probably be traced back to the lower
accuracy of the np data as compared to the pp ones and to the
weaker theoretical constraints on the long-range interaction in this channel, which is
not sensitive to the tensor part of the OPEP.\footnote{The same
  arguments apply to the np $^1$S$_0$ scattering length too, which is,
  however, additionally stabilized by constraining the value of the np
  coherent scattering length in our fits.} It is also instructive to
look at the systematics of various uncertainties with respect to the
cutoff variation. In particular, the statistical uncertainty turns out
to be largely insensitive to the value of $\Lambda$ in line with the
already discussed results for the LECs. On the other hand, 
decreasing the cutoff values
results in integrating out a part of the TPEP. This explains why our
results for soft choices of $\Lambda$ show a smaller uncertainty
from the variation of  the $\pi N$ LECs. On the other hand, lower
values of the cutoff result in increased distortions of phase shifts
at larger energies, which increases the sensitivity of the fits to the
choice of $E_{\rm lab}^{\rm max}$. Again, this is clearly reflected in
the uncertainty pattern of our predictions. For $\Lambda \ge 450$~MeV,
the uncertainty from the choice of $E_{\rm lab}^{\rm max}$ in the fits
turns out to be negligible compared to the other error sources. 

Further, we emphasize that while the uncertainties from different
sources turn out to be of a comparable size for the effective range
parameters at N$^4$LO, the uncertainty at lower chiral orders and for
higher-energy observables at N$^4$LO and N$^4$LO$^+$ is, in most
cases, dominated by the 
truncation error, see section \ref{PWD} for more details.  

Finally, given the fact that our results
for the S-wave scattering lengths are affected by the 
constrained values of the deuteron binding energy and the np coherent
scattering length, we have used in these cases a more general approach
for quantifying the statistical uncertainty based on the corresponding
allowed variation of the 
  $\chi^2/{\rm datum}$ as explained in section
  \ref{sec:StatUncertainty}. The usage of the
  covariance matrix together with the quadratic
  approximation in  Eq.~(\ref{observable-expansion}) is found to
  yield overestimated statistical uncertainties for these observables.

\subsection{Deuteron properties}
\label{Deut}

We now turn to the deuteron properties. As already pointed out before,
we use the deuteron binding energy of $B_d = 2.224575$~MeV \cite{DBD}   as
an additional constraint when performing the fits starting from
N$^3$LO. All other deuteron properties come out as predictions. Similarly to our earlier studies
in Refs.~\cite{Epelbaum:2014efa,Epelbaum:2014sza}, we find a very good convergence for the deuteron
parameters with respect to the chiral order as exemplified in Table
\ref{tab_deut2} for the case of the intermediate cutoff $\Lambda = 450$~MeV. 
\begin{table}[t]
\caption{Deuteron binding energy $B_d$, expectation value of the
  kinetic energy $\langle T_{\rm kin} \rangle$, asymptotic $S$ state
  normalization $A_S$, asymptotic $D/S$ state ratio $\eta$, radius
  $r_d$ and quadrupole moment $Q$ at various orders in the chiral expansion
  for the cutoff $\Lambda = 450$~MeV in comparison with empirical
  values. Also shown is the $D$-state
  probability $P_D$. Notice that $r_d$ and $Q_d$ are calculated
  without taking into account meson-exchange current contributions and
  relativistic corrections. The deuteron binding energy is
  calculated by solving the relativistic
  Schr\"odinger equation. 
\label{tab_deut2}}
\smallskip
\begin{tabular*}{\textwidth}{@{\extracolsep{\fill}}lllllllr}
\hline 
\hline
\noalign{\smallskip}
   &  LO & NLO  & N$^2$LO  & N$^3$LO  & N$^4$LO  & N$^4$LO$^+$  & Empirical
\smallskip
 \\
\hline \hline
\smallskip
$B_d$ (MeV)           & 2.1201 & 2.1843 & 2.2012 & 2.2246$^\star$ &
                                                                    2.2246$^\star$ & 2.2246$^\star$ & 2.224575(9) \cite{DBD} \\ 
$\langle T_{\rm kin} \rangle$ (MeV)           & 14.24 &  13.47 &
                                                                   14.44
                           & 14.35 & 14.17 & 14.22 & --- \\ 
$A_S$ (fm$^{-1/2}$) & 0.8436  & 0.8728  & 0.8787  & 0.8844  & 0.8847 &
                                                                       0.8847
                                                                &
                                                                  0.8846(8) \cite{Ericson:1982ei} \\ 
$\eta$                    & 0.0220 & 0.0236 & 0.0251 & 0.0257 &
                                                                0.0255
                                                 &   0.0255 &
                                                              0.0256(4)
  \cite{Rodning:1990zz} \\ 
$r_d$ (fm)              & 1.946 &  1.967  & 1.970 & 1.966 &  1.966 &
                                                                     1.966
                                                                &
                                                                  $1.97535(85)^\dagger$ \cite{Huber:1998zz}
  \\ 
$Q$ (fm$^2$)         & 0.227 & 0.249   & 0.268 & 0.272 &  0.269 &
                                                                  0.270
                                                                &
                                                                  0.2859(3)
  \cite{Bishop:1979zz} \\ 
$P_D$ ($\%$)          & 2.77 & 3.59       & 4.63 & 4.70 & 4.54 &  4.59
                                                                &
                                                                  --- \\[4pt]
\hline \hline
\end{tabular*}
\begin{tabular*}{\textwidth}{@{\extracolsep{\fill}}l}
$^\star$The deuteron binding energy has been
  taken as input in the fit. \\[-2pt]
$^\dagger$This value corresponds to the so-called deuteron
structure radius, which is defined as a square root of the difference
of the \\[-3pt]
deuteron, proton and neutron mean square charge radii.
\end{tabular*}
\end{table}
Notice further that contrary to those studies, we do not include in our fits
any constraints on the D-state probability $P_D$ as already explained
in section \ref{sec:fitting}. In Table \ref{tab_deut}, we compare our N$^4$LO$^+$
predictions for all considered cutoff values with each other
and with the corresponding results based on the potentials of
Ref.~\cite{Entem:2017gor}.\footnote{While the corresponding potentials
  are referred to as N$^4$LO in Ref.~\cite{Entem:2017gor}, we prefer
  to regard them as N$^4$LO$^+$ since they
  also include the four N$^5$LO contact interactions in F-waves
  given in Eq.~(\ref{LECsF}). No strict N$^4$LO potentials are
  provided in that paper.}
\begin{table}[t]
\caption{Deuteron properties calculated based on the N$^4$LO$^+$
  potentials of this work and the ones of Ref.~\cite{Entem:2017gor} for
  different cutoff values $\Lambda$ (in units of MeV). For notation see Table
  \ref{tab_deut2}. Notice that the deuteron binding energy is
  calculated by solving the nonrelativistic (relativistic)
  Schr\"odinger equation for the potentials of
  Ref.~\cite{Entem:2017gor}  (of this work).  
\label{tab_deut}}
\smallskip
\begin{tabular*}{\textwidth}{@{\extracolsep{\fill}}lllllllllr}
\hline 
\hline
\noalign{\smallskip} 
& \multicolumn{3}{c}{ ---  N$^4$LO$^+$ potentials of \cite{Entem:2017gor}
    --- } &
\multicolumn{5}{c}{ --- SMS chiral potentials at N$^4$LO$^+$, this
  work  --- } & Empirical
\\
   &   $\Lambda = 450$    &  $\Lambda = 500$   &  $\Lambda=550$   & 
$\Lambda=350$ & $\Lambda=400$  & $\Lambda=450$  & $\Lambda=500$  &
 $\Lambda=550$  &
\smallskip
 \\
\hline \hline
\smallskip
$B_d$ (MeV) & 2.2246$^\star$ & 2.2246$^\star$ & 2.2246$^\star$   &
                                                           2.2246$^\star$
                        & 2.2246$^\star$ & 2.2246$^\star$ &
                                                            2.2246$^\star$
              & 2.2246$^\star$ & 2.224575(9)  \cite{DBD}  \\ 
$\langle T_{\rm kin} \rangle$ (MeV)   & 13.05 & 13.32 &  14.04  &
                                                                  12.73
                        & 13.44  & 14.22  & 15.11  & 15.98  & --- \\ 
$A_S$ (fm$^{-1/2}$) & 0.8852 & 0.8852 &  0.8851  &      0.8848  & 0.8847  & 0.8847  & 0.8849  & 0.8851  & 0.8846(9)  \cite{Ericson:1982ei} \\ 
$\eta$ & 0.0254 & 0.0258 &    0.0257 &           0.0256 & 0.0255 & 0.0255 & 0.0257 & 0.0258 &  0.0256(4) \cite{Rodning:1990zz} \\ 
$r_d$ (fm) & 1.966 & 1.973 &   1.970  &          1.965 & 1.965 &
                                                                 1.966
                                                              & 1.967
              & 1.968 &  $1.97535(85)^\dagger$ \cite{Huber:1998zz} \\ 
$Q$ (fm$^2$)& 0.269 & 0.273 &   0.271 &             0.265 & 0.267 & 0.270 & 0.273 & 0.276 &  0.2859(3)  \cite{Bishop:1979zz} \\ 
$P_D$ ($\%$) & 4.38 & 4.10 &   4.13 &          3.58 & 4.12 & 4.59 & 5.01 & 5.35 & --- \\[4pt]
\hline \hline
\end{tabular*}
\begin{tabular*}{\textwidth}{@{\extracolsep{\fill}}l}
$^\star$The deuteron binding energy has been
  taken as input in the fit. \\[-2pt]
$^\dagger$This value corresponds to the so-called deuteron
structure radius, which is defined as a square root of the difference
of the \\[-3pt]
deuteron, proton and neutron mean square charge radii.
\end{tabular*}
\end{table}
We also show the expectation value of the kinetic energy which, while
not being 
an observable quantity, allows one to draw conclusions regarding the
strength of the short-range repulsion and perturbativeness of the
potentials. For the semilocal coordinate-space regularized potentials
of Refs.~\cite{Epelbaum:2014efa,Epelbaum:2014sza}, the
expectation value of the kinetic energy in the deuteron varies in the range of  
$\langle T_{\rm kin} \rangle = 16.13 \ldots 23.33$~MeV at N$^3$LO and 
$\langle T_{\rm kin} \rangle = 15.27 \ldots 20.59$~MeV at N$^4$LO
depending on the cutoff $R$. 
These large values are in line with the nonperturbative
nature of these potentials \cite{Hoppe:2017lok} and 
signal the appearance of a strong repulsive core in the $^3$S$_1$
channel, which can be traced back to the large numerical values of the
redundant contact interactions at order $Q^4$, see Table II of
Ref.~\cite{Epelbaum:2014efa}.  Further, we observe somewhat larger
values for the asymptotic S-state normalization $A_S$ as compared with
$A_S = 0.8843 \ldots 0.8846$~fm$^{-1/2}$ using 
the N$^4$LO potentials of Ref.~\cite{Epelbaum:2014sza}. This difference
emerges primarily from using the $\pi N$ LECs from the Roy-Steiner
equation analysis. 
The remaining deuteron properties come out
rather similar to the ones based on the coordinate-space regularized potentials
of Refs.~\cite{Epelbaum:2014sza}. 

\begin{figure}[tb]
\vskip 1 true cm
\begin{center}
\includegraphics[width=\textwidth,keepaspectratio,angle=0,clip]{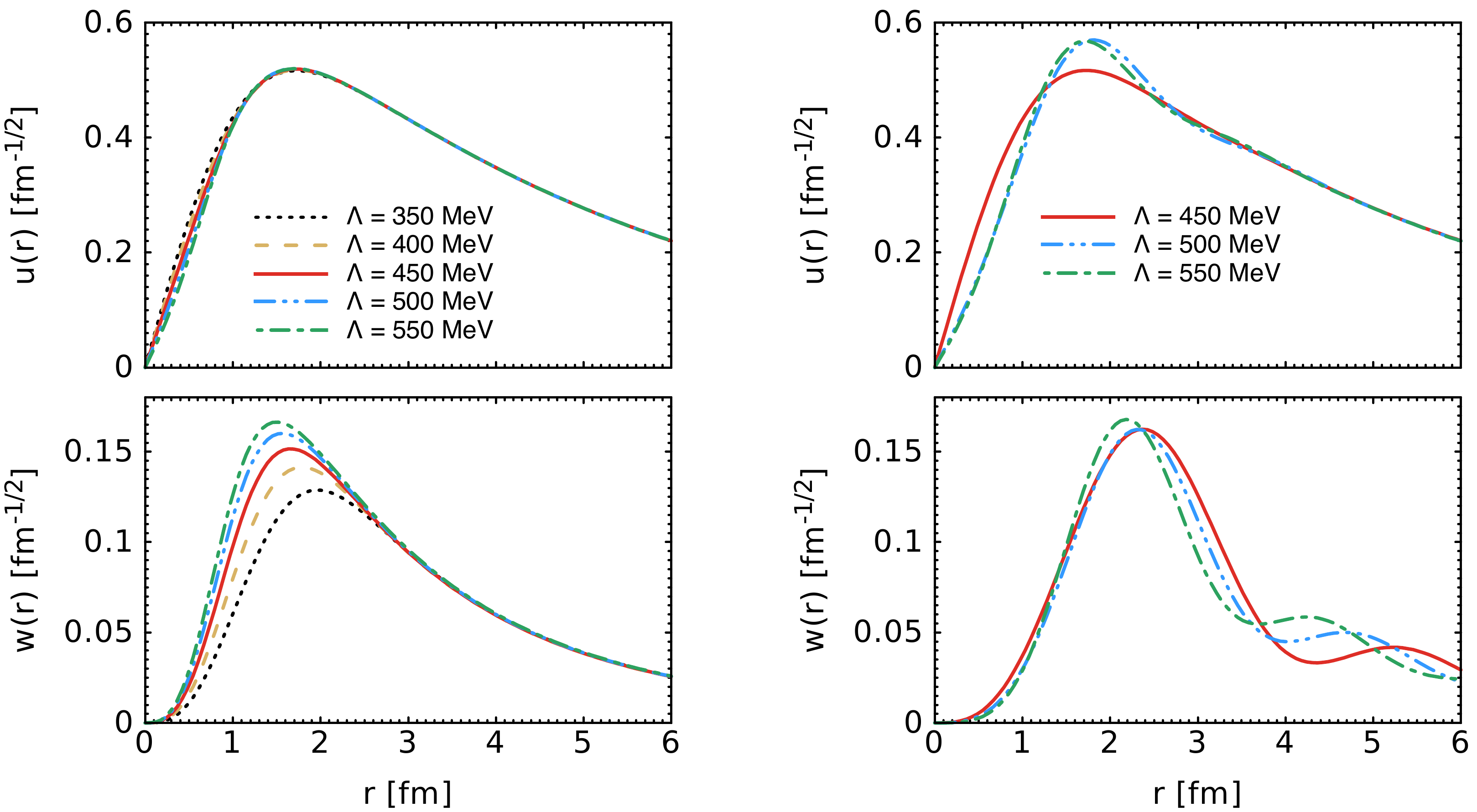}
\end{center}
    \caption{(Color online) Deuteron wave functions in coordinate
      space based on the SMS N$^4$LO$^+$ potentials of this work (left
      panel) and N$^4$LO$^+$ potentials of Ref.~\cite{Entem:2017gor}
      (right panel).
      Black dotted, brown dashed, red solid, blue dashed-dotted and
      green dashed-double-dotted lines show the results obtained using
      the cutoffs $\Lambda = 350$, $400$, $450$, $500$ and $550$~MeV,
      respectively. 
\label{fig:DeuteronWF}
}
\end{figure}

In Fig.~\ref{fig:DeuteronWF}, we show the deuteron wave functions 
at N$^4$LO$^+$ for all considered cutoff values (left panel).
The S-state wave functions appear to be very stable with respect to the
cutoff variation at distances $r \gtrsim 1$~fm and take the values of $u(r)/r
\big|_{r = 0} = 0.32 \ldots 0.55$~fm$^{-3/2}$, depending on the
cutoff, at the origin. This has to be compared with the corresponding variation for
the N$^4$LO potentials of Ref.~\cite{Epelbaum:2014sza}, $ u(r)/r \big|_{r = 0}
= -0.22 \ldots 0.14$~fm$^{-3/2}$,  and is in line with a
significantly smaller amount of short-distance repulsion in the new
potentials already mentioned in connection with the expectation value
of the kinetic energy.  The D-state wave function shows a considerably
stronger cutoff variation and becomes insensitive to the values of 
$\Lambda$ only for  $r \gtrsim 2.5$~fm. We also show in the right panel of
Fig.~\ref{fig:DeuteronWF} the corresponding wave functions from the nonlocal
potentials of Ref.~\cite{Entem:2017gor} at N$^4$LO$^+$. The
oscillations occur presumably due to the non-Gaussian form of the
regulator functions employed in that paper. Interestingly, one observes
that the maximum probability for the nucleons be in a D-state is
shifted in the potentials of Ref.~\cite{Entem:2017gor} towards considerably
larger distances. This could be a consequence of a stronger suppression
of the tensor part of the OPEP at intermediate distances for the
nonlocal regulator employed in that paper. 

Finally, we list in Table \ref{tab_errors} our results for the asymptotic S state normalization $A_S$
  and asymptotic D/S state ratio $\eta$ at N$^4$LO$^+$  along with the
  corresponding uncertainties. 
\begin{table}[t]
\caption{Predictions for the asymptotic S state normalization $A_S$
  and asymptotic D/S state ratio $\eta$ at N$^4$LO$^+$ for the cutoff values of $\Lambda =
  400$, $450$, $500$ and $550$~MeV. The first uncertainty is
  statistical, the second one corresponds to the truncation error at
  N$^4$LO, the third one estimates the sensitivity to the $\pi
  N$ LECs while the last uncertainty reflects the sensitivity to the
  choice of the maximum energy in the fits as explained in the text. 
  The statistical uncertainty of $A_S$ is
  calculated from the corresponding allowed variation of the
  $\chi^2/{\rm datum}$ as described in section
  \ref{sec:StatUncertainty}. For $\eta$, the estimations are based on the quadratic approximation
  in Eq.~(\ref{observable-expansion}) along with  the corresponding
  covariance matrices. 
\label{tab_errors}}
\smallskip
\begin{tabular*}{\textwidth}{@{\extracolsep{\fill}}lrrrrr}
\hline 
\hline
\noalign{\smallskip }
 &$\Lambda = 400$ MeV  &$\Lambda = 450$ MeV  &$\Lambda = 500$ MeV
  &$\Lambda = 550$ MeV  & Empirical
\smallskip 
 \\
\hline \hline
\noalign{\smallskip} 
$A_S$ (fm$^{-1/2}$)  &  $0.8847_{(-3)}^{(+3)} (6)(4)(4)$  &
                                                            $0.8847_{(-3)}^{(+3)} (3)(5)(1)$  & $0.8849_{(-3)}^{(+3)} (1)(7)(0)$  & $0.8851_{(-3)}^{(+3)} (3)(8)(1)$ & 0.8846(8)  \cite{Ericson:1982ei} \\ [5pt]
$\eta$ & $0.0255_{(-1)}^{(+1)} (1)(3)(2)$ & $0.0255_{(-1)}^{(+1)}
                                            (1)(4)(1)$ &
                                                         $0.0257_{(-1)}^{(+1)} (1)(5)(1)$ & $0.0258_{(-1)}^{(+1)} (1)(5)(1)$ & 0.0256(4)  \cite{Rodning:1990zz} \\[5pt]
\hline \hline
\end{tabular*}
\end{table}
Our predictions for these quantities agree with the
empirical values and with the values obtained from the
phenomenological potentials including the CD Bonn \cite{Machleidt:2000ge}, Nijm I, II and
Reid93 \cite{Stoks:1994wp} potentials. On the other hand, there is a significant disagreement with
the partial wave analysis by the Granada group of
Ref.~\cite{Perez:2013mwa}, which quotes the values of
$A_S = 0.8829(4)$~fm$^{1/2}$ and $\eta = 0.02493(8)$ (where the
uncertainties are of the purely statistical nature). 
The systematics of the various error sources of $A_S$ and $\eta$
with respect to the
cutoff variation appears to be similar to the one discussed in section \ref{ERE} in
the context of the effective range parameters. Notice that in spite of
a significant dependence of $A_S$ and $\eta$ on the values of the
$\pi N$ LECs, it is not possible to discriminate between the different
sets given the accuracy of the empirical values (and statistical and
truncation uncertainties of our results). We do not quote the uncertainty
for the remaining deuteron properties either because they are not
observable ($\langle T_{\rm kin} \rangle$, $P_D$) or since our
present calculations of them are incomplete ($r_d$,
$Q$). Interestingly, these
features are also clearly reflected in the cutoff variation for these
quantities, which reaches $\sim 23\%$ ($\sim 40\%$) for $\langle
T_{\rm kin} \rangle$ ($P_D$) for $\Lambda = 350 -550$~MeV. For $Q$
and $r_d$, the cutoff variation is smaller and amounts to $\sim 4\%$ and $0.15\%$, respectively.\footnote{The smaller cutoff
  dependence of the deuteron radius reflects the long-range nature of
  this observable as opposed to that of $Q$.}  In both cases, the
observed $\Lambda$-dependence is smaller than the deviations from the very
precisely known experimental/empirical values listed in Table
\ref{tab_deut}. These deviations amount to $\sim 0.015$~fm$^2$ and $\sim 0.009$~fm
for $Q$ and $r_d$, respectively, and are comparable with the truncation
errors for these quantities at N$^2$LO, namely $\delta Q^{(3)} = \pm (0.005 \ldots
0.011)$~fm$^2$  (depending on the cutoff) and $\delta r_d^{(3)} = \pm 0.005$~fm, which estimate the expected size of 
N$^3$LO contributions to these observables.  This is fully in line with the fact that our
calculations do not take into account the relativistic corrections and
contributions to the exchange charge operator at N$^3$LO, see
Ref.~\cite{Kolling:2009iq,Kolling:2011mt} for explicit
expressions. Our results further indicate that starting from N$^3$LO,
the theoretical uncertainty for both quantities is dominated by the
one of the $\pi N$ LECs similarly to other low-energy observables considered in
this and previous sections. For both $Q$ and $r_d$, employing the $\pi N$ LECs
from set 2 tends to increase the discrepancy with the empirical
numbers. 

\subsection{Phase shifts}
\label{PWD}

\begin{figure}[tb]
\vskip 1 true cm
\begin{center}
\includegraphics[width=1.0\textwidth,keepaspectratio,angle=0,clip]{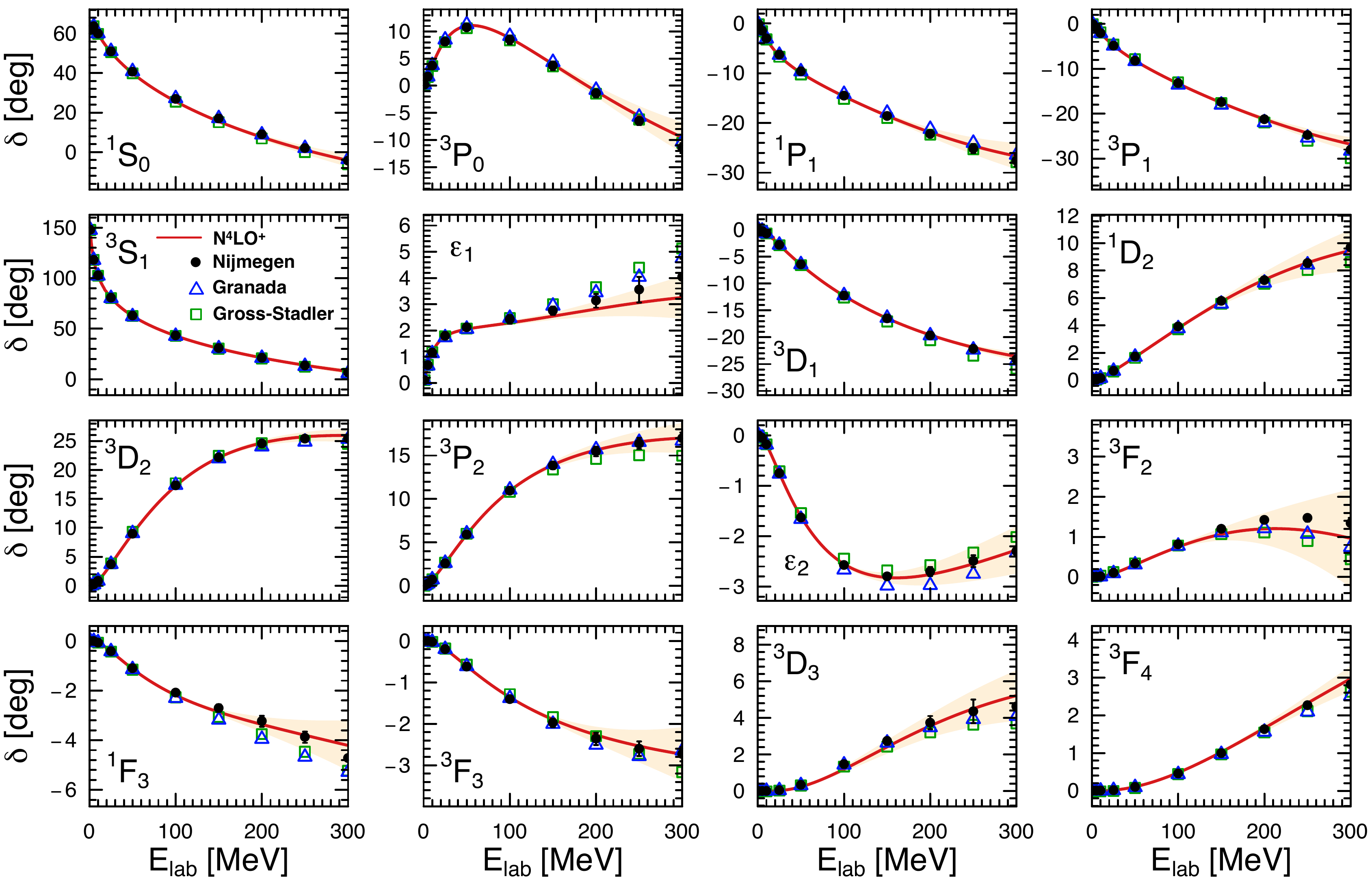}
\end{center}
    \caption{(Color online) Neutron-proton S-, P-, D- and F-wave phase
      shifts and the mixing angles $\epsilon_1$, $\epsilon_2$ and
      $\epsilon_2$ as obtained at N$^4$LO$^+$ using the cutoff
      $\Lambda = 450$~MeV (red solid lines) in comparison with  the Nijmegen \cite{Stoks:1993tb} (solid dots)
      the Granada \cite{Perez:2013mwa}  (blue open triangles) and
      Gross-Stadler \cite{Gross:2008ps}  (green open squares)
      PWA. Light shaded bands show the estimated truncation error as
      explained in appendix \ref{ResPWA}. The shown uncertainties of the
      Nijmegen PWA correspond to systematic errors estimated from the
      Nijm I, II and Reid93 potentials \cite{Stoks:1994wp} as explained in Ref.~\cite{Epelbaum:2014efa}.
\label{fig:phaseshiftsN4LO+}
}
\end{figure}

In Fig.~\ref{fig:phaseshiftsN4LO+}, 
we show the np S-, P-, D- and F-wave phase shifts and 
the mixing angles $\epsilon_1$ and $\epsilon_2$ as obtained at
N$^4$LO$^+$ for the cutoff $\Lambda = 450$~MeV along with the
estimated truncation uncertainties as explained in appendix
\ref{ResPWA}.  The resulting values for the pp and np phase shifts and
mixing angles in the fitted channels at the laboratory energies of 
$E_{\rm lab} = 1$, $5$, $10$, $25$, $50$, $100$, $150$, $200$, $250$
and $300$~MeV are listed in Tables \ref{tab:ppSP}-\ref{tab:npEps} of
this appendix together with the estimated statistical and truncation
errors as well as the uncertainties reflecting the sensitivity to the
$\pi N$ LECs and the choice of the maximum fitting energy in the
fits. Similarly to the effective range parameters and the deuteron
properties, the theoretical uncertainty of the phase shifts at very
low energy is typically dominated by the uncertainty in the $\pi N$
LECs. On the other hand, starting from $E_{\rm lab} \sim 100$~MeV, the
truncation error starts becoming dominant. With very few exceptions, the
statistical uncertainty is found to be negligibly small as compared
to other error sources. The sensitivity to the employed energy range
in the fits also appears to be small except for few cases such as
especially the $^1$F$_3$ phase shift, for which it dominates the
theoretical uncertainty. 
 
As already pointed out, our N$^4$LO$^+$ fit for $\Lambda = 450$~MeV provides a
nearly perfect description of the data and thus qualifies to be
considered as partial wave analysis. It is, therefore,  interesting to
compare our results for phase shifts with the ones from other
PWAs. For pp phase shifts and mixing angles, our results are in a 
very good agreement with the Nijmegen PWA \cite{Stoks:1993tb} except for the $^3$P$_0$
phase shift at low energy, where the differences are at the
$\sim 3\sigma$ level assuming that our uncertainty is dominated by the
statistical error. The discrepancy is slightly less pronounced for
the Granada PWA of Ref.~\cite{Perez:2013mwa}. Notice that similarly to
that paper, we find considerably smaller
statistical uncertainties as compared with the Nijmegen PWA. 
More differences are found for np phase shifts and mixing
angles. First, the already mentioned discrepancy in the $^3$P$_0$
phase shift at low energy persists in the np system, where it 
becomes even more pronounced. For example, we obtain  at $E_{\rm lab} =
50$~MeV the value of 
$\delta_{3P0}^{\rm np} = 11.02^\circ \pm 0.05^\circ$, where the uncertainty
is purely statistical, which has to be compared with the results from
the Nijmegen and Granada  PWA of $\delta_{3P0}^{\rm np} = 10.70^\circ$
and  $\delta_{3P0}^{\rm np} = 11.30^\circ  \pm 0.03^\circ$,
respectively, and with the value from the Gross-Stadler analysis of
Ref.~\cite{Gross:2008ps},  $\delta_{3P0}^{\rm np} = 10.61^\circ$. 
%Given that the np isovector phase shifts are extracted solely from 
%pp data (from both pp and np data) in the Nijmegen PWA (in our
%analysis), these results indicate that IB contributions to the nuclear
%force may play an important role in this particular case. 
Furthermore, there are significant differences between our results and
the ones of both the Granada  and Gross-Stadler PWA for the $^3$S$_1$
phase shifts at the lowest considered energy of $E_{\rm lab} = 1$~MeV.  
Since the deuteron S-state normalization $A_S$ is correlated with the
$^3$S$_1$ scattering length, this discrepancy is probably the origin
of the different predictions for this observable mention in section
\ref{Deut}. Our results for  the $^3$S$_1$
phase shift are, however, in an excellent agreement with those from the Nijmegen
PWA. We also observe some discrepancies between our results and those
from the Nijmegen \cite{Stoks:1993tb} and Granada \cite{Perez:2013mwa} PWA for the mixing angle
$\epsilon_2$ ($\epsilon_3$) at low energies (at energies $E_{\rm lab}
\gtrsim 200$~MeV). Comparison of our results with those from the
Gross-Stadler analysis of Ref.~\cite{Gross:2008ps} reveals more differences, in
particular also in the $^3$P$_1$, $^3$F$_3$ and $^3$F$_4$ channels. We
emphasize, however, that their analysis is based solely on np
scattering data and thus potentially suffers from considerably larger
statistical uncertainties, which are, however, not quoted in that
paper. Finally, our N$^4$LO$^+$ results agree with the ones from our
earlier N$^3$LO analysis in Ref.~\cite{Epelbaum:2014efa}, see Table 7 of that
paper. The largest deviation at the level of $\sim 1.5\sigma$ occurs
for the pp $^3$P$_0$ phase shift at $E_{\rm lab} = 100$~MeV.

We conclude this section with emphasizing that our error
analysis does not involve propagation of the uncertainty in the
value(s) of the $\pi N$ coupling constant which, in principle, could significantly 
affect the results for phase shifts and mixing angles at low energy. Also, a more
complete treatment of isospin-breaking effects as compared to the one
employed in our analysis would be required in order to draw final 
conclusions regarding the observed discrepancies. Work along these
lines is in progress.

\section{Alternative choice of the contact interactions at order $Q^4$}
\label{sec:Redundant}

As described in section \ref{sec:Cont}, our choice of the basis of the
independent order-$Q^4$ contact interactions in
Eqs.~(\ref{ConventionOffShell1}) and (\ref{ConventionOffShell2})
represents just one possible convention. In this section we explore an
alternative choice using the parametrization 
\beqa
\label{ConventionOffShell1Tilde}
D_{1S0}^1\,  p^2 p'^2 + D_{1S0}^2 \, (p^4 + p'^4 ) &=:& \tilde D_{1S0}
\, (p'^2 + p^2)^2 +
\tilde D_{1S0}^{\rm off}\,  (p'^2 - p^2)^2\,,\nn
D_{3S1}^1\,  p^2 p'^2 + D_{3S1}^2 \, (p^4 + p'^4 ) &=:& \tilde D_{3S1} \,
(p'^2 + p^2)^2 +
\tilde D_{3S1}^{\rm off}\,  (p'^2 - p^2)^2\,, \nn
D_{\epsilon 1}^1\,  p^2 p'^2 + D_{\epsilon 1}^2 \, p^4 &=:&
\tilde D_{\epsilon 1} \, p^2 (p'^2 + p^2)+
\tilde D_{\epsilon 1}^{\rm off}\,  p^2 (p'^2 - p^2)\,,
\eeqa
and setting 
\beq
\label{ConventionOffShell2Tilde}
\tilde D_{1S0}^{\rm off} = \tilde D_{3S1}^{\rm off} = \tilde D_{\epsilon 1}^{\rm off} = 0\,.
\eeq
%We restrict ourselves to the case of N$^4$LO$^+$
%potential.  
To have a meaningful comparison between the two choices of the 
contact interactions, we have 
redone the fits based on the alternative set of contact terms
employing the same fitting protocol as described in section
\ref{sec3}. We then find very similar results for the description of
the scattering  data.  For example, for the intermediate cutoff $\Lambda
= 450$~MeV, the $\chi^2/{\rm datum}$ for both the np and pp
cases agrees with the corresponding ones listed in Table
\ref{tab_chi2_AllCutoffs} in all considered energy bins to all given
figures. 
The similarity also extends to observables. For example, with the new
set of contact interactions, we obtain for the various
scattering lengths the values of $a_{1S0}^{\rm np} = -23.739$~fm,  
$a_{3S1} = 5.420$~fm and $a_{1S0}^{\rm pp} = -7.815$~fm, while the
corresponding effective ranges are  $r_{1S0}^{\rm np} = 2.697$~fm,  
$r_{3S1} = 1.754$~fm and $r_{1S0}^{\rm pp} = 2.770$~fm. Similarly, we
obtain $A_S=0.8847$~fm$^{-1/2}$ and $\eta = 0.0255$ for the asymptotic
S state normalization and the asymptotic D/S state ratio,
respectively. These values coincide with the ones 
given in sections \ref{ERE} and  \ref{Deut} for the same cutoff
value. Thus, the dependence of our results on the choice of the
order-$Q^4$ contact interactions is negligible at the accuracy level of 
our calculations.  We have verified that this conclusion also
holds for the phase shifts. This provides explicit numerical evidence
of the redundancy of the corresponding contact 
interactions.  

On the other hand, the off-shell behavior of the potentials does show
a dependence on the choice of the contact interactions, which may also 
significantly affect their perturbativeness. We found that the choice specified in
Eqs.~(\ref{ConventionOffShell1Tilde}) and (\ref{ConventionOffShell2Tilde}) 
leads to somewhat larger Weinberg eigenvalues and, therefore, to more nonperturbative
potentials for the hard cutoff choices as compared with  the convention employed in
our analysis and given in Eqs.~(\ref{ConventionOffShell1}) and
(\ref{ConventionOffShell2}). For example, for the cutoff of $\Lambda =
500$~MeV, switching the basis of contact interactions to that given  
in Eqs.~(\ref{ConventionOffShell1Tilde}),
(\ref{ConventionOffShell2Tilde}) increases the maximum values of the
largest in magnitude
repulsive eigenvalue in the $^1$S$_0$ ($^3$S$_1$--$^3$D$_1$) channel
as $0.72 \to 0.98$ ($1.81 \to 2.37$) at N$^3$LO,  
$0.83 \to 0.97$ ($0.89 \to 1.11$) at N$^4$LO and 
$0.80 \to 0.92$ ($0.91 \to 0.98$) at N$^4$LO$^+$ in the considered energy range of
$E_{\rm lab} = 0-300$~MeV. The effects become even more pronounced for
the largest cutoff value of $\Lambda = 550$~MeV.

\section{Comparison with other NN potentials}
\label{sec:ComparisonPot}

It is instructive to compare our results for the description of NN
scattering data with that based on the modern high-precision
phenomenological potentials. To this aim we have calculated 
 the values of the $\chi^2/{\rm
datum}$ for the reproduction of pp and np data from the 2013 Granada
database \cite{Perez:2013jpa} using the CD
Bonn \cite{Machleidt:2000ge} and the Nijm I, II and Reid93 potentials developed by the
Nijmegen group \cite{Stoks:1994wp}. 
In Table \ref{tab_chi2_Phenom}, 
\begin{table}[t]
\caption{$\chi^2/{\rm datum}$ for the description of the
  neutron-proton and proton-proton scattering data at N$^4$LO$^+$ for  $\Lambda = 450$~MeV in comparison with the
  high-precision phenomenological potentials. For each potential, the
  number of adjustable parameters is indicated in the subscript. For
  the SMS N$^4$LO$^+$ potential, $27 + 1$ refer to $27$ contact interactions
  contributing to neutron-proton and proton-proton channels as given
  in Table \ref{tab_LEC} and the
  cutoff value. 
\label{tab_chi2_Phenom}}
\smallskip
\begin{tabular*}{\textwidth}{@{\extracolsep{\fill}}cccccc}
\hline 
\hline 
\noalign{\smallskip}
 $E_{\rm lab}$ bin &  CD Bonn$_{(43)}$ \cite{Machleidt:2000ge}   &  Nijm I$^\star_{(41)}$ \cite{Stoks:1994wp}   &  Nijm II$^\star_{(47)}$
  \cite{Stoks:1994wp} &  Reid93$^\star_{(50)}$ \cite{Stoks:1994wp}  &  SMS
                                                                N$^4$LO$^+_{(27+1)}$,
                                                                this work
\smallskip
 \\
\hline 
\hline 
%\smallskip
\multicolumn{5}{l}{neutron-proton scattering data} \\ 
0--100 & 1.08 & 1.07 & 1.08 & 1.09 & 1.08 \\ 
0--200 & 1.08 & 1.07 & 1.07 & 1.09 & 1.07 \\
0--300 & 1.09 & 1.08 & 1.08 & 1.10 & 1.06 \\ [4pt]
%\smallskip
\hline 
% \smallskip
\multicolumn{5}{l}{proton-proton scattering data} \\ 
0--100 & 0.89 & 0.87 & 0.88 & 0.85 & 0.86 \\ 
0--200 & 0.98 & 0.98 & 1.00 & 0.99 & 0.95 \\ 
0--300 & 1.01 & 1.03 & 1.05 & 1.04 & 1.00  \\ [4pt]
\hline 
\hline 
\end{tabular*}
\begin{tabular*}{\textwidth}{@{\extracolsep{\fill}}l}
$^\star$Since the potentials developed by the
  Nijmegen group are only available for $j \le 9$, we have
  supplemented them by our LO  \\[-3pt]
potential for $j >
  9$.
 \end{tabular*}
\end{table}
the corresponding values of the $\chi^2/{\rm
datum}$ are compared with the ones based on our N$^4$LO$^+$ potential
for the cutoff $\Lambda = 450$~MeV  in three
energy bins of $E_{\rm lab} = 0-100$~MeV,  $0-200$~MeV and $0-300$~MeV.  
Notice that we found it to be insufficient to only include partial waves
with $j \le 9$ and have, therefore, extended the  Nijm I, II and Reid93
potentials by using for $j > 9$ partial waves the results of the
OPEP taken from our own LO potentials.  

It is difficult to precisely
benchmark our results for $\chi^2/{\rm datum}$ against the values quoted in
the original publications mainly due to the different selection of
experimental data, but also due to possible differences in the  treatment of
the long-range electromagnetic interactions. The potentials of the Nijmegen group were fitted to
their 1993 database with $1787$ pp and $2514$ np data below $E_{\rm
  lab} = 350$~MeV \cite{Stoks:1994wp}.  The CD Bonn potential was
fitted to a larger database which included additional data available in
the year 2000, featuring $2932$ pp and $3058$ data \cite{Machleidt:2000ge}. The
self-consistent 2013 Granada database includes $2996$ pp and $3717$ np
mutually consistent scattering data (including normalizations).\footnote{The usage of an improved data
  selection criterium has allowed the authors of Ref.~\cite{Perez:2013jpa} to keep
  some of the data rejected by the Nijmegen group.} 
While our $\chi^2/{\rm datum}$ values
for the Nijmegen potentials are somewhat higher than the ones quoted
in their original publication~\cite{Stoks:1994wp}, namely $\chi^2/{\rm datum}
=1.05\, (1.00),\, 1.04 \, (1.00), \, 1.04\, (1.00)$ for np (pp) scattering data below $350$~MeV
for the Nijm I, Nijm II and Reid93 potentials, respectively, we find
that these interactions are still doing a remarkably good job in reproducing the 2013
Granada database.\footnote{Increasing the maximum energy of the
  included scattering data from $E_{\rm lab} = 300$~MeV  to $350$~MeV
  adds $\sim 0.02-0.03$ to the corresponding $\chi^2/{\rm datum}$
  values for all considered
  phenomenological 
  potentials.}  Our slightly increased $\chi^2$ values 
presumably just reflect the effect of including additional data in the Granada
database. For the CD Bonn potential, our result for the reproduction
of pp data agrees with the corresponding value of $\chi^2/{\rm
  datum} = 1.01$ for energies below $350$~MeV given in 
Ref.~\cite{Machleidt:2000ge}. This is consistent with the fact 
that the number of pp data in the 
Granada database is only slightly larger than the one used in that
paper. 
On the other hand, we do observe a higher value  
for the description of np data as compared with $\chi^2/{\rm
  datum} = 1.02$ quoted in Ref.~\cite{Machleidt:2000ge}, which probably
reflects  a considerably larger number of np data included in the Granada database. 

As shown in Table \ref{tab_chi2_Phenom}, the reproduction of the
np and pp data below $300$~MeV by our N$^4$LO$^+$ potential with
$\Lambda = 450$~MeV is more accurate than by any of the considered
high-precision potentials.\footnote{The description of the data can be
  further optimized by performing the fits up to
$E_{\rm lab}^{\rm max} = 300$~MeV rather than up to $E_{\rm lab}^{\rm
  max} = 260$~MeV. We then obtain 
at N$^4$LO$^+$ the values of $\chi^2/{\rm
  datum} = 1.08$, $1.03$, $1.04$ and $1.07$ for both np and pp data
using the cutoffs of $\Lambda = 400$, $450$, $500$ and $550$~MeV,
respectively.}  This is a remarkable result, especially if
one takes into account that the number of adjustable parameters is
$\sim 40\%$ smaller as compared with the one used in these potentials. For the
sake of completeness, we also mention that the partial wave analysis
by the Granada group obtains $\chi^2/{\rm
  datum} = 1.04$ for both the np and pp data below $E_{\rm lab} =
350$~MeV 
using $46$ parameters
\cite{Perez:2013jpa}. In Ref.~\cite{Perez:2013oba}, this group was also able to
describe np and pp data below $350$~MeV with $\chi^2/{\rm
  datum} = 1.07$ using a coarse-grained NN potential involving the OPEP and
TPEP and featuring $30$ short-range parameters, $1$ cutoff and $3$
parameters corresponding to the $\pi N$ LECs $c_1$, $c_3$ and $c_4$. 

Finally, we compare in Table \ref{tab_chi2_EM} the description of the np
and pp data using the semilocal N$^4$LO$^+$ potential of this work for $\Lambda =
450$~MeV
with that based on the nonlocal N$^3$LO potentials of
Ref.~\cite{Entem:2003ft} and the recent N$^4$LO$^+$ potentials of Ref.~\cite{Entem:2017gor}.  
\begin{table}[t]
\caption{$\chi^2/{\rm datum}$ for the description of the
  np and pp scattering data at N$^4$LO$^+$ for
  $\Lambda = 450$~MeV in comparison with the N$^3$LO and N$^4$LO$^+$
  potentials of Refs.~\cite{Entem:2003ft} and \cite{Entem:2017gor},
  respectively.  All cutoff values are given in units of MeV.  
\label{tab_chi2_EM}}
\smallskip
\begin{tabular*}{\textwidth}{@{\extracolsep{\fill}}ccccccc}
\hline 
\hline 
\noalign{\smallskip}
& \multicolumn{2}{c}{ --- EM N$^3$LO \cite{Entem:2003ft}  --- }
  & \multicolumn{3}{c}{ --- EMN N$^4$LO$^+$ \cite{Entem:2017gor}  --- } & --- SMS
                                                                N$^4$LO$^+$,
                                                                this
                                                             work --- \\
 $E_{\rm lab}$ bin   & $\Lambda = 500$ & $\Lambda = 600$ & $\Lambda = 450$ & $\Lambda =
                                                          500$ &
                                                                 $\Lambda
                                                                 =
                                                                 550$
                                      & $\Lambda = 450$ 
\smallskip
 \\
\hline 
\hline 
%\smallskip
\multicolumn{5}{l}{neutron-proton scattering data} \\ 
0--100 & 1.17 & 1.35 & 1.28 & 1.11 & 1.11 & 1.08\\ 
0--200 & 1.17 & 1.33 & 1.33 & 1.18 & 1.23 & 1.07\\
0--300 & 1.23 & 1.37 & 2.48 & 1.26 & 1.35 & 1.06\\ [4pt]
%\smallskip
\hline 
% \smallskip
\multicolumn{5}{l}{proton-proton scattering data} \\ 
0--100 & 1.01 & 1.35 & 0.90 & 1.01 & 1.17 & 0.86\\ 
0--200 & 1.32 & 1.60 & 1.05 & 1.16 & 1.43 & 0.95\\ 
0--300 & 1.39 & 2.07 & 1.46 & 1.21 & 1.41 & 1.00 \\ [4pt]
\hline 
\hline 
\end{tabular*}
\end{table}
Here, our results for the $\chi^2/{\rm datum}$ values
show larger deviations from the ones quoted in the original papers. 
The increased sensitivity to the
selection of data as compared with the results based on the
high-precision potentials is probably a consequence of their less precise
nature.
Specifically, our $\chi^2$
values for pp data  in the lowest energy bin turn out to be lower than
the ones given in Refs.~\cite{Entem:2003ft,Entem:2017gor}, but the
value of $\chi^2/{\rm datum} = 1.21$ for the reproduction of 
pp data in the range of $E_{\rm lab} = 0 - 300$~MeV using the 
N$^4$LO$^+$ potential of Ref.~\cite{Entem:2017gor} with $\Lambda = 500$~MeV
agrees with the one given in that paper for the energy range of
$E_{\rm lab} = 0 - 290$~MeV. On the other hand, for np data, we find
significantly larger values in all energy bins. One should further
keep in mind that the values of the $\chi^2/{\rm datum}$ for the nonlocal potentials of
Refs.~\cite{Entem:2003ft,Entem:2017gor} given in Table
\ref{tab_chi2_EM} do not provide a very accurate measure for the
achievable accuracy of these interactions as they were fitted to
a different data set. However, our results for the phenomenological
high-precision potentials discussed above indicate that the impact of
using different databases should be fairly small. 

\begin{figure}[tb]
\vskip 1 true cm
\begin{center}
\includegraphics[width=1.0\textwidth,keepaspectratio,angle=0,clip]{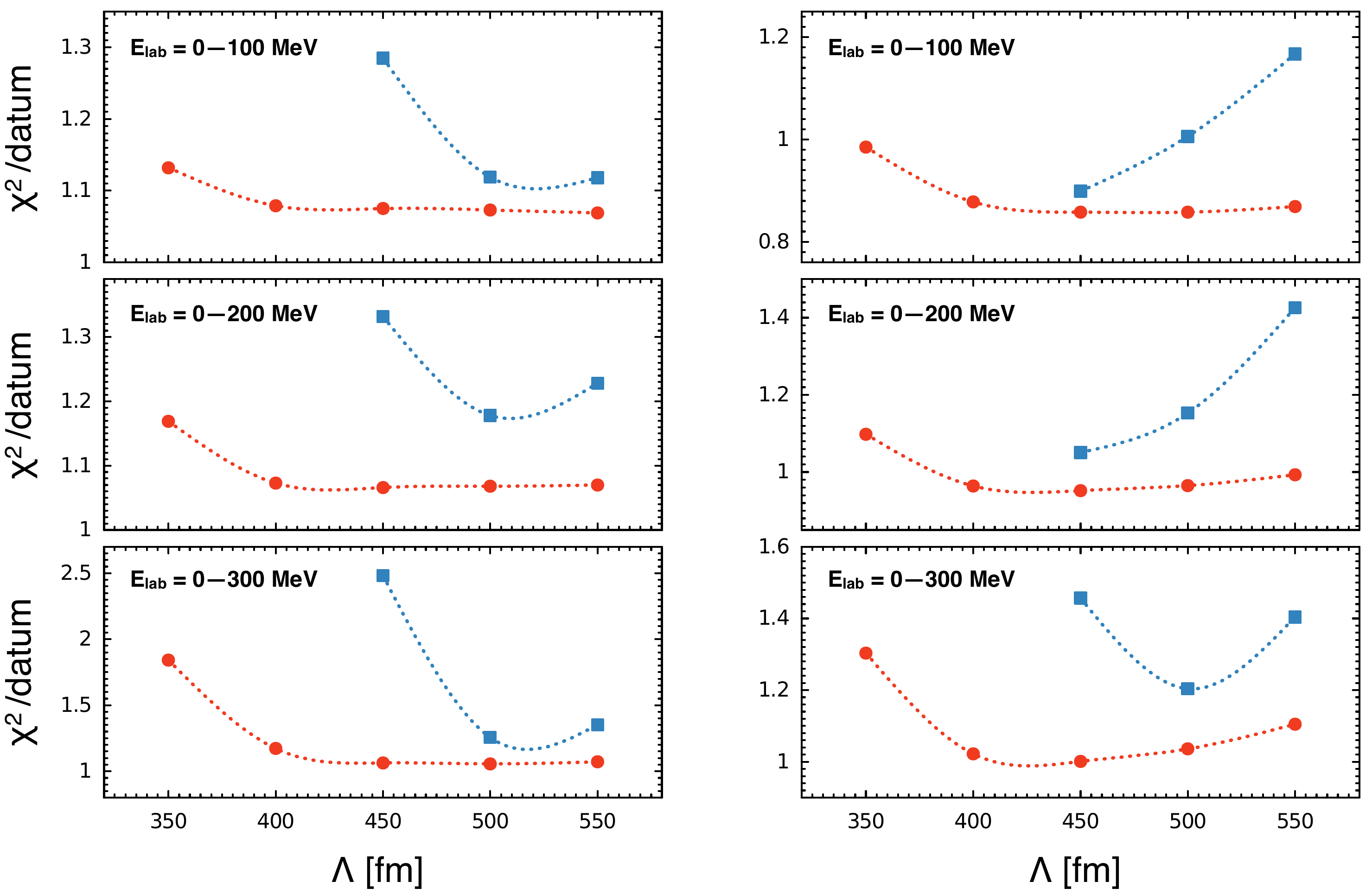}
\end{center}
    \caption{(Color online) $\chi^2/{\rm datum}$ for the description of the
  neutron-proton (left panel) and proton-proton (right panel)
  scattering data in the energy bins of $E_{\rm lab} = 0-100$~MeV
  (upper row), $E_{\rm lab} = 0-200$~MeV (middle row) and $E_{\rm lab}
  = 0-300$~MeV (bottom row) for different values of the momentum-space
  cutoff $\Lambda$. Red solid circles give the results of the SMS
  N$^4$LO$^+$ potentials of this work while solid blue squares show the
  results of the EMN N$^4$LO$^+$ potentials of
  Ref.~\cite{Entem:2017gor}. The lines are drawn to guide the eyes. 
\label{fig:Chi2}
}
\end{figure}

When looking at the results for the cutoff $\Lambda = 500$~MeV in 
Table \ref{tab_chi2_EM}, one observes that the description of the data with the old N$^3$LO 
potential of Ref.~\cite{Entem:2003ft} is almost comparable in quality
with that based on the new N$^4$LO$^+$ potential of
Ref.~\cite{Entem:2017gor}, which, however, involves additional contact
interactions in F-waves. This seemingly contradictory result can
presumably be traced back to the additional optimization steps
performed in Ref.~\cite{Entem:2003ft}, which involve fine tuning of the $\pi
N$ LECs $c_i$ and the usage of a partial wave dependent functional form
of the regulator, which effectively increase the number of adjustable parameters
in that model.\footnote{Ref.~\cite{Entem:2017gor} does not provide the
  values of the exponent $n$ when regularizing the contact
  interactions using Eq.~(\ref{RegMom}). The rationale behind their
  actual choice is not clear to us.} 
To allow for a more detailed comparison between the
nonlocal and semilocal potentials at N$^4$LO$^+$, we show
in Fig.~\ref{fig:Chi2} the $\chi^2/{\rm datum}$ values for the
reproduction of np and pp data as a function of the employed cutoff
$\Lambda$. 
Notice that the N$^4$LO$^+$ potentials of Ref.~\cite{Entem:2017gor} are only
available for the cutoff values of $\Lambda = 450$, $500$ and
$550$~MeV. While our semilocal interactions yield, in spite of having less short-range
parameters, a significantly
better description of the 2013 Granada database than the
ones of Ref.~\cite{Entem:2017gor}  in all considered
cases, the differences become especially pronounced for softer
cutoff choices. Indeed, the $\chi^2/{\rm datum}$ for the nonlocal 
N$^4$LO$^+$ potentials increases rapidly for the cutoff $\Lambda =
450$~MeV as compared to $\Lambda =
500$~MeV (except for pp data at low and intermediate energy), while
our semilocal potentials show a very small amount of finite regulator 
artefacts even for $\Lambda = 400$~MeV. We attribute the superior
performance of the semilocal potentials primarily to the
employed local regularization of the pion-exchange contributions,
which maintains the long-range part of the interaction and reduces the
amount of cutoff artefacts.

\section{Summary and conclusions}
\label{summary}

We now summarize the main results of our paper. 
\begin{itemize}
\item
We have shown that three out of fifteen N$^3$LO NN contact
interactions employed e.g.~in
Refs.~\cite{Epelbaum:2004fk,Entem:2003ft,Epelbaum:2014efa,Epelbaum:2014sza,Entem:2017gor}, 
namely the ones which contribute to the $^1$S$_0$ and $^3$S$_1$--$^3$D$_1$ channels,
are redundant as they can be eliminated by performing the short-range unitary
transformations  specified in Eq.~(\ref{ShortRangeUT}). Such unitary
transformations only affect the values of the LECs accompanying
short-range operators that need to be determined from the data anyway,
but do not change the expressions for the nuclear forces and current
operators. This unitary ambiguity 
can, in particular, be exploited to reshuffle parts of the
short-range many-body forces 
into the off-shell behavior of the NN potential at short distances, see
Ref.~\cite{Gl90} for a general discussion. 
%We also provide numerical
%evidence for this redundancy by demonstrating that the $\chi^2$ for the
%description of the NN scattering data remains essentially unchanged
%upon a variation of one of the two N$^3$LO contact interactions in the
%$^1$S$_0$ channel.
Furthermore, using the N$^4$LO potentials of Ref.~\cite{Epelbaum:2014sza}, we 
have demonstrated that  removing the
redundant interactions according to  Eq.~(\ref{ConventionOffShell2})
leads to a strong suppression of the off-diagonal matrix elements in
the $^1$S$_0$ and  $^3$S$_1$ channels at momenta of the order of $p
\sim 600~$MeV and results in much softer potentials as compared to the
ones of Ref.~\cite{Epelbaum:2014sza}. The increased perturbativeness of
the resulting potentials is verified and further quantified by
performing a Weinberg eigenvalue analysis. 
\item
We have introduced a simple momentum-space regularization approach for the long-range part
of the interaction by an appropriate modification of the pion 
propagators in the ultraviolet region. Contrary to the frequently used nonlocal regulator of
Eq.~(\ref{RegMom}), the employed regularization does not affect the
left-hand singularity structure of the potential associated with the
pion-exchange contributions at any order in the $1/\Lambda$-expansion
and thus manifestly maintains the long-range part of the interaction
(provided $\Lambda \sim \Lambda_{\rm b}$). Moreover, in contrast to 
the coordinate-space regulator used in Refs.~\cite{Epelbaum:2014efa,Epelbaum:2014sza}, our new
approach can be straightforwardly applied to regularize 
three-nucleon forces using the machinery
developed in Refs.~\cite{Golak:2009ri,Hebeler:2015wxa}.  
\item
Using the new momentum-space regularization scheme for the long-range contributions
and employing a nonlocal Gaussian regulator for the minimal set of
independent contact interactions chosen according to 
Eq.~(\ref{ConventionOffShell2}), we have developed a new family of semilocal chiral potentials
up to N$^4$LO for the cutoff values of $\Lambda = 350$, $400$, $450$,
$500$ and $550~$MeV. To this aim, we have fitted the LECs of the contact
interactions to the pp and np scattering data
of the 2013 Granada database \cite{Perez:2013jpa}, the deuteron
binding energy and world average value of the np coherent scattering
length. The determined LECs are found to be of a natural
size. The stability of the fits and convergence towards a minimum of
the $\chi^2$ are greatly improved by the removal of  the redundant
interactions. 
Using the
values of the $\pi$N LECs  from
Ref.~\cite{Hoferichter:2015tha}, obtained by matching chiral perturbation
theory to the solution of the Roy-Steiner equations, we find a continuous improvement in
the description of the data from LO to N$^4$LO. In particular, we
confirm the earlier findings of
Refs.~\cite{Epelbaum:2014efa,Epelbaum:2014sza} concerning the evidence
of the TPEP by observing a strong reduction in the $\chi^2$ when going
from NLO to N$^2$LO and from N$^3$LO to N$^4$LO. Notice that the TPEP
is predicted in a parameter-free way by the chiral symmetry of QCD and
its breaking pattern in combination with the empirical information
on the $\pi$N system.  
\item 
At N$^4$LO, the np data and the low-energy pp data
are very well described as reflected by the values of $\chi^2 / {\rm 
datum}$ of order $\sim 1$. However, the $\chi^2 / {\rm 
datum}$ increases considerably for pp data above $E_{\rm lab} \sim
150$~MeV. This can be traced back to the high precision of 
some of the experimental pp data such as especially the CO(67) data
set \cite{Cox:1968jxz}, which exceeds the accuracy of
our calculations 
at N$^4$LO. To describe these data with $\chi^2/{\rm datum}
\sim 1$ one needs to accurately reproduce the F-wave phase shifts, which  
are still predicted in a parameter-free way at N$^4$LO. We have shown that the
inclusion of the leading (i.e.~N$^5$LO) contact interactions in
the $^1$F$_3$,  $^3$F$_2$,  $^3$F$_3$ and $^3$F$_4$ channels, which
are also taken into account in the potentials of
Ref.~\cite{Entem:2017gor},    
strongly improves the description of pp data above $E_{\rm lab} \sim
150$~MeV. It is, however, important to emphasize that the
corresponding LECs come out of a natural
size, thus showing no signs of enhancement beyond 
naive dimensional analysis. Consequently, there are no indications
that these terms need to be promoted to N$^4$LO. 
Except for the lowest considered cutoff value
of $\Lambda = 350~$MeV, where significant
regulator artefacts are observed, the resulting N$^4$LO$^+$ potentials lead to
excellent description 
of both the np and pp scattering data up to $E_{\rm lab} = 300~$MeV
with $\chi^2 / {\rm 
datum} \sim 1$, which qualifies them to be regarded as partial wave analyses.   
We expect the N$^4$LO$^+$ potentials to be particularly useful for
estimating and/or verifying the truncation uncertainty at
N$^4$LO as they allow one to calculate the contributions of some of the
sixth-order terms in the potential to an observable of interest.  
\item
We have performed a comprehensive error analysis of our results. In
addition to the truncation uncertainties estimated using the algorithm
formulated in Refs.~\cite{Epelbaum:2014efa,Epelbaum:2014sza}, we have  
calculated the covariance matrix for all cutoff values and chiral
orders. This allows us to quantify the statistical
uncertainty of the LECs accompanying the contact interactions and to
propagate statistical errors when calculating observables. To estimate
the uncertainty associated with the $\pi N$ LECs, we have redone the
fits using the values of LECs $c_i$, $\bar d_i$ and $\bar e_i$ employed in 
Refs.~\cite{Epelbaum:2014efa,Epelbaum:2014sza}
and taken from Refs.~\cite{Fettes:1998ud,Buettiker:1999ap,Krebs:2012yv}. 
We found that using these two different sets of $\pi N$ LECs has, generally, 
a moderate impact on the calculated two-nucleon observables. 
%except for  very low energies. 
%the
%deuteron S-wave asymptotic normalization
%$A_S$, the D/S state ratio $\eta$ and some of the effective range parameters. 
Finally, by varying the maximum fitting energy in the range of
$E_{\rm lab}^{\rm max} = 220 \ldots 300$~MeV, we have verified that our
N$^4$LO$^+$ results are largely insensitive to the choice of $E_{\rm
  lab}^{\rm max}$. We found that the theoretical uncertainty is, in
most cases, dominated by the truncation error as conjectured in
Ref.~\cite{Epelbaum:2014efa}. However, at N$^4$LO and at very low energies, other sources of
uncertainty such as especially the one from the $\pi N$ LECs do become
comparable to the truncation error or even dominant. 
Our determination of the S-wave scattering lengths, effective range
parameters and the deuteron properties such as the S state
normalization $A_S$ and D/S state ratio $\eta$ provides new reference
values for these observables.   
\item
We have performed a Weinberg eigenvalue analysis of the new chiral
potentials to assess their perturbativeness in quantitative
terms. For $\Lambda = 350 \ldots 500~$MeV, the magnitude of the largest repulsive
Weinberg eigenvalues at positive energies does not exceed $1$ except for the $^1$S$_0$
($^3$S$_1$--$^3$D$_1$) channel at NLO (N$^3$LO) for $\Lambda =
500~$MeV, indicating that the interactions are
perturbative (except for the deuteron channel at low energies). For
the highest considered cutoff value of $\Lambda = 550~$MeV, the
potentials become nonperturbative, which is especially true for
the N$^3$LO version. 
\end{itemize}

To summarize,  we have developed a new family of semilocal
momentum-space-regularized chiral NN potentials up to N$^4$LO$^+$. 
The new regularization approach is intended to simplify and accelerate
applications beyond the two-nucleon system. Thanks to the removal of
the redundant contact interactions at order $Q^4$, the new potentials
are much softer than the ones of
Ref.~\cite{Epelbaum:2014efa,Epelbaum:2014sza}, which makes them
valuable starting points for many-body applications using {\it ab initio} methods. At
the same time, they provide an outstanding description of
the NN data. The precision of our N$^4$LO$^+$
interactions is considerably higher than that of other 
available chiral EFT interactions including the ones of Ref.~\cite{Entem:2017gor} at the same
chiral order. We attribute this feature primarily to the improved
regularization approach which, per construction,  maintains the long-range part of the
interaction. In fact, the description of the 2013 Granada database
below $E_{\rm lab} = 300$~MeV
at N$^4$LO$^+$ for $\Lambda = 400 \ldots 550$~MeV 
is comparable to that based on the  
modern high-precision phenomenological potentials including
CD-Bonn, Nijm I, II and Reid 93, while our N$^4$LO$^+$ potential with
$\Lambda = 450$~MeV is, actually, more precise than any NN interaction 
we are aware of. At the same time, the number of
adjustable parameters is reduced by $\sim 40\%$ as compared with the
high-precision phenomenological potentials, which provides yet
another clear evidence of the (parameter-free) chiral $2 \pi$
exchange. Our results leave little room for a possible improvement of 
the description of NN scattering data in a complete N$^5$LO analysis
as proposed in Ref.~\cite{Entem:2015xwa}, especially given a
large number of new contact interactions at this chiral order. 

In the future, we plan to apply the new chiral NN interactions,
supplemented with the consistently regularized three-body forces and current
operators, to study nucleon-deuteron scattering, properties of light and medium-mass
nuclei and selected electroweak reactions. Work along these lines is
in progress by the LENPIC Collaboration. It would also be interesting
to perform a comprehensive analysis of isospin-breaking contributions
to the nuclear force and to study the role of the $\Delta (1232)$
resonance, see
Refs.~\cite{Kaiser:1998wa,Krebs:2007rh,Epelbaum:2007sq,Piarulli:2014bda,Ekstrom:2017koy}
for first steps along this line.

%%%%%%%%%%%%%%%%%%%%%%%%%%%%%%%%%%%%%%%%%%%%%%%%%%%%%%%%%%%%%%%%%%%%%%%%%%%%%%%%%
\section*{Acknowledgments}

We would like to thank David Entem for providing us with the code to
generate matrix elements of the nonlocal potentials of
Ref.~\cite{Entem:2017gor} as well as Andreas Ekstr\"om, Christian
Forssen, Dick Furnstahl, Ashot Gasparyan,
Dean Lee, Ulf-G.~Mei{\ss}ner, Witold
Nazarewicz and Enrique Ruiz Arriola for sharing their insights into the
considered topics and useful discussions. We also thank  Ulf-G.~Mei{\ss}ner
for helpful comments on the manuscript. 
This work was supported by BMBF (contract No.~05P2015 - NUSTAR R\&D)
and by DFG through funds provided to the Sino-German CRC 110
``Symmetries and the Emergence of Structure in QCD'' (Grant No.~TRR110).

%%%%%%%%%%%%%%%%%%%%%%%%%%%%%%%%%%%%%%%%%%%%%%%%%%%%%%%%%%%%%%%%%%%%%%%%%%%%%%%%%
\appendix
%%%%%%%%%%%%%%%%%%%%%%%%%%%%%%%%%%%%%%%%%%%%%%%%%%%%%%%%%%%%%%%%%%%%%%%%%%%%%%%%%

\section{Relations between the different sets of LECs for the contact interactions}
\def\theequation{\Alph{section}.\arabic{equation}}
\setcounter{equation}{0}
\label{appCont}

As described in section \ref{sec:Cont}, we employ the contact-interaction
potential $V_{\rm cont}^{(0)} + V_{\rm cont}^{(2)} + V_{\rm
  cont}^{(4)}$, where  $V_{\rm cont}^{(0)}$ and  $V_{\rm cont}^{(2)}$
are given in Eq.~(\ref{Vcon}) while the adopted form of $V_{\rm
  cont}^{(4)}$ without redundant terms is specified in
Eq.~(\ref{ContQ4Min}). After performing partial wave decomposition,
the matrix elements can be written as
\beqa
\label{ContactsLSJNoRedundant}
\langle i_S, \, p' | V_{\rm cont} | i_S, \, p \rangle  &=& \tilde
C_{i_S} + C_{i_S} (p^2 + p'^2) + D_{i_S} p^2 p'^2\,, \nn [3pt]
\langle i_P, \, p' | V_{\rm cont} | i_P, \, p \rangle  &=& C_{i_P} p
p' +  D_{i_P} p p' (p^2 + p'^2)\,, \nn [3pt]
\langle i_D, \, p' | V_{\rm cont} | i_D, \, p \rangle  &=&  D_{i_D}
p^2 p'^2 \,, \nn [3pt]
\langle ^3S_1 , \, p' | V_{\rm cont} | ^3D_1 , \, p \rangle  &=&
C_{\epsilon 1} p^2 + D_{\epsilon 1} p^2 p'^2
\,, \nn [3pt]
\langle ^3P_2 , \, p' | V_{\rm cont} | ^3F_2 , \, p \rangle  &=&
D_{\epsilon 2} p^3 p' \,, 
\eeqa
where $i_S = \big\{1S0, \; 3S1 \big\}$, $i_P = \big\{1P1, \; 3P0 , \;
3P1, \; 3P2 \big\}$, $i_D = \big\{1D2, \; 3D1 , \;
3D2, \; 3D3 \big\}$ and 
\beqa
\tilde C_{1S0} &=&4 \pi  (C_S-3 C_T)\,, \nn
\tilde C_{3S1} &=&4 \pi  (C_S+C_T) \,, \nn
C_{1S0} &=&\pi  \left(4 C_1+C_2-12 C_3-3 C_4-4 C_6-C_7\right) \,, \nn
C_{3S1} &=&\frac{\pi }{3}  \left(12 C_1+3 C_2+12 C_3+3 C_4+4
  C_6+C_7\right) \,, \nn
C_{3P0} &=&-\frac{2\pi }{3}  \left(4 C_1-C_2+4 C_3-C_4+4 C_5-12 C_6+3
  C_7\right) \,, \nn
C_{1P1} &=&-\frac{2\pi }{3}   \left(4 C_1-C_2-12 C_3+3 C_4-4
  C_6+C_7\right) \,, \nn
C_{3P1} &=&-\frac{2 \pi }{3}  \left(4 C_1-C_2+4 C_3-C_4+2 C_5+8 C_6-2
  C_7\right) \,,\nn
C_{3P2} &=&-\frac{2 \pi}{3}  \left(4 C_1-C_2+4 C_3-C_4-2 C_5\right)\,,
\nn
C_{\epsilon 1} &=&\frac{2 \sqrt{2} \pi }{3} \left(4 C_6+C_7\right) \,,
\nn
D_{1S0} &=&\frac{4\pi }{3}  \left(16 D_1'+D_2'+2 D_3'-3 \left(16 D_4'+D_5'+2 D_6'\right)+2 D_{10}'+2 D_{11}'-2
   D_{12}'\right) \,, \nn
D_{3S1} &=&\frac{4\pi}{9} \left(48 D_1'+3 \left(D_2'+2 D_3'+16 D_4'+D_5'+2 D_6'\right)-2 D_{10}'-2 D_{11}'+2
   D_{12}'\right) \,, \nn
D_{3P0} &=&-\frac{\pi}{3}   \left(16 D_1'-D_2'+16 D_4'-D_5'+8 D_7'+2
  D_8'+16 D_9'-12 D_{10}'-4 D_{11}'\right)\,, \nn
D_{1P1} &=&-\frac{\pi}{3}   \left(16 D_1'-D_2'-48 D_4'+3 D_5'+8
  D_9'-4 D_{10}'-4 D_{11}'\right)\,, \nn
D_{3P1} &=&-\frac{\pi}{3}   \left(16 D_1'-D_2'+16 D_4'-D_5'+4
  D_7'+D_8'-12 D_9'+8 D_{10}'+4 D_{11}'\right) \,, \nn
D_{3P2} &=&-\frac{\pi}{15} \left(80 D_1'-5 \left(D_2'-16
    D_4'+D_5'+4 D_7'+D_8'\right)-4 D_9'+4 D_{11}'\right) \,, \nn
D_{3D1} &=&\frac{2\pi}{45}  \left(48 D_1'+3 \left(D_2'-4 D_3'+16 D_4'+D_5'-4 D_6'+12 D_7'-3 D_8'\right)+20 D_{10}'+20
   D_{11}'-32 D_{12}'\right) \,, \nn
D_{1D2} &=&\frac{2\pi}{15}  \left(16 D_1'+D_2'-4 D_3'-48 D_4'-3
  D_5'+12 D_6'+8 D_{10}'+8 D_{11}'+4 D_{12}'\right) \,, \nn
D_{3D2} &=&\frac{2\pi}{15}  \left(16 D_1'+D_2'-4 D_3'+16 D_4'+D_5'-4 D_6'+4 D_7'-D_8'-12 D_{10}'-12 D_{11}'+8
   D_{12}'\right) \,, \nn
D_{3D3} &=&\frac{2\pi}{15}  \left(16 D_1'+D_2'-4 D_3'+16 D_4'+D_5'-4
  D_6'-8 D_7'+2 D_8'-4 D_{12}'\right) \,,\nn
D_{\epsilon 1} &=&-\frac{8 \sqrt{2} \pi }{9} \left(2
  \left(D_{10}'+D_{11}'\right)+D_{12}'\right) \,, \nn
D_{\epsilon 2} &=&\frac{8\pi }{5} \sqrt{\frac{2}{3}}  \left(D_9'-D_{11}'\right)\,.
\eeqa

\section{Subtraction constants in the long-range part of the two-pion exchange potential}
\def\theequation{\Alph{section}.\arabic{equation}}
\setcounter{equation}{0}
\label{app1}

In this appendix we give the various functions $C_i (\mu)$ which
enter the spectral integrals for the TPEP and ensure its regular
behavior at short-distances as discussed in section \ref{sec:reg}. To
determine $C_i (\mu)$, we make use of the relations between the
coordinate- and momentum-space potentials utilizing the conventions of
Eqs.~(\ref{2PEdecMom}),   (\ref{2PEdecCoord}), see e.g.~Ref.~\cite{Veerasamy:2011ak}
\beqa
X_{C, \, \Lambda} (r) &=& \frac{1}{2 \pi^2} \int q^2 dq \, j_0 (q
r) \, X_{C, \, \Lambda} (q)\,, \nn
X_{T, \, \Lambda} (r) &=& - \frac{1}{2 \pi^2} \int q^4 dq \, j_2 (q
r) \, X_{T, \, \Lambda} (q)\,, \nn
X_{S, \, \Lambda} (r) &=&  \frac{1}{2 \pi^2} \int q^2 dq \, j_0 (q
r) \, X_{S, \, \Lambda} (q) + \frac{1}{6 \pi^2} \int q^4 dq \, j_0 (q
r) \, X_{T, \, \Lambda} (q) \,, \nn
X_{LS, \, \Lambda} (r) &=& - \frac{1}{\pi^2 r} \int q^3 dq \, j_1 (q
r) \, X_{LS, \, \Lambda} (q)\,,
\eeqa
where $X = \{V, \, W \}$ and $j_i(x)$ are the spherical Bessel functions
of the first kind. For the  functions $C_i^1 (\mu)$ entering the
single-subtracted spectral integrals, we obtain the expressions 
\beqa
C_{T}^{1}(\mu) &=&-\frac{2 \Lambda \left( 15 \Lambda ^6-3 \Lambda ^4 \mu ^2+ \Lambda ^2 \mu ^4-
  \mu ^6 \right) + \sqrt{2 \pi } \mu ^7 e^{\frac{\mu ^2}{2 \Lambda ^2}} \text{erfc}\left(\frac{\mu }{\sqrt{2} \Lambda
   }\right)}{30 \Lambda ^7}\,, \nn
C_{LS}^{1}(\mu) &=& -\frac{2 \Lambda \left( 3 \Lambda ^4- \Lambda ^2
    \mu ^2 + 
  \mu ^4 \right) - \sqrt{2 \pi } \mu ^5 e^{\frac{\mu ^2}{2 \Lambda ^2}} \text{erfc}\left(\frac{\mu }{\sqrt{2} \Lambda
   }\right)}{6 \Lambda ^5}\,.
\eeqa
Next, the functions $C_i^{2} (\mu )$ appearing in double-subtracted spectral integrals
have the form 
\beqa
C_{C,1}^{2}(\mu) &=&\frac{2 \Lambda \mu^2 \left(2 \Lambda ^4-4 \Lambda ^2 \mu ^2-
   \mu ^4 \right) + \sqrt{2 \pi } \mu ^5 e^{\frac{\mu ^2}{2 \Lambda ^2}} \left(5 \Lambda ^2+\mu ^2\right)
   \text{erfc}\left(\frac{\mu }{\sqrt{2} \Lambda }\right)}{4 \Lambda ^5}
\,, \nn
C_{C,2}^{2}(\mu) &=&- \frac{2 \Lambda \left( 6 \Lambda ^6-2 \Lambda ^2 \mu ^4-  
   \mu ^6 \right) +\sqrt{2 \pi } \mu ^5 e^{\frac{\mu ^2}{2 \Lambda ^2}} \left(3 \Lambda ^2+\mu ^2\right)
   \text{erfc}\left(\frac{\mu }{\sqrt{2} \Lambda }\right)}{12 \Lambda ^7}\,, \nn
C_{S,1}^{2}(\mu) &=&\frac{2 \Lambda \mu^2 \left( 2 \Lambda ^4-4 \Lambda ^2 \mu ^2-
   \mu ^4 \right) + \sqrt{2 \pi } \mu ^5 e^{\frac{\mu ^2}{2 \Lambda ^2}} \left(5 \Lambda ^2+\mu ^2\right)
   \text{erfc}\left(\frac{\mu }{\sqrt{2} \Lambda }\right)}{6 \Lambda ^5}\,, \nn
C_{S,2}^{2}(\mu) &=& -\frac{2 \Lambda \left( 15 \Lambda ^6- \Lambda ^4 \mu ^2-3 \Lambda ^2
   \mu ^4-2  \mu ^6 \right) +\sqrt{2 \pi } \mu ^5 e^{\frac{\mu ^2}{2 \Lambda ^2}} \left(5 \Lambda ^2+2 \mu ^2\right)
   \text{erfc}\left(\frac{\mu }{\sqrt{2} \Lambda }\right)}{30 \Lambda
   ^7}\,, \nn
C_{T,1}^{2}(\mu) &=&-\frac{2 \Lambda  \mu ^2 \left(-30 \Lambda ^8+12
   \Lambda ^6 \mu ^2-6 \Lambda ^4 \mu ^4+8 \Lambda ^2 \mu ^6+\mu
   ^8\right) -\sqrt{2 \pi } \mu ^9 e^{\frac{\mu ^2}{2 \Lambda ^2}} \left(9 \Lambda ^2+\mu ^2\right)
   \text{erfc}\left(\frac{\mu }{\sqrt{2} \Lambda }\right)}{60 \Lambda ^9}\,, \nn
C_{T,2}^{2}(\mu) &=&\frac{2 \Lambda  \left(-210 \Lambda ^{10}+6 \Lambda ^6 \mu ^4-4 \Lambda ^4 \mu ^6+6 \Lambda ^2 \mu
   ^8+\mu ^{10}\right)-\sqrt{2\pi } \mu ^9 e^{\frac{\mu ^2}{2 \Lambda ^2}} \left(7 \Lambda ^2+\mu
   ^2\right) \text{erfc}\left(\frac{\mu }{\sqrt{2} \Lambda
   }\right)}{420  \Lambda ^{11}}\,, \nn
C_{LS, \, 1}^{2}(\mu) &=&  \frac{2 \Lambda  \mu ^2 \left(6 \Lambda ^6-4 \Lambda ^4 \mu ^2+6 \Lambda ^2 \mu ^4+\mu ^6\right)-\sqrt{2
   \pi } \mu ^7 e^{\frac{\mu ^2}{2 \Lambda ^2}} \left(7 \Lambda ^2+\mu ^2\right)
   \text{erfc}\left(\frac{\mu }{\sqrt{2} \Lambda }\right)}{12 \Lambda ^7}        \,, \nn
C_{LS, \, 2}^{2}(\mu) &=&  - \frac{2 \Lambda  \left(30 \Lambda ^8-2 \Lambda ^4 \mu
   ^4+4 \Lambda ^2 \mu ^6+\mu ^8\right) - \sqrt{2 \pi } \mu ^7 e^{\frac{\mu ^2}{2 \Lambda ^2}} \left(5 \Lambda ^2+\mu ^2\right)
   \text{erfc}\left(\frac{\mu }{\sqrt{2} \Lambda }\right)}{60 \Lambda ^9}       \,. 
\eeqa
The functions $C_i^{3} (\mu)$ entering the triple-subtracted spectral integrals are given by
\beqa
C_{C,1}^{3}(\mu) &=&
\frac{2 \Lambda  \mu ^4 \left(-8 \Lambda ^6+24 \Lambda ^4 \mu ^2+13 \Lambda ^2 \mu ^4+\mu
   ^6\right)-\sqrt{2 \pi } \mu ^7 e^{\frac{\mu ^2}{2 \Lambda ^2}} \left(35 \Lambda ^4+14 \Lambda ^2 \mu
   ^2+\mu ^4\right) \text{erfc}\left(\frac{\mu }{\sqrt{2} \Lambda
   }\right)}{16 \Lambda ^7}\,, \nn
C_{C,2}^{3}(\mu) &=&-\frac{2 \Lambda  \mu ^2 \left(-12 \Lambda
   ^8+12 \Lambda ^4 \mu ^4+11 \Lambda ^2 \mu ^6+\mu ^8\right) -\sqrt{2 \pi } \mu ^7 e^{\frac{\mu ^2}{2 \Lambda ^2}} \left(21 \Lambda ^4+12 \Lambda ^2 \mu ^2+\mu
   ^4\right) \text{erfc}\left(\frac{\mu }{\sqrt{2} \Lambda }\right)}{24
 \Lambda ^9}\,, \nn
C_{C,3}^{3}(\mu) &=&\frac{2 \Lambda  \left(-120 \Lambda ^{10}+8 \Lambda ^4 \mu ^6+9 \Lambda ^2 \mu ^8+\mu
   ^{10}\right)-\sqrt{2 \pi } \mu ^7 e^{\frac{\mu ^2}{2 \Lambda ^2}} \left(15 \Lambda ^4+10 \Lambda ^2 \mu
   ^2+\mu ^4\right) \text{erfc}\left(\frac{\mu }{\sqrt{2} \Lambda
   }\right)}{240 \Lambda ^{11}} \,, \nn
C_{S,1}^{3}(\mu) &=&\frac{2\Lambda  \mu ^4 \left(-8 \Lambda ^6+24 \Lambda ^4 \mu ^2+13 \Lambda ^2 \mu ^4+\mu
   ^6\right)-\sqrt{2 \pi } \mu ^7 e^{\frac{\mu ^2}{2 \Lambda ^2}} \left(35 \Lambda ^4+14 \Lambda ^2 \mu
   ^2+\mu ^4\right) \text{erfc}\left(\frac{\mu }{\sqrt{2} \Lambda
   }\right)}{24 \Lambda ^7}\,, \\
C_{S,2}^{3}(\mu) &=&\frac{2 \Lambda \mu^2 \left(30 \Lambda ^8-4 \Lambda ^6 \mu
   ^2-18 \Lambda ^4 \mu ^4-21 \Lambda ^2 \mu ^6-2 \mu
   ^{8} \right) +  \sqrt{2 \pi } \mu ^7 e^{\frac{\mu ^2}{2 \Lambda ^2}} \left(35 \Lambda ^4+23 \Lambda ^2 \mu ^2+2 \mu
   ^4\right) \text{erfc}\left(\frac{\mu }{\sqrt{2} \Lambda }\right)}{60 \Lambda ^9}\,, \nn
C_{S,3}^{3}(\mu) &=&\frac{2 \Lambda  \left(-420 \Lambda ^{10}+4 \Lambda ^6 \mu ^4+16 \Lambda ^4 \mu ^6+25 \Lambda ^2
   \mu ^8+3 \mu ^{10}\right)-\sqrt{2\pi } \mu ^7 e^{\frac{\mu ^2}{2 \Lambda ^2}} \left(35 \Lambda ^4+28
   \Lambda ^2 \mu ^2+3 \mu ^4\right) \text{erfc}\left(\frac{\mu }{\sqrt{2} \Lambda }\right)}{840
   \Lambda ^{11}}\,. \nonumber
\eeqa
Finally, the constants $C_1^2$ and $C_2^2$ entering Eq.~(\ref{TPEPRelPoles})
are given by:
\beqa
C_1^2 &=& - \frac{4 \Lambda 
   \left(\Lambda ^2+M_\pi^2\right) -\sqrt{2 \pi } M_\pi e^{\frac{2 M_\pi^2}{\Lambda ^2}} \left(5 \Lambda ^2+4
   M_\pi^2\right) \text{erfc}\left(\frac{\sqrt{2} M_\pi}{\Lambda }\right)}{2 \Lambda ^5}\,, \nn
C_2^2 &=& \frac{2 \Lambda  \left(\Lambda ^2+2 M_\pi^2\right)-\sqrt{2 \pi } M_\pi e^{\frac{2
   M_\pi^2}{\Lambda ^2}} \left(3 \Lambda ^2+4 M_\pi^2\right) \text{erfc}\left(\frac{\sqrt{2}
   M_\pi}{\Lambda }\right)}{6 \Lambda ^7}\,.
\eeqa

%%%%%%%%%%%%%%%%%%%%%%%%%%%%%%%%%%%%%%%%%%%%%%%%%%%%%%%%%%%%%%%%%%%%%%%%%%%%%%%%%
\section{Calculation of scattering observables}
\def\theequation{\Alph{section}.\arabic{equation}}
\setcounter{equation}{0}
\label{app2}

In order to perform fits to experimental scattering data, we have to
calculate observables from the nuclear potential. We do this by first
solving the Lippmann-Schwinger equation 
for the partial-wave projected T-matrix for the total angular momentum $j$,
total spin $s$ and orbital angular momenta $l$, $l'$
\begin{align}
	T_{l'l}^{sj}(p',p;k^2) = V_{l'l}^{sj}(p',p) + m_N \sum_{l''}
  \int_{0}^{\infty} dq q^2 \frac{V_{l'l''}^{sj}(p',q)
  T_{l''l}^{sj}(q,p;k^2)}{k^2-q^2+i\epsilon} \,,
\end{align}
which corresponds to the relativistic Schr\"odinger equation cast
in the nonrelativistic form, see e.g.~Eq.~(20) of
Ref.~\cite{Epelbaum:2014efa}. 
Further, $V_{l'l}^{sj}(p',p)$ is
the corresponding partial-wave projected NN potential,
whose contact-interaction part is explicitly given in
section \ref{sec:Cont}, while the long-range part is described in
detail in section \ref{sec:reg}. The average nucleon mass $m_N$ 
corresponds to $m_N = 2m_pm_n/(m_n+m_p)$, $m_N=m_p$ or $m_N=m_n$ in
the case of np, pp or
nn scattering, respectively. Finally, $k\equiv |\vec{k}|$ denotes the
relative momentum of the two nucleons in the cms frame, which is related
to the kinetic energy of the particles in the laboratory frame via 
\begin{align}
	k^2= \frac{m_p^2 E_{\rm lab}\left( E_{\rm lab} + 2m_n\right)}{\left( m_n + m_p \right)^2 + 2E_{\rm lab}m_p},
\end{align}
for np scattering
and via
\begin{align}
	k^2 = \frac{1}{2}m_N E_{\rm lab}
\end{align}
for pp (nn) scattering. 
The on-shell partial-wave S-Matrix can then be obtained as
\begin{align}
	S_{l'l}^{sj}(k) = \delta_{l'l} - i\pi km_N T_{l'l}^{sj}(k,k,k^2).
\end{align}
The partial-wave S-matrix needs to be modified  to account for e.m.~interactions as explained in
section \ref{sec2}. In particular, for
pp scattering, we first calculate $S_{l'l}^{sj}$ with
respect to Coulomb functions. The resulting S-matrix elements are 
then corrected in the $^1S_0$ channel  according to
Eq.~\eqref{1S0pp-corrections} and finally multiplied with the 
appropriate e.m.~interaction partial wave
S-matrices as given in Eq.~\eqref{smatrix-em-adjustment}.  
After the adjustment for the e.m.~interactions, all partial-wave 
S-matrices with total angular momentum $j \le 20$ are summed
up in the singlet-triplet representation of the scattering amplitude
according to 
\begin{align}
  \label{pwd_amplitude}
  M_{m'_s,m_s}^{s} =& \frac{1}{iq} \sum_{j l' l}
                      C(l',s,j;m_s-m'_s,m'_s,m_s)
                      Y_{l'(m'_s-m_s)}(\hat{q}) i^{l-l'} \left(
                      S_{l'l}^{js} - \delta_{l'l} \right) \nonumber\\
                    &\times C(l,s,j;0,m_s,m_s) \sqrt{\pi (2l+1)}
                      \left[ 1-(-1)^{l+s+t} \right]. 
\end{align}
Here, $t$ denotes the total isospin, $C(j_1,j_2,j_3;m_1,m_2m_3)$ 
are Clebsch-Gordan coefficients while $Y_{lm}\left(\hat{q}\right)$ are
the spherical harmonics. The nuclear scattering amplitude is then 
combined with the amplitudes of the long-range electromagnetic
interactions given in section \ref{sec2}. We express the
full scattering amplitude $M$ in terms of the Wolfenstein-like
amplitudes $a$ - $f$, often called Saclay amplitudes, whose definition
and relation to the  Singlet-Triplet amplitude are given e.g.~in Ref.~\cite{LaFrance:1980}.
Expressions for observables in terms of the Saclay amplitudes
can be found in Table III of \cite{LaFrance:1980}, where we make use
of those expressions with the rotation angles $\alpha = 0$ and $\beta
= \pi/2$, corresponding to the non-relativistic 
%equal-mass 
limit and thereby neglecting Wigner spin rotations.

%%%%%%%%%%%%%%%%%%%%%%%%%%%%%%%%%%%%%%%%%%%%%%%%%%%%%%%%%%%%%%%%%%%%%%%%%%%%%%%%%
\section{Results for np and pp phase shifts and mixing angles}
\def\theequation{\Alph{section}.\arabic{equation}}
\setcounter{equation}{0}
\label{ResPWA}

%%%%%%%%%%%%%%%%%%%%%%%%%%%%%%%%%%%%%%%%%%%%%%%%%%%%%%%%%%%%%%%%%%%%%%%%%%%%%%%%%%
\begin{table}[tb]
	\caption{Proton-proton S- and P-wave phase shifts in degrees as
          obtained at N$^4$LO$^+$ for the cutoff
          $\Lambda = 450$~MeV as function of the laboratory energy
          $E_{\rm lab}$ (in MeV). The phase shifts are calculated with
          respect to Coulomb functions. The first uncertainty is
  statistical, the second one corresponds to the truncation error at
  N$^4$LO, the third one estimates the sensitivity to the $\pi
  N$ LECs while the last uncertainty reflects the sensitivity to the
  choice of the maximum energy in the fits as explained in the
  text. The truncation errors are estimated assuming the breakdown
  scale of $\Lambda_{\rm b} = 650$~MeV. Errors smaller than
  $0.005^\circ$ are not shown. \label{tab:ppSP}}
	\smallskip
	\begin{tabular*}{\textwidth}{@{\extracolsep{\fill}}rr@{\extracolsep{4pt}}l@{\extracolsep{\fill}}r@{\extracolsep{4pt}}l@{\extracolsep{\fill}}r@{\extracolsep{4pt}}l@{\extracolsep{\fill}}r@{\extracolsep{4pt}}l}
		\hline 
		\hline 
		\noalign{\smallskip}
		$E_{\rm lab}$ & \multicolumn{2}{l}{$^1$S$_0$} & \multicolumn{2}{l}{$^3$P$_0$} & \multicolumn{2}{l}{$^3$P$_1$} & \multicolumn{2}{l}{$^3$P$_2$}\\[2pt]
		\hline
		\hline
		\\[-9pt]
		 1 & $32.79$ & ${}_{(-0)}^{(+1)} (0) (0) (0)$ & $0.14$ &  & $-0.08$ &  & $0.01$ & \\[4pt]
		 5 & $54.88$ & ${}_{(-1)}^{(+1)} (2) (4) (0)$ & $1.61$ &  & $-0.89$ &  & $0.22$ & ${}_{(-0)}^{(+0)} (0) (1) (0)$\\[4pt]
		 10 & $55.26$ & ${}_{(-2)}^{(+2)} (2) (6) (1)$ & $3.81$ & ${}_{(-1)}^{(+1)} (1) (1) (1)$ & $-2.04$ & ${}_{(-0)}^{(+0)} (0) (1) (0)$ & $0.67$ & ${}_{(-0)}^{(+0)} (0) (2) (0)$\\[4pt]
		 25 & $48.75$ & ${}_{(-3)}^{(+4)} (2) (6) (1)$ & $8.77$ & ${}_{(-3)}^{(+3)} (2) (1) (2)$ & $-4.88$ & ${}_{(-1)}^{(+1)} (0) (2) (1)$ & $2.51$ & ${}_{(-0)}^{(+0)} (1) (5) (0)$\\[4pt]
		 50 & $39.05$ & ${}_{(-5)}^{(+5)} (4) (1) (2)$ & $11.74$ & ${}_{(-5)}^{(+5)} (2) (4) (3)$ & $-8.27$ & ${}_{(-2)}^{(+2)} (1) (1) (3)$ & $5.82$ & ${}_{(-1)}^{(+1)} (1) (3) (0)$\\[4pt]
		 100 & $25.18$ & ${}_{(-6)}^{(+6)} (23) (12) (4)$ & $9.61$ & ${}_{(-7)}^{(+7)} (14) (11) (1)$ & $-13.32$ & ${}_{(-3)}^{(+3)} (8) (4) (4)$ & $10.88$ & ${}_{(-2)}^{(+2)} (12) (10) (1)$\\[4pt]
		 150 & $15.11$ & ${}_{(-8)}^{(+8)} (62) (20) (6)$ & $4.80$ & ${}_{(-8)}^{(+8)} (45) (9) (8)$ & $-17.54$ & ${}_{(-4)}^{(+5)} (27) (5) (1)$ & $13.90$ & ${}_{(-3)}^{(+3)} (37) (13) (2)$\\[4pt]
		 200 & $7.19$ & ${}_{(-12)}^{(+12)} (1.30) (23) (8)$ & $-0.20$ & ${}_{(-10)}^{(+11)} (1.00) (3) (17)$ & $-21.18$ & ${}_{(-8)}^{(+8)} (58) (5) (4)$ & $15.63$ & ${}_{(-5)}^{(+5)} (75) (1) (4)$\\[4pt]
		 250 & $0.72$ & ${}_{(-18)}^{(+18)} (2.10) (21) (11)$ & $-4.79$ & ${}_{(-16)}^{(+16)} (1.80) (22) (29)$ & $-24.26$ & ${}_{(-14)}^{(+15)} (99) (25) (12)$ & $16.58$ & ${}_{(-7)}^{(+7)} (1.20) (22) (5)$\\[4pt]
		 300 & $-4.64$ & ${}_{(-25)}^{(+26)} (3.20) (13) (13)$ & $-8.79$ & ${}_{(-22)}^{(+23)} (2.80) (48) (42)$ & $-26.74$ & ${}_{(-22)}^{(+23)} (1.50) (57) (22)$ & $17.03$ & ${}_{(-8)}^{(+8)} (1.70) (49) (6)$\\[4pt]
		\hline 
		\hline 
	\end{tabular*}
\end{table}
%%%%%%%%%%%%%%%%%%%%%%%%%%%%%%%%%%%%%%%%%%%%%%%%%%%%%%%%%%%%%%%%%%%%%%%%%%%%%%%%%%
\begin{table}[tb]
	\caption{Proton-proton $^1$D$_2$ phase shifts and the mixing
          angle $\epsilon_2$ in degrees as
          obtained  at N$^4$LO$^+$ for the cutoff
          $\Lambda = 450$~MeV as function of the laboratory energy
          $E_{\rm lab}$ (in MeV). For the notation see Table \ref{tab:ppSP}. \label{tab:ppD}}
	\smallskip
	\begin{tabular*}{\textwidth}{@{\extracolsep{\fill}}rr@{\extracolsep{4pt}}l@{\extracolsep{\fill}}r@{\extracolsep{4pt}}l@{\extracolsep{\fill}}}
		\hline 
		\hline 
		\noalign{\smallskip}
		$E_{\rm lab}$ & \multicolumn{2}{l}{$^1$D$_2$} & \multicolumn{2}{l}{$\epsilon_2$} \\[2pt]
		\hline
		\hline
		\\[-9pt]
		 1 & $0.00$ &  & $0.00$ & \\[4pt]
		 5 & $0.04$ &  & $-0.05$ & \\[4pt]
		 10 & $0.17$ &  & $-0.20$ & \\[4pt]
		 25 & $0.70$ & ${}_{(-0)}^{(+0)} (0) (1) (0)$ & $-0.81$ & \\[4pt]
		 50 & $1.69$ & ${}_{(-0)}^{(+0)} (1) (3) (0)$ & $-1.70$ & \\[4pt]
		 100 & $3.72$ & ${}_{(-1)}^{(+1)} (7) (4) (1)$ & $-2.64$ & ${}_{(-1)}^{(+1)} (3) (1) (0)$\\[4pt]
		 150 & $5.61$ & ${}_{(-2)}^{(+2)} (24) (3) (2)$ & $-2.89$ & ${}_{(-2)}^{(+2)} (10) (3) (0)$\\[4pt]
		 200 & $7.19$ & ${}_{(-4)}^{(+4)} (55) (16) (3)$ & $-2.81$ & ${}_{(-3)}^{(+3)} (21) (4) (1)$\\[4pt]
		 250 & $8.42$ & ${}_{(-5)}^{(+5)} (97) (30) (4)$ & $-2.58$ & ${}_{(-4)}^{(+4)} (34) (3) (1)$\\[4pt]
		 300 & $9.33$ & ${}_{(-6)}^{(+6)} (1.50) (42) (5)$ & $-2.29$ & ${}_{(-4)}^{(+4)} (47) (2) (1)$\\[4pt]
		\hline 
		\hline 
	\end{tabular*}
\end{table}
%%%%%%%%%%%%%%%%%%%%%%%%%%%%%%%%%%%%%%%%%%%%%%%%%%%%%%%%%%%%%%%%%%%%%%%%%%%%%%%%%%
\begin{table}[htb]
	\caption{Proton-proton F-wave phase shifts in degrees as
          obtained at N$^4$LO$^+$ for the cutoff
          $\Lambda = 450$~MeV as function of the laboratory energy
          $E_{\rm lab}$ (in MeV). The second uncertainty corresponds to
          the  N$^5$LO truncation error.  
For the remaining notation see Table \ref{tab:ppSP}. \label{tab:ppF}}
	\smallskip
	\begin{tabular*}{\textwidth}{@{\extracolsep{\fill}}rr@{\extracolsep{4pt}}l@{\extracolsep{\fill}}r@{\extracolsep{4pt}}l@{\extracolsep{\fill}}r@{\extracolsep{4pt}}l@{\extracolsep{\fill}}}
		\hline 
		\hline 
		\noalign{\smallskip}
		$E_{\rm lab}$ & \multicolumn{2}{l}{$^3$F$_2$} & \multicolumn{2}{l}{$^3$F$_3$} & \multicolumn{2}{l}{$^3$F$_4$}\\[2pt]
		\hline
		\hline
		\\[-9pt]
		1 & $0.00$ &  & $0.00$ &  & $0.00$ & \\[4pt]
		5 & $0.00$ &  & $-0.01$ &  & $0.00$ & \\[4pt]
		10 & $0.01$ &  & $-0.03$ &  & $0.00$ & \\[4pt]
		25 & $0.11$ &  & $-0.23$ &  & $0.02$ & \\[4pt]
		50 & $0.34$ & ${}_{(-0)}^{(+0)} (0) (1) (0)$ & $-0.68$ &  & $0.12$ & ${}_{(-0)}^{(+0)} (0) (1) (0)$\\[4pt]
		100 & $0.81$ & ${}_{(-1)}^{(+1)} (4) (2) (1)$ & $-1.47$ & ${}_{(-1)}^{(+1)} (2) (2) (0)$ & $0.51$ & ${}_{(-0)}^{(+0)} (1) (2) (1)$\\[4pt]
		150 & $1.14$ & ${}_{(-2)}^{(+2)} (15) (3) (3)$ & $-2.03$ & ${}_{(-2)}^{(+2)} (8) (2) (1)$ & $1.06$ & ${}_{(-1)}^{(+1)} (4) (3) (3)$\\[4pt]
		200 & $1.29$ & ${}_{(-4)}^{(+4)} (39) (2) (6)$ & $-2.41$ & ${}_{(-5)}^{(+5)} (19) (1) (2)$ & $1.70$ & ${}_{(-3)}^{(+3)} (10) (1) (7)$\\[4pt]
		250 & $1.26$ & ${}_{(-7)}^{(+7)} (74) (1) (10)$ & $-2.68$ & ${}_{(-9)}^{(+9)} (38) (6) (3)$ & $2.37$ & ${}_{(-4)}^{(+5)} (20) (4) (12)$\\[4pt]
		300 & $1.08$ & ${}_{(-11)}^{(+11)} (1.20) (6) (14)$ & $-2.87$ & ${}_{(-14)}^{(+14)} (64) (14) (5)$ & $3.02$ & ${}_{(-7)}^{(+7)} (33) (11) (18)$\\[4pt]
		\hline 
		\hline 
	\end{tabular*}
\end{table}
%%%%%%%%%%%%%%%%%%%%%%%%%%%%%%%%%%%%%%%%%%%%%%%%%%%%%%%%%%%%%%%%%%%%%%%%%%%%%%%%%%
\begin{table}[htb]
	\caption{Neutron-proton S-wave phase shifts in degrees as
          obtained at N$^4$LO$^+$ for the cutoff
          $\Lambda = 450$~MeV as function of the laboratory energy
          $E_{\rm lab}$ (in MeV). For the notation see Table
          \ref{tab:ppSP}. \label{tab:npS}}
	\smallskip
	\begin{tabular*}{\textwidth}{@{\extracolsep{\fill}}rr@{\extracolsep{4pt}}l@{\extracolsep{\fill}}r@{\extracolsep{4pt}}l@{\extracolsep{\fill}}}
		\hline 
		\hline 
		\noalign{\smallskip}
	$E_{\rm lab}$ & \multicolumn{2}{l}{$^1$S$_0$} & \multicolumn{2}{l}{$^3$S$_1$}\\[2pt]
	\hline
	\hline
	\\[-9pt]
	1 & $62.02$ & ${}_{(-1)}^{(+5)} (2) (5) (4)$ & $147.74$ & ${}_{(-1)}^{(+2)} (1) (1) (0)$\\[4pt]
	5 & $63.51$ & ${}_{(-4)}^{(+6)} (4) (13) (11)$ & $118.17$ & ${}_{(-2)}^{(+3)} (1) (2) (1)$\\[4pt]
	10 & $59.78$ & ${}_{(-6)}^{(+8)} (6) (19) (17)$ & $102.59$ & ${}_{(-2)}^{(+4)} (1) (2) (1)$\\[4pt]
	25 & $50.58$ & ${}_{(-12)}^{(+13)} (9) (28) (29)$ & $80.57$ & ${}_{(-4)}^{(+5)} (1) (1) (2)$\\[4pt]
	50 & $39.95$ & ${}_{(-18)}^{(+19)} (11) (36) (45)$ & $62.64$ & ${}_{(-6)}^{(+6)} (3) (8) (2)$\\[4pt]
	100 & $25.67$ & ${}_{(-28)}^{(+29)} (23) (43) (70)$ & $42.99$ & ${}_{(-8)}^{(+8)} (14) (15) (3)$\\[4pt]
	150 & $15.52$ & ${}_{(-35)}^{(+36)} (62) (51) (91)$ & $30.50$ & ${}_{(-11)}^{(+12)} (40) (10) (6)$\\[4pt]
	200 & $7.59$ & ${}_{(-42)}^{(+43)} (1.20) (63) (1.10)$ & $21.20$ & ${}_{(-16)}^{(+17)} (86) (7) (8)$\\[4pt]
	250 & $1.14$ & ${}_{(-49)}^{(+50)} (2.10) (78) (1.20)$ & $13.83$ & ${}_{(-23)}^{(+24)} (1.60) (35) (13)$\\[4pt]
	300 & $-4.19$ & ${}_{(-56)}^{(+57)} (3.20) (97) (1.40)$ & $7.80$ & ${}_{(-30)}^{(+31)} (2.50) (70) (19)$\\[4pt]
		\hline 
		\hline 
	\end{tabular*}
\end{table}
%%%%%%%%%%%%%%%%%%%%%%%%%%%%%%%%%%%%%%%%%%%%%%%%%%%%%%%%%%%%%%%%%%%%%%%%%%%%%%%%%%
\begin{table}[htb]
	\caption{Neutron-proton P-wave phase shifts in degrees as
          obtained  at N$^4$LO$^+$ for the cutoff
          $\Lambda = 450$~MeV as function of the laboratory energy
          $E_{\rm lab}$ (in MeV). For the notation see Table \ref{tab:npS}. \label{tab:npP}}
	\smallskip
	\begin{tabular*}{\textwidth}{@{\extracolsep{\fill}}rr@{\extracolsep{4pt}}l@{\extracolsep{\fill}}r@{\extracolsep{4pt}}l@{\extracolsep{\fill}}r@{\extracolsep{4pt}}l@{\extracolsep{\fill}}r@{\extracolsep{4pt}}l}
		\hline 
		\hline 
		\noalign{\smallskip}
	$E_{\rm lab}$ & \multicolumn{2}{l}{$^3$P$_0$} & \multicolumn{2}{l}{$^1$P$_1$} & \multicolumn{2}{l}{$^3$P$_1$} & \multicolumn{2}{l}{$^3$P$_2$}\\[2pt]
	\hline
	\hline
	\\[-9pt]
	1 & $0.18$ &  & $-0.19$ &  & $-0.11$ &  & $0.02$ & \\[4pt]
	5 & $1.65$ & ${}_{(-0)}^{(+0)} (0) (1) (0)$ & $-1.52$ & ${}_{(-0)}^{(+0)} (0) (2) (1)$ & $-0.92$ &  & $0.26$ & ${}_{(-0)}^{(+0)} (0) (1) (0)$\\[4pt]
	10 & $3.72$ & ${}_{(-1)}^{(+1)} (1) (1) (1)$ & $-3.11$ & ${}_{(-1)}^{(+1)} (1) (4) (2)$ & $-2.03$ & ${}_{(-0)}^{(+0)} (0) (1) (0)$ & $0.73$ & ${}_{(-0)}^{(+0)} (1) (2) (0)$\\[4pt]
	25 & $8.33$ & ${}_{(-3)}^{(+3)} (2) (1) (2)$ & $-6.48$ & ${}_{(-3)}^{(+3)} (2) (12) (8)$ & $-4.81$ & ${}_{(-1)}^{(+1)} (0) (2) (1)$ & $2.60$ & ${}_{(-1)}^{(+1)} (1) (5) (0)$\\[4pt]
	50 & $11.02$ & ${}_{(-5)}^{(+5)} (2) (4) (3)$ & $-9.92$ & ${}_{(-7)}^{(+7)} (4) (21) (16)$ & $-8.19$ & ${}_{(-2)}^{(+2)} (1) (1) (3)$ & $5.90$ & ${}_{(-1)}^{(+1)} (1) (3) (0)$\\[4pt]
	100 & $8.76$ & ${}_{(-7)}^{(+7)} (14) (11) (1)$ & $-14.66$ & ${}_{(-13)}^{(+14)} (16) (15) (26)$ & $-13.30$ & ${}_{(-3)}^{(+3)} (8) (5) (4)$ & $10.94$ & ${}_{(-2)}^{(+2)} (12) (11) (1)$\\[4pt]
	150 & $3.95$ & ${}_{(-8)}^{(+8)} (46) (9) (8)$ & $-18.56$ & ${}_{(-18)}^{(+19)} (45) (14) (19)$ & $-17.56$ & ${}_{(-5)}^{(+5)} (28) (5) (1)$ & $13.92$ & ${}_{(-3)}^{(+3)} (37) (13) (2)$\\[4pt]
	200 & $-1.02$ & ${}_{(-11)}^{(+11)} (1.00) (3) (18)$ & $-21.88$ & ${}_{(-24)}^{(+25)} (94) (56) (26)$ & $-21.24$ & ${}_{(-8)}^{(+9)} (59) (6) (4)$ & $15.63$ & ${}_{(-5)}^{(+5)} (75) (0) (4)$\\[4pt]
	250 & $-5.57$ & ${}_{(-16)}^{(+16)} (1.80) (23) (29)$ & $-24.60$ & ${}_{(-34)}^{(+36)} (1.70) (1.10) (43)$ & $-24.33$ & ${}_{(-15)}^{(+15)} (1.00) (27) (12)$ & $16.57$ & ${}_{(-7)}^{(+7)} (1.20) (23) (5)$\\[4pt]
	300 & $-9.53$ & ${}_{(-23)}^{(+23)} (2.90) (49) (42)$ & $-26.66$ & ${}_{(-50)}^{(+54)} (2.60) (1.60) (82)$ & $-26.81$ & ${}_{(-23)}^{(+24)} (1.50) (60) (23)$ & $17.01$ & ${}_{(-8)}^{(+8)} (1.70) (51) (6)$\\[4pt]
		\hline 
		\hline 
	\end{tabular*}
\end{table}
%%%%%%%%%%%%%%%%%%%%%%%%%%%%%%%%%%%%%%%%%%%%%%%%%%%%%%%%%%%%%%%%%%%%%%%%%%%%%%%%%%
\begin{table}[htb]
	\caption{Neutron-proton D-wave phase shifts in degrees at N$^4$LO$^+$ for the cutoff
          $\Lambda = 450$~MeV as function of the laboratory energy
          $E_{\rm lab}$ (in MeV). For the notation see Table \ref{tab:npS}. \label{tab:npD}}
	\smallskip
	\begin{tabular*}{\textwidth}{@{\extracolsep{\fill}}rr@{\extracolsep{4pt}}l@{\extracolsep{\fill}}r@{\extracolsep{4pt}}l@{\extracolsep{\fill}}r@{\extracolsep{4pt}}l@{\extracolsep{\fill}}r@{\extracolsep{4pt}}l}
		\hline 
		\hline 
		\noalign{\smallskip}
		$E_{\rm lab}$ & \multicolumn{2}{l}{$^3$D$_1$} & \multicolumn{2}{l}{$^1$D$_2$} & \multicolumn{2}{l}{$^3$D$_2$} & \multicolumn{2}{l}{$^3$D$_3$}\\[2pt]
		\hline
		\hline
		\\[-9pt]
		1 & $-0.01$ &  & $0.00$ &  & $0.01$ &  & $0.00$ & \\[4pt]
		5 & $-0.18$ &  & $0.04$ &  & $0.22$ &  & $0.00$ & \\[4pt]
		10 & $-0.68$ & ${}_{(-0)}^{(+0)} (0) (1) (0)$ & $0.16$ &  & $0.85$ &  & $0.00$ & \\[4pt]
		25 & $-2.82$ & ${}_{(-1)}^{(+1)} (0) (2) (1)$ & $0.68$ & ${}_{(-0)}^{(+0)} (0) (1) (0)$ & $3.72$ & ${}_{(-0)}^{(+0)} (0) (3) (0)$ & $0.03$ & ${}_{(-0)}^{(+0)} (1) (2) (1)$\\[4pt]
		50 & $-6.47$ & ${}_{(-2)}^{(+2)} (1) (2) (1)$ & $1.69$ & ${}_{(-0)}^{(+0)} (1) (3) (0)$ & $8.97$ & ${}_{(-1)}^{(+1)} (1) (12) (1)$ & $0.24$ & ${}_{(-1)}^{(+1)} (3) (7) (3)$\\[4pt]
		100 & $-12.28$ & ${}_{(-4)}^{(+4)} (4) (3) (2)$ & $3.79$ & ${}_{(-1)}^{(+1)} (7) (4) (1)$ & $17.23$ & ${}_{(-6)}^{(+6)} (8) (25) (4)$ & $1.20$ & ${}_{(-4)}^{(+4)} (16) (13) (14)$\\[4pt]
		150 & $-16.47$ & ${}_{(-8)}^{(+8)} (10) (8) (10)$ & $5.73$ & ${}_{(-2)}^{(+2)} (25) (3) (2)$ & $22.12$ & ${}_{(-10)}^{(+10)} (25) (20) (8)$ & $2.42$ & ${}_{(-8)}^{(+8)} (37) (7) (31)$\\[4pt]
		200 & $-19.57$ & ${}_{(-15)}^{(+16)} (20) (13) (22)$ & $7.33$ & ${}_{(-4)}^{(+4)} (56) (16) (3)$ & $24.65$ & ${}_{(-14)}^{(+14)} (50) (3) (11)$ & $3.56$ & ${}_{(-13)}^{(+13)} (64) (11) (51)$\\[4pt]
		250 & $-21.88$ & ${}_{(-24)}^{(+25)} (43) (16) (37)$ & $8.56$ & ${}_{(-5)}^{(+5)} (98) (31) (4)$ & $25.72$ & ${}_{(-17)}^{(+17)} (81) (16) (13)$ & $4.51$ & ${}_{(-17)}^{(+18)} (1.00) (36) (69)$\\[4pt]
		300 & $-23.57$ & ${}_{(-35)}^{(+38)} (82) (18) (55)$ & $9.47$ & ${}_{(-6)}^{(+6)} (1.50) (43) (5)$ & $25.94$ & ${}_{(-19)}^{(+19)} (1.20) (32) (15)$ & $5.22$ & ${}_{(-21)}^{(+22)} (1.40) (61) (85)$\\[4pt]
		\hline 
		\hline 
	\end{tabular*}
\end{table}
%%%%%%%%%%%%%%%%%%%%%%%%%%%%%%%%%%%%%%%%%%%%%%%%%%%%%%%%%%%%%%%%%%%%%%%%%%%%%%%%%%
\begin{table}[htb]
	\caption{Neutron-proton F-wave phase shifts in degrees as
          obtained at N$^4$LO$^+$ for the cutoff
          $\Lambda = 450$~MeV as function of the laboratory energy
          $E_{\rm lab}$ (in MeV). The second uncertainty corresponds to the  N$^5$LO truncation error.  For the remaining notation see Table \ref{tab:npS}. \label{tab:npF}}
	\smallskip
	\begin{tabular*}{\textwidth}{@{\extracolsep{\fill}}rr@{\extracolsep{4pt}}l@{\extracolsep{\fill}}r@{\extracolsep{4pt}}l@{\extracolsep{\fill}}r@{\extracolsep{4pt}}l@{\extracolsep{\fill}}r@{\extracolsep{4pt}}l}
		\hline 
		\hline 
		\noalign{\smallskip}
		$E_{\rm lab}$ & \multicolumn{2}{l}{$^3$F$_2$} & \multicolumn{2}{l}{$^1$F$_3$} & \multicolumn{2}{l}{$^3$F$_3$} & \multicolumn{2}{l}{$^3$F$_4$}\\[2pt]
		\hline
		\hline
		\\[-9pt]
		1 & $0.00$ &  & $0.00$ &  & $0.00$ &  & $0.00$ & \\[4pt]
		5 & $0.00$ &  & $-0.01$ &  & $0.00$ &  & $0.00$ & \\[4pt]
		10 & $0.01$ &  & $-0.07$ &  & $-0.03$ &  & $0.00$ & \\[4pt]
		25 & $0.09$ &  & $-0.42$ &  & $-0.20$ &  & $0.02$ & \\[4pt]
		50 & $0.31$ & ${}_{(-0)}^{(+0)} (0) (1) (0)$ & $-1.12$ & ${}_{(-0)}^{(+0)} (0) (1) (1)$ & $-0.61$ &  & $0.11$ & ${}_{(-0)}^{(+0)} (0) (1) (0)$\\[4pt]
		100 & $0.75$ & ${}_{(-1)}^{(+1)} (4) (2) (1)$ & $-2.18$ & ${}_{(-1)}^{(+1)} (3) (3) (5)$ & $-1.36$ & ${}_{(-1)}^{(+1)} (2) (2) (0)$ & $0.48$ & ${}_{(-0)}^{(+0)} (1) (2) (1)$\\[4pt]
		150 & $1.07$ & ${}_{(-2)}^{(+2)} (16) (3) (3)$ & $-2.86$ & ${}_{(-4)}^{(+4)} (12) (5) (17)$ & $-1.89$ & ${}_{(-2)}^{(+2)} (8) (2) (1)$ & $1.03$ & ${}_{(-1)}^{(+1)} (4) (3) (3)$\\[4pt]
		200 & $1.20$ & ${}_{(-4)}^{(+4)} (39) (2) (6)$ & $-3.36$ & ${}_{(-8)}^{(+8)} (31) (5) (38)$ & $-2.27$ & ${}_{(-5)}^{(+5)} (20) (1) (2)$ & $1.66$ & ${}_{(-3)}^{(+3)} (10) (1) (7)$\\[4pt]
		250 & $1.16$ & ${}_{(-7)}^{(+7)} (75) (1) (10)$ & $-3.80$ & ${}_{(-14)}^{(+14)} (61) (3) (66)$ & $-2.54$ & ${}_{(-9)}^{(+9)} (39) (7) (3)$ & $2.32$ & ${}_{(-5)}^{(+5)} (20) (4) (12)$\\[4pt]
		300 & $0.96$ & ${}_{(-11)}^{(+11)} (1.20) (6) (15)$ & $-4.21$ & ${}_{(-20)}^{(+21)} (1.00) (2) (1.00)$ & $-2.73$ & ${}_{(-14)}^{(+14)} (65) (15) (5)$ & $2.97$ & ${}_{(-7)}^{(+7)} (34) (12) (18)$\\[4pt]
		\hline 
		\hline 
	\end{tabular*}
\end{table}
%%%%%%%%%%%%%%%%%%%%%%%%%%%%%%%%%%%%%%%%%%%%%%%%%%%%%%%%%%%%%%%%%%%%%%%%%%%%%%%%%%
\begin{table}[htb]
	\caption{Neutron-proton mixing angles $\epsilon_1$,
          $\epsilon_2$ and $\epsilon_3$ as obtained at N$^4$LO$^+$ for the cutoff
          $\Lambda = 450$~MeV as function of the laboratory energy
          $E_{\rm lab}$ (in MeV). For the notation see Table \ref{tab:npS}. \label{tab:npEps}}
	\smallskip
	\begin{tabular*}{\textwidth}{@{\extracolsep{\fill}}rr@{\extracolsep{4pt}}l@{\extracolsep{\fill}}r@{\extracolsep{4pt}}l@{\extracolsep{\fill}}r@{\extracolsep{4pt}}l@{\extracolsep{\fill}}}
		\hline 
		\hline 
		\noalign{\smallskip}
		$E_{\rm lab}$ & \multicolumn{2}{l}{$\epsilon_1$} & \multicolumn{2}{l}{$\epsilon_2$} & \multicolumn{2}{l}{$\epsilon_3$}\\[2pt]
		\hline
		\hline
		\\[-9pt]
		1 & $0.11$ &  & $0.00$ &  & $0.00$ & \\[4pt]
		5 & $0.67$ & ${}_{(-1)}^{(+1)} (0) (2) (0)$ & $-0.05$ &  & $0.01$ & \\[4pt]
		10 & $1.16$ & ${}_{(-1)}^{(+1)} (1) (6) (1)$ & $-0.18$ &  & $0.08$ & \\[4pt]
		25 & $1.77$ & ${}_{(-3)}^{(+3)} (1) (14) (3)$ & $-0.75$ &  & $0.55$ & \\[4pt]
		50 & $2.05$ & ${}_{(-5)}^{(+5)} (3) (24) (5)$ & $-1.61$ &  & $1.61$ & \\[4pt]
		100 & $2.29$ & ${}_{(-7)}^{(+7)} (8) (31) (9)$ & $-2.55$ & ${}_{(-1)}^{(+1)} (3) (1) (0)$ & $3.46$ & ${}_{(-0)}^{(+0)} (1) (1) (0)$\\[4pt]
		150 & $2.54$ & ${}_{(-10)}^{(+10)} (14) (24) (11)$ & $-2.82$ & ${}_{(-2)}^{(+2)} (11) (3) (0)$ & $4.71$ & ${}_{(-0)}^{(+0)} (2) (2) (0)$\\[4pt]
		200 & $2.81$ & ${}_{(-15)}^{(+15)} (29) (6) (11)$ & $-2.76$ & ${}_{(-3)}^{(+3)} (21) (4) (1)$ & $5.47$ & ${}_{(-0)}^{(+0)} (6) (2) (1)$\\[4pt]
		250 & $3.06$ & ${}_{(-24)}^{(+24)} (52) (21) (14)$ & $-2.55$ & ${}_{(-4)}^{(+4)} (34) (3) (1)$ & $5.84$ & ${}_{(-1)}^{(+1)} (13) (1) (3)$\\[4pt]
		300 & $3.26$ & ${}_{(-36)}^{(+35)} (82) (53) (23)$ & $-2.27$ & ${}_{(-4)}^{(+4)} (47) (2) (1)$ & $5.93$ & ${}_{(-2)}^{(+2)} (24) (1) (6)$\\[4pt]
		\hline 
		\hline 
	\end{tabular*}
\end{table}
%%%%%%%%%%%%%%%%%%%%%%%%%%%%%%%%%%%%%%%%%%%%%%%%%%%%%%%%%%%%%%%%%%%%%%%%%%%%%%%%%%

In this appendix we collect the results for the pp and np phase shifts
and mixing angles from our partial wave analysis, which are given in
Tables \ref{tab:ppSP}--\ref{tab:ppF} and
\ref{tab:npS}--\ref{tab:npEps}, respectively. 
We restrict
ourselves to the highest considered order (N$^4$LO$^+$) and to the
intermediate cutoff $\Lambda = 450$~MeV which yields the best 
description of the scattering data from the 2013 Granada database
below $E_{\rm lab} = 300$~MeV.  Furthermore, we only show the results
for the channels involving contact interactions that have been fitted
to the data. Except for F-waves, the truncation uncertainties correspond to N$^4$LO and
have been estimated using the algorithm described in section
\ref{sec:ErrorTrunc} without relying on the explicit 
knowledge of the N$^4$LO$^+$ contributions. 
Motivated by the findings of Ref.~\cite{Furnstahl:2015rha}, see also \cite{Melendez:2017phj} for a
related discussion, we have assumed a slightly higher value for the
breakdown scale of $\Lambda_{\rm b} = 650$~MeV as compared with 
$\Lambda_{\rm b} = 600$~MeV adopted in Refs.~\cite{Epelbaum:2014efa,Epelbaum:2014sza}. Since our
results for F-waves take into account the order-$Q^6$ contact interactions, 
we show the N$^5$LO truncation error for the F-wave phase shifts. 

In order to verify the consistency of our approach to uncertainty
quantifications, we have performed several tests. Specifically,   
by regarding our predicted values for the phase shifts and mixing
angles from Tables \ref{tab:npS}--\ref{tab:npEps} along with the
corresponding statistical and truncation uncertainties as ``synthetic
data'', we have followed the approach similar to that of Ref.~\cite{Epelbaum:2014efa}
and calculated the $\tilde \chi^2 /{\rm datum}$ for their reproduction using
some of our potentials. The statistical and truncation errors 
listed in Tables \ref{tab:npS}--\ref{tab:npEps} were
then added in quadrature to define the corresponding ``synthetic error
bars''.  To avoid a possible confusion with the $\chi^2$
for the description of experimental data considered in the main part of the
manuscript, we use a symbol $\tilde \chi^2$ for such an
approach. Further, we restrict ourselves to np phase shifts and mixing
angles and to the
energies of $E_{\rm lab} = 1$, $5$, $10$, $25$, $50$, $100$, $150$,
$200$, $250$ and $300$~MeV. 

To test our estimated N$^4$LO truncation
errors in the S-, P- and D-waves and the mixing angles
$\epsilon_{1,2,3}$, we have calculated the corresponding $\tilde \chi^2 /{\rm
  datum}$ for the reproduction of the ``synthetic data'' 
using the N$^4$LO potential with the same cutoff value
of $\Lambda = 450$~MeV, which turns out to be $\tilde \chi^2 /{\rm
  datum} = 1.04$. Notice that the number of data is determined by the number of
channels and energies. To also verify the consistency of the estimated 
N$^5$LO truncation uncertainties in F-waves, we have used our
N$^4$LO$^+$ potentials for different cutoff choices. 
We find $\tilde \chi^2 /{\rm
  datum} = 0.70$ ($\tilde \chi^2 /{\rm
  datum} = 0.50$) for the cutoff choice of  $\Lambda = 400$~MeV
($\Lambda = 500$~MeV) for the
reproduction of our ``synthetic data'' in the  S-, P-, D and F-waves
and the mixing angles $\epsilon_{1,2,3}$. Notice that we do not
include the theoretical uncertainty
corresponding to these potentials when
calculating the $\tilde \chi^2$. For $\Lambda = 550$~MeV, we
find a larger value of $\tilde \chi^2 /{\rm datum} = 2.29$, which is
consistent with the larger truncation errors for this cutoff choice.
%not included in the calculation of the  $\tilde \chi^2$. 

We believe that the estimated errors provide a realistic account of
the theoretical uncertainty of our calculations. It should, however,
be emphasized that we do not estimate errors from the uncertainty in
the $\pi N$ coupling constant and its possible charge
dependence. Also our incomplete treatment of isospin-breaking
corrections may potentially affect the results for the phase shifts and mixing
angles at low energy.

%%%%%%%%%%%%%%%%%%%%%%%%%%%%%%%%%%%%%%%%%%%%%%%%%%%%%%%%%%%%%%%%%%%%%%%%%%%%%%%%%%

\end{document}